\documentclass[prd,amsmath,amssymb,twocolumn,tightenlines,floatfix,notitlepage,nofootinbib,preprintnumbers]{revtex4}

\usepackage{graphicx}
\usepackage{mathrsfs}

\newcommand{\nn}{\nonumber}
\newcommand{\bn}{{\bar n}}

\newcommand{\be}{\begin{equation}}
\newcommand{\ee}{\end{equation}}

\newcommand{\SCETplus}{\mbox{${\rm SCET}_{+}$}}
\newcommand{\SCETplusplus}{\mbox{${\rm SCET}_{++}$}}

\newcommand{\minus}{\!-\!}
\newcommand{\plus}{\!+\!}

\newcommand{\vect}[1]{\mathbf{#1}}
\newcommand{\abs}[1]{\left\lvert #1\right\rvert}
\newcommand{\bra}[1]{\left\langle #1\right\rvert}
\newcommand{\ket}[1]{\left\lvert #1\right\rangle}
\newcommand{\wt}{\widetilde}

\newcommand{\as}{\alpha_s}
\newcommand{\MSbar}{\overline{\text{MS}}}

\newcommand{\cu}{\mathrm{c}}
\newcommand{\e}{\mathrm{e}}

\newcommand{\GeV}{\text{ GeV}}

\newcommand{\eg}{\emph{e.g.},\ }
\newcommand{\incl}{\text{incl.}}
\newcommand{\alg}{\text{alg.}}
\newcommand{\cone}{\text{cone}}
\newcommand{\kT}{\text{k$_T$}}
\newcommand{\taumax}{\tau_{\text{max}}}

\newcommand{\eq}[1]{Eq.~\eqref{eq:#1}}
\newcommand{\eqs}[2]{Eqs.~\eqref{eq:#1} and \eqref{eq:#2}}
\newcommand{\eqss}[3]{Eqs.~\eqref{eq:#1}, \eqref{eq:#2}, and \eqref{eq:#3}}
\renewcommand{\sec}[1]{Sec.~\ref{sec:#1}}
\newcommand{\ssec}[1]{Sec.~\ref{ssec:#1}}

\newcommand{\appx}[1]{App.~\ref{app:#1}}
\newcommand{\fig}[1]{Fig.~\ref{fig:#1}}

\newcommand{\cO}{\mathcal{O}}

\newcommand{\cL}{\mathcal{L}}

\newcommand{\cK}{\mathcal{K}}

\allowdisplaybreaks[1]

\DeclareMathOperator{\Tr}{Tr}

\DeclareMathOperator{\Li}{Li}

\newcommand{\LP}{\mathscr{L}}
\newcommand{\iLP}{\mathscr{L}^{-1}}

\newcommand{\LANL}{Theoretical Division, Group T-2, MS B283, Los Alamos National Laboratory, P.O. Box 1663,
Los Alamos, NM 87545, USA}

\begin{document}

\title{A Soft-Collinear Mode for Jet Cross Sections in Soft Collinear Effective Theory}
\author{Yang-Ting Chien}
{\affiliation{\LANL}
\author{Andrew Hornig}
{\affiliation{\LANL}
\author{Christopher Lee}
{\affiliation{\LANL}
\date{December 11, 2015}
\preprint{LA-UR-15-26983}
\preprint{INT-PUB-15-060}

\begin{abstract}
We propose the addition of a new ``soft-collinear'' mode to soft collinear effective theory (SCET) below the usual soft scale to factorize and resum logarithms of jet radii $R$ in jet cross sections. We consider exclusive 2-jet cross sections in $e^+e^-$ collisions with an energy veto $\Lambda$ on additional jets. The key observation is that there are actually two pairs of energy scales whose ratio is $R$: the transverse momentum $QR$ of the energetic particles inside jets and their total energy $Q$, and the transverse momentum $\Lambda R$ of soft particles that are cut \emph{out} of the jet cones and their energy $\Lambda$. The soft-collinear mode is necessary to factorize and resum logarithms of the latter hierarchy. We show how this factorization occurs in the jet thrust cross section for cone and $k_T$-type algorithms at $\cO(\as)$ and using the thrust cone algorithm at $\cO(\as^2)$. We identify the presence of hard-collinear, in-jet soft, global (veto) soft, and soft-collinear modes in the jet thrust cross section. We also observe here that the in-jet soft modes measured with thrust are actually the ``csoft'' modes of the theory \SCETplus. We dub the new theory with both csoft and soft-collinear modes ``\SCETplusplus''.  We go on to explain the relation between the ``unmeasured'' jet function appearing in total exclusive jet cross sections and the hard-collinear and csoft functions in measured jet thrust cross sections. We do not resum logs that are non-global in origin, arising from the ratio of the scales of soft radiation whose thrust is measured at $Q\tau/R$ and of the soft-collinear radiation at $2\Lambda R$. Their resummation would require the introduction of additional operators beyond those we consider here. The steps we outline here are a necessary part of summing logs of $R$ that are global in nature and have not been factorized and resummed beyond leading-log level previously.
\end{abstract}

\maketitle

\section{Introduction}
\label{sec:intro}

Hadronic jets play a crucial role in testing the dynamics of the strong interaction described by Quantum Chromodynamics (QCD), probing the constituents of protons and other hadrons, and  searching for signatures of physics beyond the Standard Model. Reliable predictions of jet cross sections require advanced tools in QCD, such as factorization of hard, collinear, and soft dynamics \cite{Sterman:1995fz}, the resummation of large logarithms of separated energy scale ratios in perturbation theory \cite{Luisoni:2015xha}, and the accounting of nonperturbative effects, identifying universal properties when possible \cite{Lee:2006nr}.

Many jet observables have been resummed to very high accuracy, especially with the advent of effective field theory (EFT) techniques applicable to jets with soft collinear effective theory (SCET) \cite{Bauer:2000ew,Bauer:2000yr,Bauer:2001ct,Bauer:2001yt,Bauer:2002nz}. Two-jet event shapes in $e^+e^-$ collisions \cite{Dasgupta:2003iq} such as thrust, hemisphere jet masses, and $C$-parameter have now been resummed to next-to-next-to-next-to-leading logarithmic (N$^3$LL) accuracy \cite{Becher:2008cf,Abbate:2010xh,Chien:2010kc,Hoang:2014wka}, with a plethora of other observables now resummed to NNLL accuracy. The resummation of jet observables dependent on the definition of an algorithm have also been studied, such as angularity jet shapes \cite{Ellis:2009wj,Ellis:2010rw,Larkoski:2014uqa}, jet masses \cite{Dasgupta:2012hg,Chien:2012ur,Jouttenus:2013hs,Liu:2014oog} and other jet substructure observables \cite{Larkoski:2015zka,Larkoski:2015kga}. One class of logarithms that have not yet been fully resummed are logs of the angular size $R$ of jets in the definition of these algorithms. Fixed-order dependence on $R$ in resummed cross sections in another variable has been computed, but a generic procedure to resum logs of $R$ themselves has not yet been given. Ref.~\cite{Dasgupta:2014yra} has recently applied generating functional methods to resum leading logs of $R$ in QCD in several jet observables. The logs of $R$ in jet cross sections and the jet shape $\Psi(r/R)$ \cite{Ellis:1991vr,Ellis:1992qq} were partially resummed in a soft approximation in \cite{Seymour:1997kj}, and the resummation of the jet shape was also explored recently using EFT techniques \cite{Chien:2014nsa}.

In this paper we explore steps toward an EFT-based method to factorize and resum logs of jet radii in jet thrust and exclusive jet cross sections. We consider specifically two-jet thrust and total cross sections in $e^+e^-$ collisions. For example, with the Sterman-Weinberg (S-W) jet algorithm \cite{Sterman:1977wj}, one obtains the total 2-jet cross section to $\cO(\as)$,
\begin{equation}
\label{eq:SW1loop}
\frac{\sigma_2^{\text{SW}}}{ \sigma_0} = 1 + \frac{\alpha_s C_F}{\pi}\left( -4\ln 2\beta\ln\delta - 3\ln\delta - \frac{\pi^2}{3} + \frac{5}{2}\right)\,,
\end{equation}
where $\delta$ is the half-angle of each cone and $\beta$ is the fraction of total final-state energy lying outside the cones. Because S-W cones are not fixed to be exactly back-to-back, at $\cO(\as)$ this is equivalent to the $k_T$-type recombination algorithms \cite{Catani:1991hj,Catani:1993hr,Ellis:1993tq,Dokshitzer:1997in,Cacciari:2008gp}.
For a fixed cone algorithm (\eg \cite{Salam:2007xv}) with cones of half-angle $r$ and energy veto $\Lambda$ on particles outside, one obtains instead
\begin{equation}
\label{eq:cone1loop}
\begin{split}
\frac{\sigma_2^{\text{cone}}}{ \sigma_0} = 1 + \frac{\alpha_s C_F}{\pi}\biggr(\! \minus 4\ln\frac{2\Lambda}{Q}\ln R - 3\ln R  - \frac{1}{2} + 3\ln 2\biggr),
\end{split}
\end{equation}
where $R \equiv \tan(r/2)$. In both \eqs{SW1loop}{cone1loop}, $\sigma_0$ is the Born cross section. Below we will identify $\beta = \Lambda/Q$ and $\delta = R$ in discussing the two classes of algorithms. Meanwhile, for a measurement of jet thrust, defined in \eq{jetthrust}, the $\cO(\as)$ cross section integrated from 0 to $\tau$ is given by
\begin{align}
\label{eq:jetthrust1loop}
\frac{1}{\sigma_0}\sigma_\cu^\alg(\tau) &=  1 + \frac{\as C_F}{2\pi} \Bigl[ - 2 \ln^2\tau - 3\ln\tau \\
&\quad - 8\ln R \ln\frac{2\Lambda R}{Q\tau} - 1 \Bigr]+ 2\Delta\sigma^\alg_\cu(\tau) \,, \nn
\end{align}
for $0<\tau<R^2$ for fixed cone algorithm and  $0<\tau<R^2/4$ for S-W or \kT. $\Delta\sigma^\alg_c$ is a nonsingular contribution given in \eqs{Deltacone}{DeltakT}.

It has been a challenge to resum the logarithms in cross sections like \eqss{SW1loop}{cone1loop}{jetthrust1loop} because it has not been clear how to relate the logs of $R$ to a hierarchy of energy scales in an EFT. Numerous groups have now applied SCET to the study of jet cross sections dependent on a jet algorithm, \eg \cite{Cheung:2009sg,Ellis:2009wj,Ellis:2010rw,Trott:2006bk,Kelley:2011aa,vonManteuffel:2013vja,Chay:2015ila,Chay:2015dva}. None of these have yet achieved a complete factorization of the logs of $R$ in \eqs{SW1loop}{cone1loop}, as they make use of a single soft mode to describe the radiation with energy at the scale $\Lambda = Q\beta$. Ref.~\cite{Ellis:2010rw} did introduce the notion of refactorizing the soft function for a cross section with multiple measurements on jets (such as different jet masses along with the jet veto) but stopped short of identifying different modes for soft particles carrying the jet veto energy.

In this paper we show that the logs of $R$ can only be factorized completely if an additional mode is added to SCET below the ordinary soft scale, namely, a \emph{soft-collinear} mode, with soft energy but collinear in its directionality, with light-cone momenta $p_{sc} = (n\cdot p,\bn\cdot p, p_\perp)$ scaling as  $\sim \Lambda(1,R^2,R)$ or $\Lambda(R^2,1,R)$. We show the appearance of this mode in the one- and two-loop contributions to the jet thrust and total 2-jet cross sections and how it leads to more complete separation of the scale dependence at these orders, thus making possible resummation of a larger set of logs of $R$ than before.

We do not deal here with the factorization and resummation of logs that are non-global in nature \cite{Dasgupta:2001sh}, that is, from correlated soft emissions into regions measured with disparate scales, \eg the jet thrust $Q\tau$ and the veto energy $\Lambda$. The scale $R$ also enters these ratios, as found in \cite{Banfi:2010pa,Hornig:2011tg,Kelley:2011aa,vonManteuffel:2013vja}. Recent breakthrough progress has been made in factorizing and resumming these observables in the framework of SCET \cite{Larkoski:2015zka,Larkoski:2015kga,Neill:2015nya} through the realization that to factorize and resum non-global logs (NGLs), more differential measurements of subjets, from which soft radiation can be emitted, must first be made. New modes and scales associated with the boundary of the full jet appear and play a crucial role in the resummation of the NGLs. Similar ideas have just recently been outlined in \cite{Becher:2015hka} in the context of some of the same two-jet cross sections we consider here, also identifying the soft-collinear mode (with the less mellifluous name ``coft'') and the necessity of additional operators accounting for multiple eikonal directions within jets that can emit this radiation.

Resummation of the NGLs and thus a full resummation of all logs containing dependence on the algorithm parameters $\Lambda$ and $R$, and a full realization of the power of the soft-collinear mode, will indeed require treatment of emissions from multiple collinear sources within jets. Such a treatment, being pursued by several groups, is one we leave out of the scope of the present paper. We deal here with all the issues of implementation of the soft-collinear mode at the first step of the full treatment, namely, with one eikonal source within the full jets. We identify those logs that can newly be factorized and resummed at this level with the introduction of this mode. We study both the jet thrust and total 2-jet cross sections. In the process we tie together several dangling ideas in previous literature:
\begin{itemize}
\item The soft mode for radiation inside jets measured with a thrust-like variable is in fact the collinear-soft (csoft) mode (not to be confused with soft-collinear!) introduced in \cite{Bauer:2011uc} in the context of multi-jet cross sections in which two of the jets approach each other in angle, increasing the virtuality of the exchanged soft radiation due to increased collinearity. This theory of SCET plus csoft modes was called \SCETplus. We point out here that the same thing happens to soft radiation contributing to thrust when confined to a cone. The increase in this soft scale by a factor $1/R$ was noted in \cite{Ellis:2010rw}; the identification of the in-jet soft mode with the csoft mode of  \cite{Bauer:2011uc} is new. The theory of \SCETplus plus our soft-collinear modes in order to factor logs of the ratio between $\Lambda$ and $\Lambda R$ is then called \SCETplusplus.

\item The ``unmeasured'' jet function introduced in \cite{Ellis:2009wj,Ellis:2010rw} for collinear radiation in a jet not measured with any other variable (as occurs in total jet cross sections) can be constructed from a convolution of the measured jet function and the in-jet measured soft function in the jet thrust or other measured jet cross section. In this paper we determine that hard-collinear modes at the scale $Q\sqrt{\tau}$ and the in-jet measured soft modes at the scale $Q\tau/R$ become one and the same mode when $\tau$ is integrated up to its maximum value in the jet cone, $R^2$ for a cone algorithm. This explains an observation in \cite{Chay:2015ila} about how to reshuffle terms between the $\cO(\as)$ measured jet and soft functions so that they integrate up to the unmeasured jet function. Our observations generalize all of these previous computations of unmeasured jet functions at $\cO(\as)$ and procedures relating them to measured jet functions to $\cO(\as^2)$ and beyond.
    This can be viewed as the ``0th'' step in a procedure to resum logs in jet cross sections by first identifying subjets with some kind of substructure measurement as in \cite{Larkoski:2015kga,Larkoski:2015zka,Neill:2015nya}.

\item Previously it has been noted that logs of $R$ in soft functions dependent on a jet veto \cite{Ellis:2009wj,Ellis:2010rw,Hornig:2011tg,Kelley:2011aa,vonManteuffel:2013vja} appear to be proportional to the cusp anomalous dimension \cite{Korchemsky:1987wg}. While the circumstantial evidence from fixed-order computations was compelling, justification was lacking. By the introduction of the soft-collinear mode, we find that both the global soft and soft-collinear functions contain a $\Gamma_{\text{cusp}}\ln(\mu/(2\Lambda))$ and  $\Gamma_{\text{cusp}}\ln(\mu/(2\Lambda R))$ pieces which combine to give $\Gamma_{\text{cusp}} \ln R$.  Because the individual anomalous dimensions are $\ln \mu$-dependent we can conclude that their coefficient, and thus that of the $\ln R$ in the unfactorized veto soft function, really is the cusp anomalous dimension, establishing the validity of the earlier conjectures.

\item We identify and resum all those logs of $R$ not associated with non-global radiation, distinguishing those coming from the hard-collinear scale $QR$ and the soft-collinear scale $2\Lambda R$. This is a necessary step to deal with global logs of $R$ before resumming also the NGLs.

\item The physical origin of the argument, $Q\tau/(2\Lambda R^2)$, of the NGLs in the jet thrust cross section now becomes clear as the ratio of the csoft scale of measured soft radiation $Q\tau/R$ and the scale of soft-collinear radiation $2\Lambda R$, the only modes that are sensitive the the jet boundary.
\end{itemize}

The necessity of additional modes such as csoft and soft-collinear modes can be traced to the measurement of multiple observables that set the collinearity of jets or softness of radiation emitted therefrom, in this case, the jet thrust $\tau$, the jet radius $R$, and soft veto energy $\Lambda$. Some modes are sensitive to only one of these measurements, such as the hard-collinear modes to $\tau$. Others are sensitive to more than one measurement at the same time, such as csoft to $\tau$ and $R$, or soft-collinear to $\Lambda$ and $R$. The emergence of modes simultaneously sensitive to multiple measurements has been explored in studies of multi-differential cross sections, \eg \cite{Bauer:2011uc,Larkoski:2015zka,Procura:2014cba}. Hard-collinear modes sensitive to $R$ were introduced in \cite{Ellis:2009wj,Ellis:2010rw}. Modes sensitive both to a soft veto $\Lambda$ and to $R$, however, are introduced here for the first time, as is the connection between csoft modes and $R$.

Particularly for recombination algorithms such as the $\kT$-type algorithms, jet cross sections begin to exhibit ``clustering logs'' \cite{Hornig:2011tg,Kelley:2012kj,Kelley:2012zs}  at $\cO(\as^2)$, even in the Abelian part, that spoil predictions of higher-order logs based on renormalization-group (RG) evolution of lower-order logs. Thus the generality of our proposed ideas will depend on the choice of algorithm. We leave a discussion of these sorts of logs out of the scope of our treatment here. At $\cO(\as^2)$ we will stick to the thrust cone algorithm as in   \cite{Kelley:2011aa,vonManteuffel:2013vja,Becher:2015hka}, which exhibits good factorization properties at least to $\cO(\as^2)$. As we have emphasized, the introduction of the soft-collinear mode is a necessary, though not always sufficient, step towards resummation of logs of $R$. We will find it already illuminates several outstanding issues, especially those outlined above.

In \sec{hierarchy} we will explore the appearance of multiple soft scales  at $\cO(\as)$ in jet cross sections and introduce the soft-collinear mode. In \sec{phasespace} we will dissect the phase space in the $\cO(\as)$ 2-jet cross sections and construct the separate global soft and soft-collinear functions at this order.  In \sec{integrate} we will review the $\cO(\as)$ jet thrust cross section and integrate it to obtain the 2-jet cross section, leading to insight into how to relate factorization theorems for the two cross sections to higher orders. In this context the complete EFT of \SCETplusplus\ makes its first appearance. In \sec{S2} we will construct the form that the $\cO(\as^2)$ soft function for the jet thrust cross section must take if it factorizes into csoft, soft and soft-collinear pieces as we propose, and compare this form to its explicit computation in \cite{vonManteuffel:2013vja}. We find the latter result  does indeed factor into pieces dependent on the global soft and soft-collinear scales, and we extract the anomalous dimensions of the individual csoft, soft and soft-collinear factors to $\cO(\as^2)$ and even $\cO(\as^3)$. In \sec{twoloop} we obtain the $\cO(\as^2)$ jet thrust cross section implied by our refactorized formula, and relate the measured jet and in-jet measured soft functions to the unmeasured jet function, obtaining its form and anomalous dimension to $\cO(\as^2)$. We also obtain the terms in the 2-jet rate that can be computed from integrating the jet thrust distribution with one collinear source of soft radiation per jet and compare to the recent prediction of \cite{Becher:2015hka}. In \sec{EERAD} we compare some of the terms in the $\cO(\as^2)$ 2-jet cross section with the numerical predictions of the program {\tt EERAD3} \cite{Ridder:2014wza}.  In \sec{NNLL} we present a formula for the resummed jet thrust distribution using the ingredients we extracted earlier. In \sec{results} we collect in summary form the main new results of the paper for easy reference. In \sec{summary} we conclude. In the Appendices we provide a number of technical formulae and known results that we need in the main part of our discussion.

\section{Hierarchy of Hierarchies}
\label{sec:hierarchy}

The jet and jet thrust cross sections in \eqss{SW1loop}{cone1loop}{jetthrust1loop} differ in one important respect from something like the global thrust or hemisphere jet mass distribution, which exhibit double and single logs of the single measurement parameter $\tau$ or $m/Q$. For thrust and jet mass, both collinear and soft radiation are measured with the same parameter, \eg $\tau$, and $\tau$ acts as a physical regulator on both the soft and collinear divergences in QCD amplitudes, generating logs of $\tau$. Factorizing logs of $\tau$, then, involves identifying  which ones come from collinear divergences and which ones from soft.  Instead, in the jet rates  \eqs{SW1loop}{cone1loop}, the collinear and soft divergences are regulated by \emph{different} parameters, in principle completely independent from each other. The soft divergences are regulated by the parameter $\Lambda$, while collinear divergences are regulated by $R$. These parameters control the amount of phase space around the divergent regions contributing to the cross section. Thus, while collinear and soft modes relevant for a global thrust or jet mass cross section are connected by $\tau$, \eg $p_n \sim Q(1,\tau,\sqrt{\tau})$ and $p_s\sim Q(\tau,\tau,\tau)$, giving the usual relation $\mu_c^2 = Q\mu_s$ amongst hard, collinear, and soft scales, the collinear and soft modes for the jet rates \eqs{SW1loop}{cone1loop} are in principle decoupled in their scaling.

Normally, to describe a jet cross section in an EFT, we integrate out of QCD hard modes of momenta
\begin{equation}
p_h \sim (Q,Q,Q)
\end{equation}
and match at a scale $\mu_H = Q$ onto a theory (SCET) of energetic collinear modes and soft modes. For 2-jet cross sections \eqs{SW1loop}{cone1loop}, these modes have light-cone momenta scaling as
\begin{equation}
\label{eq:SCET}
\begin{split}
p_n &\sim Q(1,R^2,R) \\
\text{SCET:}\quad p_{\bn} &\sim Q(R^2,1,R) \\
p_s &\sim (\Lambda,\Lambda,\Lambda).
\end{split}
\end{equation}
Trying to factor a jet thrust or two-jet cross-section like \eqss{SW1loop}{cone1loop}{jetthrust1loop} into jet and soft functions using this separation of scales yields a soft function containing logs of multiple scales that cannot be minimized by a single choice of soft factorization scale $\mu_S$ \cite{Cheung:2009sg,Ellis:2010rw}. From \cite{Ellis:2010rw}, one obtains the prediction
\be
\label{eq:naivefactorization}
\frac{\sigma_2^{\alg}}{\sigma_0} = H(Q^2,\mu) J_{\text{un}}^\alg (QR,\mu)^2S_{\text{veto}}^\cu(\Lambda,R,\mu)\,.
\ee
Here $H = \abs{C}^2$ is the hard function, built from the Wilson coefficient $C$ appearing in the matching of the QCD current $j^\mu_{\text{QCD}} = \bar q\gamma^\mu q$ on the SCET 2-jet operator $\cO_2^\mu = \bar\chi_n\gamma_\perp^\mu\chi_\bn$, through the relation $\langle j^\mu_{\text{QCD}}\rangle = C_2(\mu)\langle \cO_2\rangle (\mu)$, with the expectation value computed in some suitable external state overlapping with the operator \cite{Bauer:2002nz,Bauer:2003di,Manohar:2003vb}. The functions $J_{\text{un}}^\alg$ containing effects of the hard collinear radiation inside jets were dubbed ``unmeasured jet functions'' in \cite{Ellis:2009wj,Ellis:2010rw} to distinguish them from more usual jet functions in which some property like the jet mass is probed, \eg \cite{Bauer:2003pi,Hornig:2009vb}. The soft function $S_{\text{veto}}^\cu$ (the $^\cu$ indicates integrated, or cumulative, distribution in energy $E_{\text{out}}$ outside the cones) contains the effects of soft radiation from jets outside the cones in the veto region.

At one loop, the individual hard \cite{Bauer:2003pi}, jet, and soft \cite{Ellis:2010rw} functions in \eq{naivefactorization} are given by,
\begin{subequations}
\label{eq:HJS1}
\begin{align}
\label{eq:H1}
H(Q^2,\mu) &= 1 + \frac{\as}{4\pi} \Bigl( \Gamma_H^0 \ln^2 \frac{\mu}{Q} + \gamma_H^0 \ln\frac{\mu}{Q} + c_H^1\Bigr) \\
\label{eq:J1}
J_{\text{un.}}^\alg(QR,\mu) &= 1 + \frac{\as}{4\pi}\Bigl( \Gamma_J^0 \ln^2 \!\!\frac{\mu}{QR} \plus \gamma_J^0 \ln\!\frac{\mu}{QR} \plus c_{J\text{un}}^{1\alg}\Bigr)  \\
\label{eq:S1}
S_{\text{veto}}^\cu(\Lambda,R,\mu) &= 1 + \frac{\as}{4\pi}\Bigl( \Gamma_S^0 \ln R \ln\frac{\mu^2}{4\Lambda^2R} + c_S^1\Bigr)\,,
\end{align}
\end{subequations}
where $\as \equiv \as(\mu)$, and $\Gamma_F^0,\gamma_F^0$ are the 1-loop coefficients of the cusp and non-cusp parts of the anomalous dimensions, and $c_F^1$ are constants. The cusp parts of the anomalous dimensions are proportional to the universal cusp anomalous dimension to all orders in $\as$, which has the expansion given in \eq{gammaexpansion} with coefficients given up to three loops by \eq{cuspcoeffs}. The cusp coefficients in \eq{HJS1} are proportional to the cusp anomalous dimension in \eq{gammaexpansion} according to:
\be
\Gamma_H = -2\Gamma_{\text{cusp}}\,,\quad  \Gamma_{J\text{un}} = \Gamma_{\text{cusp}}\,,\quad \Gamma_S^0 = 2\Gamma_{\text{cusp}}^0\,.
\ee
In the form \eq{S1}, we do not actually know that the coefficient $\Gamma_S^0$ of the double log is proportional to the cusp anomalous dimension beyond $\cO(\as)$: it multiplies only a single log of $\mu$ and thus technically is part of the non-cusp anomalous dimension in the form \eq{S1}. It has previously been conjectured to be proportional to the actual $\Gamma_{\text{cusp}}$ to all orders \cite{Ellis:2010rw}, and confirmed to be so to $\cO(\as^2)$ by explicit computation  \cite{Kelley:2011aa,vonManteuffel:2013vja}. We will show below that the introduction of the soft-collinear mode enables its identification as the cusp anomalous dimension.

The non-cusp parts of the anomalous dimensions are given by the expansion in \eq{gammaexpansion}, and have coefficients for the hard and jet functions given by \eqs{gammaH}{gammaJ}.
Meanwhile, the constant $c_{J\text{un}}^{1\alg}$ depends on the particular algorithm chosen and was computed for cone and $k_T$/Sterman-Weinberg algorithms in \cite{Ellis:2010rw},
\be
\label{eq:cJalg}
\begin{split}
c_{J\text{un}}^{1\text{cone}} &=\Bigl( 7 + 6\ln 2 - \frac{5\pi^2}{6}\Bigr)C_F \,, \\
c_{J\text{un}}^{1\text{S-W}} &= \Bigl( 13 - \frac{3\pi^2}{2}\Bigr)C_F\,.
\end{split}
\ee
The hard and soft constants are given by
\be
c_H^1 = \Bigl( \frac{7\pi^2}{3} - 16\Bigr) C_F \,,\quad c_S^1= -\frac{2\pi^2}{3} C_F\,.
\ee

In \eqs{H1}{J1}, it is evident that the dependence on the hard scale $Q$ and the jet scale $QR$ have been properly isolated, and logs in the fixed-order hard or jet functions can be minimized by choosing $\mu=Q$ or $\mu=QR$, respectively. In the soft function in \eq{S1}, it may appear that the scale $\mu = 2\Lambda\sqrt{R}$ could be chosen to do the same (as suggested in \cite{Ellis:2010rw}). However, this choice, as we will see, does not succeed in minimizing logs at $\cO(\as^2)$ and beyond, and the anomalous dimension of $S_{\text{veto}}^c$ in \eq{S1} still contains an undesired large $\ln R$:
\be
\gamma_S \equiv \frac{d\ln S_{\text{veto}}^c}{d\ln\mu} = \frac{\alpha_s}{2\pi} \Gamma_S^0 \ln R.\,,
\ee
to $\cO(\as)$.
As we will show, there is actually dependence on \emph{two} scales, $2\Lambda$ and $2\Lambda R$, which must be separated out to factor the soft function $S_{\text{veto}}^c$ into pieces each dependent on a single scale.

We introduce a mode in the soft sector, with collinear scaling with respect to the soft energy $\Lambda$:
\begin{equation}
\label{eq:SCETR}
\begin{split}
p_s &\sim \Lambda(1,1,1) \\
p_{sc}^n &\sim \Lambda(1,R^2,R) \\
p_{sc}^\bn &\sim \Lambda(R^2,1,R)
\end{split}
\end{equation}
This mimics the scaling of the hard-collinear modes in the original SCET \eq{SCET}. Once the hard and hard-collinear modes are integrated out, one is left with a theory of soft modes, obeying the full QCD Lagrangian \cite{Bauer:2000yr,Bauer:2001yt}, so it is as if one is starting over from the full theory, but at a lower energy scale and one where soft gluons are sourced by Wilson lines instead of physical quarks and gluons. Within this soft QCD, one repeats the matching onto collinear modes with respect to the lower energy scale. The second theory is a ``soft SCET,'' and we name the new modes \emph{soft-collinear} modes.

Again, the soft-collinear modes in \eq{SCETR} are to be distinguished from the ``csoft'' modes of \cite{Bauer:2011uc}, which are soft modes whose virtuality is \emph{higher} than the global soft scale because the \emph{small} light-cone component is fixed to be the same as the soft while the other momentum components are forced to have collinear scaling. In \cite{Bauer:2011uc} this happened when two collinear jets whose combined invariant mass is measured grow close to one another in angle, causing the exchanged soft radiation to inherit a degree of collinearity. Below, we will see they also arise when the soft radiation from a jet whose small light-cone momentum is measured (with thrust or mass) is confined to be inside a cone. In contrast,  in \eq{SCETR}, the soft-collinear radiation lives at a scale \emph{below} the global soft scale, because its \emph{large} light-cone component of momentum is fixed by measurement of the energy $k^0 = (k^++k^-)/2$ to be equal to the soft scale $\Lambda$, while the other components scale collinearly.

The Lagrangian for the soft-collinear modes looks identical to that for the hard-collinear modes in the first SCET. They do in principle couple to a set of softer soft modes, \eg with scaling $\Lambda (R^2,R^2,R^2)$ in all components, which can be decoupled from the $sc$ modes by the BPS field redefinition \cite{Bauer:2001yt}, but such modes do not contribute anything to the jet cross sections we consider (we introduce no measurement sensitive to this scale).

\begin{figure*}[t]
\vspace{-5mm}
\begin{center}
\includegraphics[width=.65\columnwidth]{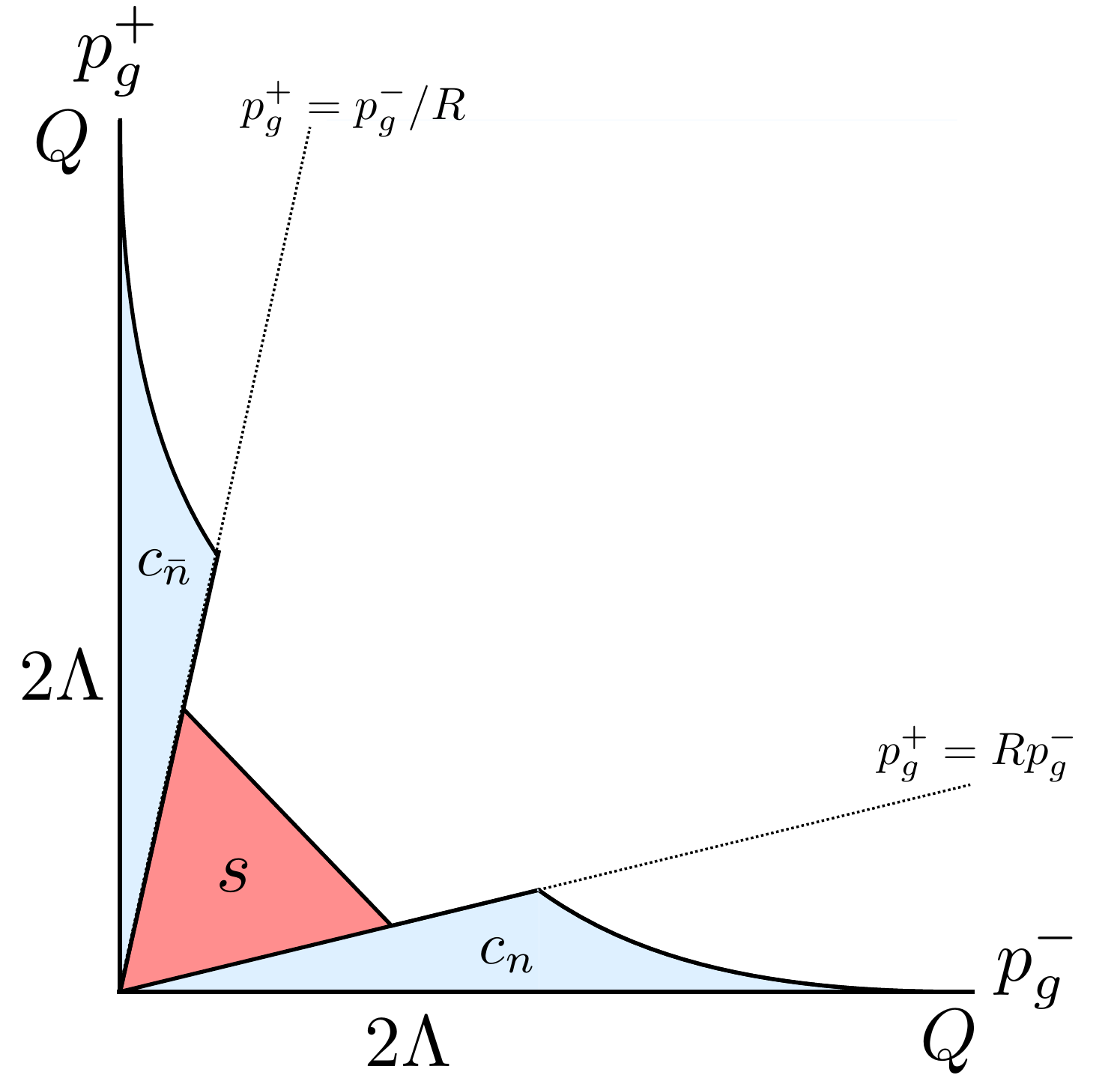} \quad \includegraphics[width=.65\columnwidth]{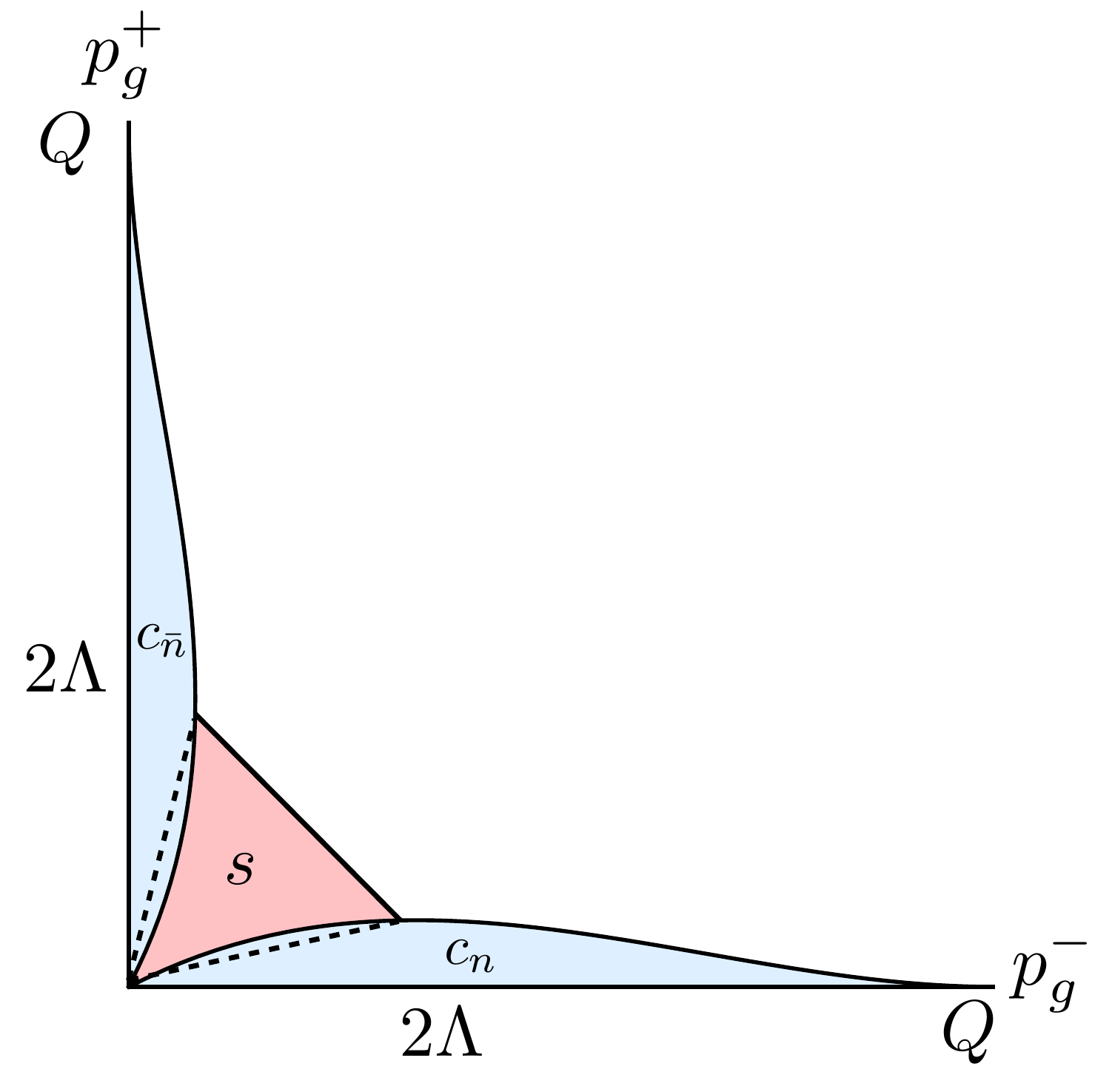}
\end{center}
\vspace{-5mm}
\caption{Phase space for cone and \kT/Sterman-Weinberg jets. The phase space for a collinear (blue) or soft (red) gluon emitted from a quark and antiquark in the computation of the $\cO(\as)$ 2-jet rate in the cone and \kT-type/Sterman-Weinberg algorithms are shown.}
\label{fig:regions}
\end{figure*}

\section{Phase Space and EFT Modes}
\label{sec:phasespace}

\begin{figure*}[t]
\vspace{-5mm}
\begin{center}
\includegraphics[width=1.7\columnwidth]{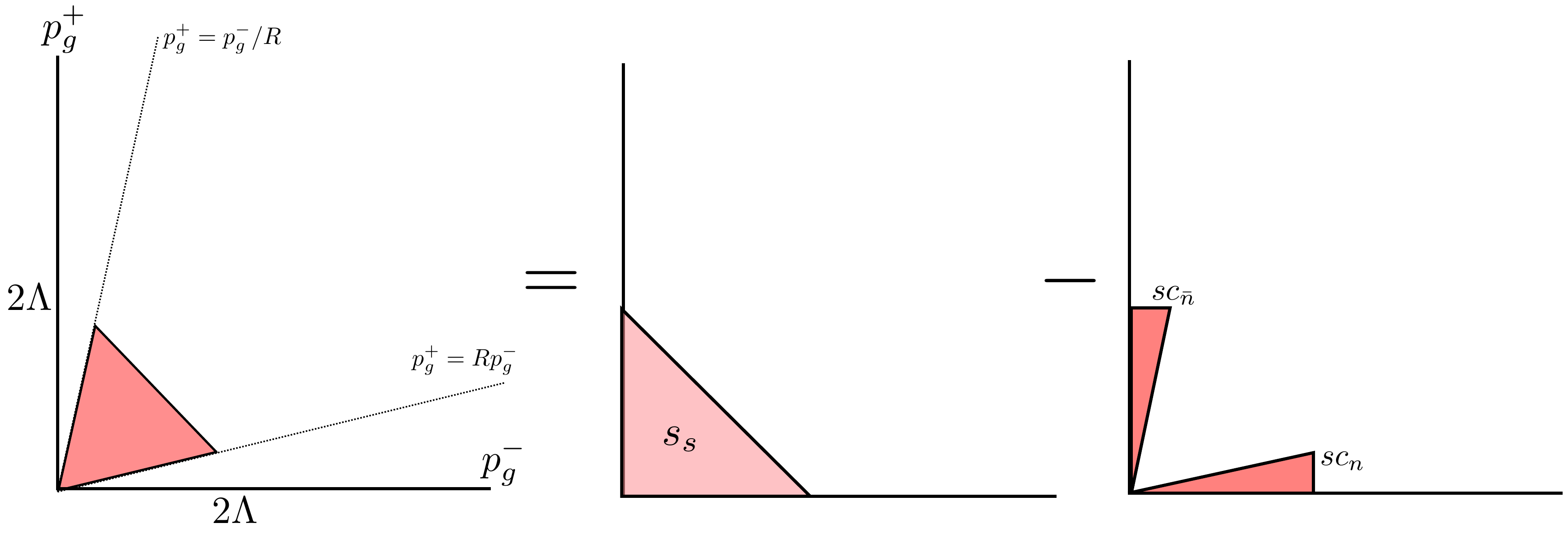}
\end{center}
\vspace{-5mm}
\caption{Soft phase space for one gluon. The phase space for one soft gluon emission is the same for both cone and \kT-type/Sterman-Weinberg algorithms. The original soft phase space on the left covers the region outside both jets where radiation is vetoed by the energy cut $E_g<\Lambda$. This region is actually sensitive to two distinct physical scales. On the right-hand side, this region is re-expressed in terms of region sensitive to one physical scale at a time. The purely soft (or ``global'') region covers all angles and is sensitive to the scale $2\Lambda$. The ``soft-collinear'' regions cover gluons of energy $\Lambda$ within the jet cones of angle $R$ and are sensitive to the scale $2\Lambda R$. }
\label{fig:soft}
\end{figure*}

In this section we consider the scaling of phase space constraints for the cone and $k_T$/Sterman-Weinberg algorithms at $\cO(\as)$ and match the full QCD calculation onto a theory of the collinear, soft, and soft-collinear modes defined above. These constraints and their collinear and soft scaling for several algorithms were extensively studied in \cite{Cheung:2009sg}.

The two-jet rate in QCD can be computed from
\be
\label{eq:QCDcs}
\begin{split}
\sigma_{\text{2-jet}}(\Lambda,R) &= \frac{1}{2Q^2} \sum_X \abs{L_\mu\bra{X}j^\mu\ket{0}}^2 \Theta_\alg(X) \\
&\quad \times (2\pi)^4\delta^4(q-p_X)\,,
\end{split}
\ee
where $j^\mu = \bar q\gamma^\mu q$ is the quark current, $L_\mu$ is the leptonic part of the amplitude (see, \eg \cite{Fleming:2007xt,Bauer:2008dt}), $p_X$ is the momentum of the final state and $q$ is the $e^+e^-$ total momentum, in the CM frame, $q = (Q,\vect{0})$, and $\Theta_\alg$ imposes the phase space constraints for the chosen jet algorithm. For a final state with hadrons $i$ of momentum $k_i$,
\begin{align}
\Theta_\cone(X) &= \sum_{i\in X}\bigl[ \theta(\theta_{i\vect{\hat n}}<r) +\theta(\theta_{i\vect{\hat n}}>\pi-r)  \\
& \quad  + \theta(r < \theta_{i\vect{\hat n}} <\pi- r)   \theta(k^0 < \Lambda) \bigr]\,, \nn
\end{align}
where $\theta_{i\vect{\hat n}}$ is the angle between the particle $i$ and the cone axis $\vect{\hat n}$ (determined, \eg by the thrust axis). The two theta functions allows particles to have any energy if inside the cone of radius $r$, and to be limited by the jet veto energy $\Lambda$ if outside. For the $k_T$ or Sterman-Weinberg algorithms, the theta functions are more complicated, having to account for different possible recombinations of particles. We will write their specific form for a definite number of collinear or soft particles below.

To $\cO(\as)$, the cross section \eq{QCDcs}  receives contributions from purely virtual, hard graphs with a $q\bar q$ final state, and from the collinear and soft gluon limits of the $q\bar q g$ final states, up to corrections suppressed by $R$ or $\Lambda/Q$,
\be
\label{eq:sigmajet1}
\frac{\sigma_{\text{2-jet}}(\Lambda,R)}{\sigma_0} = 1 + \frac{\alpha_s}{4\pi} (H^{(1)} + 2J^{(1)} + S^{(1)})\,.
\ee
$H^{(1)}$ comes from purely virtual diagrams for $q\bar q$ final states, and corresponds to the squared modulus of the one-loop part of the matching coefficient of $j^\mu$ onto the SCET two-jet operator $\cO_2^\mu$ \cite{Bauer:2002nz,Bauer:2003di,Manohar:2003vb}. The result is given by \eq{H1}.

$J^{(1)}$ and $S^{(1)}$ come from the collinear and soft limit of the QCD amplitudes and phase space constraints for one-gluon emission. These amplitudes and phase space constraints are given, \eg in \cite{Ellis:2010rw}. They are given by matrix elements of collinear and soft operators in SCET. For example, the soft function in \eq{naivefactorization} is given by \cite{Ellis:2010rw}
\be
\label{eq:Svetocdef}
S^\cu_{\text{veto}}(\Lambda,R,\mu) = \frac{1}{N_C}\!\Tr\sum_{X_s}\! \abs{\bra{X_s}T[Y_n Y_\bn]\ket{0}}^2\Theta_\alg^s\!(X_s)\,,
\ee
where $Y_{n,\bn}$ are Wilson lines of soft gluons \cite{Bauer:2001yt}, and $\Theta_\alg^s$ are the algorithm-imposed constraints in the soft limit for each final state $X_s$. We give its 1-particle form below.

The phase space constraints on a $q\bar q g$ final state in the limit of a gluon collinear to the $n$-collinear quark is
\be
\begin{split}
\label{eq:ThetaQ}
\Theta_\cone^n& = \theta\Bigl(R^2 - \frac{p_g^+}{p_g^-}\Bigr)\theta\Bigl(R^2 - \frac{p_q^+}{p_q^-}\Bigr) \,,
\end{split}
\ee
and similarly for $\Theta_\cone^\bn$. We recall $R=\tan(r/2)$.
After integrating over the momentum conserving delta functions in \eq{QCDcs}, these theta functions translate into a single set of constraints on the gluon momentum,
\be
\Theta_\cone^n = \theta\Bigl(R^2 - \frac{p_g^+}{p_g^-}\Bigr)\theta\Bigl(R^2 -  \frac{p_g^- p_g^+}{(Q-p_g^-)^2}\Bigr) \,,
\ee
and similarly for $\Theta_\cone^\bn$.
For the Sterman-Weinberg or \kT ~algorithms, the collinear gluon phase space is given by \cite{Ellis:2010rw},
\be
\Theta_\kT^n = \theta\Bigl(R^2 - \frac{Q^2 p_g^+}{p_g^-(Q-p_g^-)^2}\Bigr)\,,
\ee
and similarly for $\Theta_\kT^\bn$.
These regions are illustrated in blue in \fig{regions}, and  give the contribution $J^{(1)}$ in \eq{sigmajet1}, given by \eq{J1}.

Technically, the collinear gluon phase space also includes the region outside the jet cones/regions, where its energy would be capped by the veto energy $\Lambda$. However, this double counts the region of phase space covered by the soft gluon, and must be subtracted out \cite{Manohar:2006nz}. The contribution of this region of the collinear phase space was shown to cancel against this subtraction in \cite{Ellis:2010rw}. Thus it is not drawn in \fig{regions}. In addition, such a soft/zero-bin subtraction must also be made from the blue regions; however, these give scaleless integrals in dimensional regularization (DR), and so we do not explicitly include them. They are necessary, however, in properly interpreting any $1/\epsilon$ divergences in DR as being of IR or UV origin (in addition to virtual diagram contributions) \cite{Manohar:2006nz,Cheung:2009sg,Hornig:2009kv,Chay:2015ila}. For jet functions with finite $R$ and an additional measurement such as jet mass, the zero-bin subtractions are nonzero even in DR and are essential to obtain correct results \cite{Jouttenus:2009ns,Ellis:2010rw}. We refer the reader to these references for appropriate discussion.

In the limit that the gluon is soft, $p_g^0\sim \Lambda \ll Q$, the emitting quark or antiquark determines the jet axis and thus automatically lies inside the jet, and these constraints reduce to
\begin{align}
\label{eq:Thetas}
\Theta_\cone^s &= \theta(\theta_{qg} > r)\theta(\theta_{\bar q g}>r)\theta(p_g^0 < \Lambda)
\end{align}
Technically there is also a term for soft gluons inside the jets with no energy constraint, but in the soft limit the integrals over the soft amplitude are scaleless and zero in DR.  This region is illustrated in red in \fig{regions} and gives the contribution $S^{(1)}$ in \eq{S1}.

Going back to the soft phase space given by \eq{Thetas} and examining \eq{S1} and \fig{regions}, we see that this phase space is still sensitive to two distinct physical scales. We separate these out in \fig{soft}. The first scale is, of course, the soft jet veto energy $\Lambda$. The other, however, is the soft-collinear scale $\Lambda R$. Although the cross section does not contain any particles of this energy scale inside the jet cones, it is still sensitive to such modes because we have taken them \emph{out} of the phase space. This is evident in \fig{soft}, and is made transparent by rewriting the theta function in \eq{Thetas}:
\be
\label{eq:Thetasc}
\Theta_\cone^s = \theta(p_g^0 < \Lambda) [1 - \theta(\theta_{qg} < r) - \theta(\theta_{\bar qg} < r)]\,.
\ee
Each of the three terms, represented by the $s_s$ and $sc_{n,\bn}$ regions in \fig{soft} is now sensitive to a single physical scale. 
The contributions sum to give
\be
\label{eq:S1refact}
S^{(1)} = S_s^{(1)} + 2S_{sc}^{(1)}\,,
\ee
corresponding to the three terms in \eq{Thetasc}.

Here we can give explicit results for the $\cO(\as)$ contributions in \eq{S1refact}. We will consider them at $\cO(\as^2)$ in \sec{S2}.
The full veto-dependent soft function in \eq{Svetocfact} was computed in Ref.~\cite{Ellis:2010rw} from matrix elements of the Wilson lines in \eq{Svetocdef}, integrating over the real gluon phase space in \eq{Thetasc}. The computation there was organized in such a way that we can easily read off the results corresponding to the individual global soft and soft-collinear pieces in \eq{S1refact}, whose all-orders operator definitions we give in  \eqs{Ssdef}{Sscdef}. Working in $d=4-2\epsilon$ dimensions to regulate the divergences, in the $\MSbar$ scheme, we find that
\begin{subequations}
\label{eq:BareSoftResults}
\begin{align}
S_s^{(1), \rm{bare}} &= \frac{\alpha_s C_F}{\pi} \frac{e^{\gamma_E \epsilon}}{\Gamma(1-\epsilon)}\Big(\frac{\mu}{2\Lambda }\Big)^{\! 2\epsilon} \bigg(\frac{1}{\epsilon^2} -\frac{\pi^2}{6}\bigg) \\
S_{sc}^{(1), \rm{bare}} &= -\frac{\alpha_s C_F}{2 \pi} \frac{e^{\gamma_E \epsilon}}{\Gamma(1-\epsilon)}\Big(\frac{\mu}{2\Lambda R}\Big)^{\!2 \epsilon} \frac{1}{\epsilon^2}
\,.\end{align}
\end{subequations}
Arguments similar to those in Refs.~\cite{Hornig:2009kv,Hornig:2009vb} can be used to identify these poles as being of UV origin. Each component can be renormalized separately, and we find for the renormalized functions
\begin{subequations}
\label{eq:SS1SSc1}
\begin{align}
S_s^{(1)} &= \Gamma_S^0 \ln^2\frac{\mu}{2\Lambda} + c_{ss}^1 \\
S_{sc}^{(1)} &= \Gamma_{sc}^0 \ln^2 \frac{\mu}{2\Lambda R} + c_{sc}^1\,,
\end{align}
\end{subequations}
where
\be
\begin{split}
\label{eq:SsSsc1coeffs}
\Gamma_{sc}^0 &= -\frac{1}{2}\Gamma_S^0 = -\Gamma_{\text{cusp}}^0\,, \\
c_{ss}^1 &= -\pi^2 C_F \,,\quad c_{sc}^1 = \frac{\pi^2}{6} C_F\,.
\end{split}
\ee
Now the dependence of the cross section on the separate scales $2\Lambda$ and $2\Lambda R$ has been properly separated.

The phase space division in \eq{Thetasc} suggests the all-orders generalizations of the purely soft, or global soft, function and soft-collinear functions for veto-scale radiation:
\be
\label{eq:Ssdef}
S_s(E,\mu) =  \frac{1}{N_C}\sum_{X_s} \abs{\bra{X_s}T[Y_n^s Y_\bn^s]\ket{0}}^2\delta\Bigl(E - \sum_{i\in X_s} E_i\Bigr)\,,
\ee
where $Y_{n,\bn}^s$ are Wilson lines of the soft gluons scaling as $p_s$ in \eq{SCETR}, and $X_s$ are final states in the soft mode Hilbert space. The phase space constraint \eq{Thetas} for these soft modes reduces to unity since they cannot resolve the small angle $R$. The global soft function \eq{Ssdef} with a measurement of the total energy is equivalent to the timelike soft function that appears in Drell-Yan, \eg in \cite{Becher:2007ty}.

Meanwhile, the soft-collinear function is given by
\be
\label{eq:Sscdef}
\begin{split}
S_{sc}^n(E,R,\mu) &=  \frac{1}{N_C}\sum_{X_{sc}} \abs{\bra{X_{sc}}T[Y_n^{sc} Y_\bn^{sc}]\ket{0}}^2 \\
&\quad \times \delta\biggl(E - \sum_{i\in X_{sc}} \frac{\bn\cdot p_i}{2} \theta(\theta_{i\vect{\hat n}} > r)\biggr)\,,
\end{split}
\ee
where $Y_{n,\bn}^{sc}$ are Wilson lines of soft-collinear gluons in \eq{SCETR}, $X_{sc}$ are final states in the ($n$) soft-collinear Hilbert space, and similarly for the $\bn$ soft-collinear function (which is equal to $S_{sc}^n$). For these modes, the soft-collinear gluons in the $n$ direction only resolve the angle with respect to the $n$ jet axis; the phase space is the complement of the $sc_n$ triangle in \fig{soft}. Similarly the $\bn$-soft-collinear modes resolve only the angle with respect to the $\bn$ jet axis. Integrating a soft-collinear gluon over all angles gives a scaleless integral in DR, so we know the contribution of one soft-collinear gluon in the region $\theta_{i\vect{\hat n}}>r$ is minus the contribution over $\theta_{i\vect{\hat n}}<r$, hence the minus signs in \eq{Thetasc}.

The total veto soft function is given by
\begin{align}
\label{eq:Svetocfact}
&S_{\text{veto}}^\cu(\Lambda,R,\mu) = \int_0^\Lambda \! dE\!\! \int \! dE_s dE_n dE_\bn \delta( E \minus E_s \minus E_n \minus E_\bn) \nn \\
&\qquad\quad\times S_s(E_s,\mu) S_{sc}^n(E_n,R,\mu) S_{sc}^\bn(E_\bn,R,\mu)\,,
\end{align}
which in this form is cumulative (integrated) in $E$ (up to $\Lambda$).
The division of this soft function into contributions from global soft radiation and soft radiation confined to the cones, and their corresponding scale dependence, was anticipated in the breakup of its calculation at $\cO(\as)$ in \cite{Ellis:2010rw}. However the identification of each piece as the soft function for a different mode and RG running to two different scales $2\Lambda$ and $2\Lambda R$ was not made there.

\section{Integrating the One-Loop Jet Thrust Distribution}
\label{sec:integrate}

Although the one-loop cross sections \eqs{SW1loop}{cone1loop} are well known, we can cross check them by integrating the jet thrust distribution. This exercise will also lend us insight into a way to cross check the two-loop prediction for the jet rate from our calculations against known results for the two-loop jet thrust distribution, which we will carry out in \sec{S2}.

The jet thrust distribution identifies two jets with an algorithm just as for the total 2-jet rates \eqs{SW1loop}{cone1loop}, but makes an additional differential measurement of the jet thrust, one of the jet angularity shapes defined in \cite{Ellis:2010rw}. The jet thrust is defined by
\be
\label{eq:jetthrust}
\tau(J) = \frac{1}{2E_J} \sum_{i\in J} \abs{\vect{p}_T^i} \e^{-\eta_i}\,,
\ee
where $J$ is a jet found by the chosen algorithm, the sum is over particles $i$ within the jet, $E_J$ is the total jet energy, and the transverse momentum $\vect{p}_T$ and (pseudo)rapidity $\eta_i$ are measured with respect to the jet axis. We will consider both the double-differential distribution in the thrust of the two separate jets, $\tau_{1,2}$, and the single differential distribution in the total thrust $\tau=\tau_1+\tau_2$. Following the derivation in \cite{Ellis:2010rw}, we find the leading power contributions to the jet thrust distribution can be computed from the factorized cross section,
\begin{align}
\label{eq:jetthrustcs}
&\frac{1}{\sigma_0}\frac{d\sigma(\Lambda,R)}{d\tau_1 d\tau_2} = H(Q^2,\mu) \! \int\!  dt_n dt_\bn dk_n dk_\bn \\
&\qquad\times \delta\Bigl(\tau_1 - \frac{t_n}{Q^2} - \frac{k_n}{Q}\Bigr)\delta\Bigl(\tau_2 - \frac{t_\bn}{Q^2} - \frac{k_\bn}{Q}\Bigr) \nn \\
&\qquad \times J_n^\alg(t_n,R,\mu)J_\bn^\alg(t_\bn,R,\mu)S(k_n,k_\bn,\Lambda,R,\mu)\,, \nn
\end{align}
where $J_{n,\bn}^\alg$ are algorithm-dependent jet shape functions \cite{Ellis:2010rw,Jouttenus:2009ns} and $S$ is the jet shape soft function, equal to at least $\cO(\as)$ for cone and \kT-type algorithms. The total jet thrust distribution is then given by
\be
\label{eq:jetthrustsum}
\frac{1}{\sigma_0}\frac{d\sigma(\Lambda,R)}{d\tau}  = \frac{1}{\sigma_0}\int d\tau_1d\tau_2 \delta(\tau-\tau_1-\tau_2)\frac{d\sigma(\Lambda,R)}{d\tau_1 d\tau_2} \,.
\ee

The measurement of jet thrust in \eqs{jetthrustcs}{jetthrustsum} induces sensitivity to a different set of collinear and soft modes than those in \eqs{SCET}{SCETR}. The hard modes are integrated out the same way to give the hard function $H$ in \eq{jetthrustcs}, but the EFT we match onto is SCET with an extra set of ``collinear-soft'' (csoft) modes, a theory that was dubbed ``\SCETplus'' \cite{Bauer:2011uc}:
\begin{align}
\label{eq:csscaling}
p_{cn} & \sim Q(\tau,1,\sqrt{\tau}) \,, \quad p_{c\bn} \sim Q(1,\tau,\sqrt{\tau})\,, \nn \\
\text{SCET}_+:\quad p_{cs}  & \sim Q\Bigl(\tau,\frac{\tau}{R^2},\frac{\tau}R\Bigr)\text{ or }Q\Bigl(\frac{\tau}{R^2},\tau,\frac{\tau}R\Bigr) \,, \nn \\
p_s^\Lambda &\sim \Lambda(1,1,1)\,.
\end{align}
The csoft scale $Q\tau/R$ can be identified from the explicit computation of the soft function, given to $\cO(\as)$ in \cite{Ellis:2010rw} and below in \eq{softkalg}. Physically, it arises because the measured soft radiation with small light-cone component $\sim Q\tau$ inside the jet cone is being confined to angular region of size $R$, increasing its collinearity and virtuality---in the global thrust cross section, the soft scale would just be $Q\tau$. Confining such radiation to a cone then requires the large light-cone component to be $\sim Q\tau/R^2$ and $p_\perp\sim Q\tau/R$, as in \eq{csscaling}. The rescaling of the measured soft scale by $1/R$ was identified in \cite{Ellis:2010rw}, and here we assert further that they are in fact the csoft modes in \cite{Bauer:2011uc}, with an overall soft energy $\sim Q\tau$ but with relative collinear scaling of the components in \eq{csscaling}. In \cite{Bauer:2011uc} csoft modes arise because
 soft radiation is exchanged between two collinear jets whose angular separation grows small. Here they arise because the measured soft radiation from one jet is itself being confined to a smaller cone.

\begin{figure*}[t]
\begin{center}
\includegraphics[width=1.75\columnwidth]{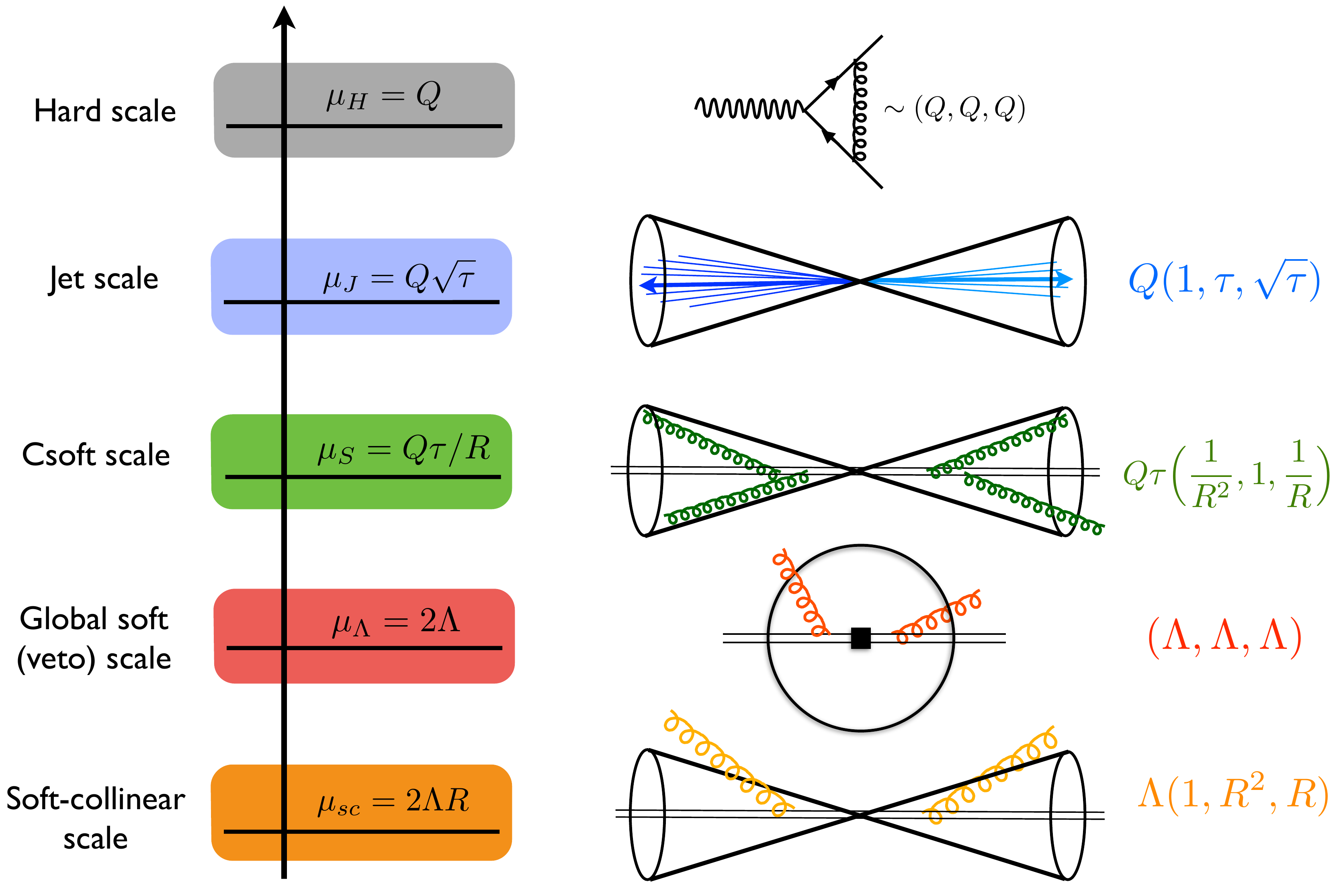}
\end{center}
\vspace{-5mm}
\caption{Scales in \SCETplusplus\ for the jet thrust cross section. The scaling of the light-cone  components of momentum $(p^\pm,p^\mp,p_\perp)$ for each mode is shown. Hard virtual modes of scale $Q$ are integrated out. The jet scale is the same as the global thrust distribution; the usual soft scale is increased by $1/R$ due to the restriction of measured soft radiation to a cone of radius $R$, like the csoft mode of \cite{Bauer:2011uc}. The soft veto on the energy $\Lambda$ of additional jets induces a global soft veto mode that cannot resolve the angle $R$ as well as the soft-collinear modes that can, see \eq{SCETR}. The csoft scale here could also be below the soft veto and/or soft-collinear scales.}
\label{fig:allscales}
\end{figure*}

The jet thrust cross section exhibits dependence, of course, on the (veto) soft and soft-collinear scales in \eq{SCETR}. We will deal with the refactorization of the soft function into pieces dependent on one of these scales at a time in the next section. The whole hierarchy of relevant scales is illustrated in \fig{allscales}. The complete EFT with hard-collinear, csoft, global (veto) soft, and soft-collinear modes we dub \SCETplusplus:
\begin{align}
\label{eq:SCETplusplus}
p_{cn} & \sim Q(\tau,1,\sqrt{\tau}) \,, \quad p_{c\bn} \sim Q(1,\tau,\sqrt{\tau})\,, \nn \\
\text{\SCETplusplus}:\quad p_{cs}  & \sim Q\Bigl(\tau,\frac{\tau}{R^2},\frac{\tau}R\Bigr)\text{ or }Q\Bigl(\frac{\tau}{R^2},\tau,\frac{\tau}R\Bigr) \,, \nn \\
p_s^\Lambda &\sim \Lambda(1,1,1)\,, \nn \\
p_{sc} &\sim \Lambda(R^2,1,R)\text{ or } \Lambda(1,R^2,R)\,.
\end{align}

To one-loop order, the jet functions $J_n^\alg = J_\bn^\alg$ in \eq{jetthrustcs} are given by
\begin{align}
\label{eq:Jalg}
J_n^\alg (t_n,R,\mu) = J^{\text{incl}}(t_n,\mu) + \Delta J^\alg(t_n,R),
\end{align}
where $J^{\text{incl}}$ is the usual inclusive jet function \cite{Bauer:2003pi,Bosch:2004th},
\begin{align}
&J^{\text{incl}}(t,\mu) = \delta(t)\Bigl[ 1 + \frac{\as C_F}{4\pi} ( 7-\pi^2)\Bigr] \\
&\quad+ \frac{\as C_F}{4\pi}\biggl\{ - \frac{3}{\mu^2}\Bigl [ \frac{\theta(t)\mu^2}{t}\Bigr]_+ + \frac{4}{\mu^2}\Bigl[ \frac{\theta(t)\ln(t/\mu^2)}{t/\mu^2}\Bigr]_+\biggr\} \,, \nn
\end{align}
where the plus distributions are defined in \appx{plus}, and $\Delta J^\alg$ is an additional contribution dependent on the algorithm. These were computed in \cite{Jouttenus:2009ns,Ellis:2010rw} for cone algorithms and \cite{Ellis:2010rw} for \kT/Sterman-Weinberg algorithms, with the result:
\be
\label{eq:DeltaJcone}
\begin{split}
\Delta J^\cone(t,R) &= \frac{\as C_F}{4\pi} \biggl[ \theta(t)\theta(Q^2 R^2 - t) \frac{6}{t + Q^2 R^2} \\
&\quad+ \frac{\theta(t - Q^2 R^2)}{t}\Bigl( 4 \ln\frac{ t}{Q^2R^2} + 3\Bigr)\biggr]\,,
\end{split}
\ee
and,
\be
\label{eq:DeltaJkT}
\begin{split}
\Delta J^\kT (t,R) &= \frac{\as C_F}{4\pi} \biggl\{ \frac{\theta(t)\theta\bigl(\frac{Q^2 R^2}{4} - t \bigr)}{t} \\
&\quad\times \Bigl[ 6x_1 + 4\ln\Bigl(\frac{1-x_1}{x_1}\frac{t}{Q^2 R^2}\Bigr)\Bigr]  \\
&\quad+ \frac{\theta\bigl(t - \frac{Q^2 R^2}{4}\bigr)}{t}\Bigl( 4 \ln\frac{ t}{Q^2R^2} + 3\Bigr)\biggr]\,,
\end{split}
\ee
where
\be
x_1 = \frac{1}{2} \biggl( 1 - \sqrt{1 - \frac{4t}{Q^2 R^2}} \ \biggr)\,.
\ee

Strictly to leading order in $t/(QR)^2$, that is, $\tau\ll R^2$ in \eq{jetthrustcs}, the algorithm-dependent corrections \eqs{DeltaJcone}{DeltaJkT} are power suppressed relative to $J^\incl$. Thus in that limit, as in \cite{Ellis:2010rw}, the jet thrust cross section is independent of the algorithm to $\cO(\as)$. However, to obtain the total 2-jet rate later, we will integrate the cross section up to kinematically maximum allowed value of $\taumax = R^2$ (cone) or $\taumax=R^2/4$ (\kT), where $\Delta J^\alg$ is no longer power suppressed. Thus we keep the contributions of both $J^\incl$ and $\Delta J^\alg$ in what follows.

The soft function in \eq{jetthrustcs}, meanwhile, receives contributions from soft gluons emitted within the jets and contributing to the jet thrust, and those outside and encountering the jet veto $\Lambda$ but not contributing to $\tau$. Putting together these different contributions at $\cO(\as)$ from \cite{Ellis:2010rw} for two measured jets in the final state, we obtain
\begin{align}
\label{eq:softkalg}
&S(k_n,k_\bn,\Lambda,R,\mu) = \delta(k) \Bigl[ 1 + \frac{\as C_F}{4\pi} \Bigl(4\ln R \ln \frac{\mu^2}{4\Lambda^2 R} \nn\\
&\quad - \frac{\pi^2}{3}\Bigr)\Bigr]   - \sum_{i=n,\bn}\frac{2\as C_F}{\pi} \frac{1}{\mu R}\biggl[ \frac{\theta(k_i)\mu R}{k_i}\ln\frac{k_i}{\mu R}\biggr]_+\,,
\end{align}
which in this form is differential in $k_{n,\bn}$ but cumulative (integrated up to $\Lambda$) in the energy veto.
Note that the parts of $\Delta J^\alg$ in \eqs{DeltaJcone}{DeltaJkT} for $t>Q^2\taumax$ cancel the total contribution of $J^\incl$ and $S$ in \eq{jetthrustcs} above $\tau=\taumax$ \cite{Ellis:2010rw}.

Putting together the hard function \eq{H1}, the jet function \eq{Jalg} including the contributions \eqs{DeltaJcone}{DeltaJkT}, and the soft function \eq{softkalg}, the prediction of \eqs{jetthrustcs}{jetthrustsum} for the (total) jet thrust cross section to $\cO(\as)$, presented in integrated form, is
\begin{align}
\label{eq:jetthrustint}
\sigma_\cu^\alg(\tau) &= \frac{1}{\sigma_0}\int_0^\tau d\tau' \frac{d\sigma(\Lambda,R)}{d\tau'} \\
&= \theta(\tau)\theta(\taumax- \tau) \biggl\{ 1 + \frac{\as C_F}{2\pi} \Bigl[ - 2 \ln^2\tau - 3\ln\tau \nn \\
&\quad - 8\ln R \ln\frac{2\Lambda R}{Q\tau} - 1 \Bigr]+ 2\Delta\sigma^\alg_c(\tau) \biggr\} \nn \\
&\quad + \theta(\tau-\taumax)\sigma_\cu^\alg(\taumax)\,, \nn
\end{align}
where $\Delta\sigma_\cu^\alg$ is given by
\be
\Delta\sigma_\cu^\alg(\tau) = \theta(\taumax-\tau)\int_0^{Q^2\tau}\!\! dt \,  \Delta J^\alg(t,R)\,.
\ee
As noted above, the contribution of $\Delta J^\alg(\tau)$ above $\tau=\taumax$ cancels against the sum of the inclusive jet function and the soft function contributions, so the integrated distribution plateaus at its constant value at $\tau=\taumax$. The final term $\sigma_c^\alg(\taumax)$ in \eq{jetthrustint} is then simply the expression in braces for $\tau<\taumax$ evaluated at $\tau=\taumax$. For the cone and \kT/S-W algorithms, the integrals of the algorithm-dependent contributions \eqs{DeltaJcone}{DeltaJkT} are given by
\be
\label{eq:Deltacone}
\Delta \sigma^\cone_\cu(\tau) = \theta(R^2 - \tau) \frac{\as C_F}{4\pi}6 \ln \frac{\tau+ R^2}{R^2}\,,
\ee
and
\begin{align}
\label{eq:DeltakT}
&\Delta\sigma^\kT_\cu(\tau) = \theta\Bigl(\frac{R^2}{4} - \tau\Bigr) \frac{\as C_F}{4\pi} \biggl\{6\biggl[ 1 - \sqrt{1 - \frac{4\tau}{R^2}} \\
&\quad + \ln\biggl( \frac{1}{2} + \frac{1}{2}\sqrt{1 - \frac{4\tau}{R^2}}\biggr) \biggr] + 4\ln^2\biggl(\frac{1}{2} + \frac{1}{2}\sqrt{1 - \frac{4\tau}{R^2}}\biggr) \nn \\
&\quad - 8\Li_2 \biggl(\frac{1}{2} - \frac{1}{2}\sqrt{1 - \frac{4\tau}{R^2}}\biggr)\biggr\}\,. \nn
\end{align}
At the maximum values $\tau=\taumax$, these simplify to
\begin{align}
\Delta\sigma^\cone_\cu(\tau=R^2) &= \frac{\as C_F}{4\pi} 6\ln 2 \\
\Delta\sigma^\kT_\cu\Bigl(\tau=\frac{R^2}{4}\Bigr) &= \frac{\as C_F}{4\pi} \Bigl(6  - \frac{2\pi^2}{3} - 6\ln 2 + 8\ln^2 2\Bigr)\,. \nn
\end{align}
Thus the predictions of \eq{jetthrustint} at $\tau=\taumax$ are
\begin{align}
\label{eq:conecs1}
\sigma_\cu^\cone(\tau \!=\! R^2) = 1 + \frac{\as C_F}{2\pi} \Bigl( &- 8\ln R\ln\frac{2\Lambda}{Q} - 6\ln R \nn \\
& + 6\ln 2 -1\Bigr)\,,
\end{align}
and
\begin{align}
\sigma_\cu^\kT\Bigl(\tau\!=\! \frac{R^2}{4}\Bigr) = 1 + \frac{\as C_F}{2\pi} \Bigl( &- 8\ln R\ln\frac{2\Lambda}{Q} - 6\ln R \nn \\
& + 5 - \frac{2\pi^2}{3}\Bigr)\,,
\end{align}
agreeing with \eqs{SW1loop}{cone1loop}.

Note that the differential jet functions $J^\alg(t,R,\mu)$ when integrated up to $t = Q^2\taumax$ do not by themselves reproduce the unmeasured jet functions $J(R,\mu)$ in \eq{J1}---the coefficient of the double log differs. However, after combining  $J_{n,\bn}^\alg(t_{n,\bn},R,\mu)$ with $S(k_n,k_\bn,\Lambda,R)$ through \eq{jetthrustcs}, the total $\tau$ distribution integrated up to $\tau=\taumax$ does equal the total 2-jet rate \eq{SW1loop} or \eq{cone1loop}. This is because the wide-angle radiation in a measured jet with jet thrust $\tau$ belongs in the csoft sector in \eq{jetthrustcs}---it cannot become very energetic and preserve a small value of $\tau$ for the jet---while in an unmeasured jet, hard collinear radiation can go all the way up to the angle $R$ (or $R/2$) and remain in the allowed collinear jet phase space. This difference was noted in \cite{Ellis:2010rw}. Thus some of the logs appear in a different sector---csoft or hard-collinear---depending on whether $\tau$ is measured or not, but their total contribution to the 2-jet cross section remains the same. The transition between the two descriptions is relevant to formulating a complete all-orders form of the factorization theorem that resums all logs, global and non-global, in the 2-jet cross section (\emph{cf.} \cite{Larkoski:2015kga,Larkoski:2015zka,Neill:2015nya}).

A slightly different approach was taken in \cite{Chay:2015ila}, where a definition of the measured (or, there, ``unintegrated'') jet function was adopted so that it does integrate up to the unmeasured (or, ``integrated'') jet function. This was done by adopting the power counting $R\sim\tau$ from the beginning, whereas \cite{Ellis:2010rw} began with $\tau\ll R$. Ref.~\cite{Chay:2015ila} effectively combined the measured jet function $J_n^\alg(t_n,R,\mu)$ in \eq{Jalg} with the part of the soft function \eq{softkalg} depending on the in-jet measured radiation with momenta $k_{n}$ into a single object that by itself integrates up to $J_{\text{un}}^\alg$. Their soft function then retains no dependence on the measurement $\tau$ and is the same for cross sections with measured or unmeasured jets. We have shown above that keeping the scales associated with $\tau$ and $R$ separate ($\tau\ll R$ from the beginnning but then integrating the total cross section up to $\tau\sim R$) does in fact reproduce the total 2-jet cross section to leading power in $R$. Both approaches lead to the correct fixed-order total 2-jet cross section. However, in the approach of \cite{Chay:2015ila},  one could not study the resummation of logs of the jet thrust $\tau$ itself, although as a path to the total 2-jet cross section it is essentially equivalent to ours. The $\mu$-dependent logs in the jet and soft functions of \cite{Chay:2015ila} are only those of $\mu/(QR)$ or $\mu/(2\Lambda)$ (no soft-collinear scale was identified), with no $\mu$-dependent logs containing the jet mass $p^2$ or soft momentum $k$ relevant to the resummation of logs of $\tau$ remaining. For the goal of simply obtaining the total fixed-order two-jet rate, however, their approach succeeds in organizing the pieces of the jet thrust cross section in a way that can be integrated simply to give the total two-jet rate. We will explore the generalization of the relation between measured and unmeasured jet/soft functions to $\cO(\as^2)$ and higher in \ssec{unmeasured2}.

\section{Two-Loop Jet Thrust Soft Function}
\label{sec:S2}

In this section we explore the factorization of the two-loop soft function for the jet thrust cross section in \eq{jetthrustcs} into csoft, global (veto) soft, and soft-collinear functions. The total jet thrust soft function was calculated to $\cO(\as^2)$ in \cite{vonManteuffel:2013vja}. Ideally, one should compute the individual factors from their operator definitions, \eg \eqs{Ssdef}{Sscdef}, the results of which to $\cO(\as)$ we gave in \eq{BareSoftResults}, and verify that they reproduce the result of \cite{vonManteuffel:2013vja}. We do not do the EFT computation explicitly at $\cO(\as^2)$ in this paper, leaving it for future work. Rather, we show that the result in \cite{vonManteuffel:2013vja} does indeed factor into pieces sensitive individually to csoft, global soft, and soft-collinear scales, and that these agree with the generic form these functions should take if they satisfy the appropriate RG evolution equations (RGEs). There are leftover non-global pieces sensitive to multiple scales, and we show these remaining pieces agree with known coefficients for leading and subleading NGLs at $\cO(\as^2)$. From \cite{vonManteuffel:2013vja} we can then extract the anomalous dimensions of the individual global factors to $\cO(\as^2)$. 

Beyond this, we give generic arguments for how the csoft, global soft, and soft-collinear anomalous dimensions should be related to each other and to other soft functions in the literature (such as those for hemisphere thrust and Drell-Yan)  to all orders in $\as$, and show our extracted anomalous dimensions indeed, nontrivially, satisfy these relations. The generic relations allow us to extract the anomalous dimensions of all the soft functions even to $\cO(\as^3)$. These results are already highly nontrivial and are a strong consistency check of our refactorization framework. From an explicit EFT computation of the csoft, global soft, and soft-collinear functions from their operator definitions such as \eqs{Ssdef}{Sscdef}, one should be able to confirm our extracted anomalous dimensions from their UV divergent behavior and reproduce the results for the renormalized functions we give in this section.

\subsection{Refactorization and RG Evolution}

In the soft function for jet thrust in \eq{softkalg}, there is dependence both on the csoft scale of measured in-jet soft radiation $k/R$ as well as the scales $2\Lambda$ and $2\Lambda R$ associated with the energy veto outside the jets, illustrated in \fig{allscales}. We notice the dependence on the jet veto scale in \eq{softkalg} is the same as in the soft function \eq{S1} for the total 2-jet cross section, and thus separates into soft and soft-collinear contributions as in \eqs{S1refact}{SS1SSc1}. In fact, the whole 1-loop jet thrust soft function \eq{softkalg} can be separated into \emph{three} categories of terms:
\begin{align}
\label{eq:Sk1refact}
S(k_n,k_\bn,\Lambda,R,\mu) = \delta(k) \bigl[ 1 &+  S_s^{(1)}(\Lambda,\mu) + 2S_{sc}^{(1)}(\Lambda R,\mu)\bigr] \nn\\
& + \sum_{i=n,\bn}S_{\text{in}}^{(1)}(k_i/R,\mu)\,,
\end{align}
where $S_{s,sc}^{(1)}$ were given in \eq{SS1SSc1}, and where $S_{\text{in}}^{(1)}$ is given by \cite{Ellis:2010rw},
\be
\label{eq:Sin1}
S_{\text{in}}^{(1)}(k/R,\mu) = \frac{\as}{4\pi} \biggl\{ \delta(k) c_{\text{in}}^1 - \frac{2\Gamma_0}{\mu R}\biggl[ \frac{\theta(k) \ln(k/(\mu R))}{k/(\mu R)} \biggr]_+\biggr\}\,,
\ee
where
\be
\label{eq:cin1}
c_{\text{in}}^1 = \frac{\pi^2}{6}C_F\,.
\ee
The form \eq{Sk1refact} suggests a generalization to an all-orders form of a refactorized jet thrust soft function,
\begin{align}
\label{eq:Skrefact}
&S(k_n,k_\bn,\Lambda,R,\mu)  = \int_0^\Lambda dE\, S_{\text{in}}(k_n/R,\mu) S_{\text{in}}(k_\bn/R,\mu)  \nn \\
&\qquad\times S_s(E,\mu) \otimes S_{sc}^2(E R,\mu)\otimes S_{ng}(k_{n,\bn},E,R)\,,
\end{align}
where the $\otimes$ are convolutions in the jet veto energy measurement variables and/or in $k_{n,\bn}$. $S_{ng}$ contains the leftover fixed-order NGLs, which begin at $\cO(\as^2)$, and are arguments of the ratios of the various soft scales \cite{Banfi:2010pa,Hornig:2011tg}. They come from correlated emissions into separated regions of phase space. ($S_{sc}^2$ itself is a convolution of two identical pieces.) For the soft function for the total jet thrust cross section in \eq{jetthrustsum},
\begin{align}
\label{eq:Sksumrefact}
S(k,\Lambda,R,\mu)  = \int \! dk_n dk_\bn\delta(k \minus k_n \minus k_\bn) S(k_n,k_\bn,\Lambda,R,\mu)\,.
\end{align}
We will distinguish the two soft functions \eqs{Skrefact}{Sksumrefact} by the number of their arguments.
The refactorization of the $k$-dependent $S_{\text{in}}$ from the jet veto-dependent parts was already proposed in \cite{Ellis:2010rw}, and the separation of $S_{ng}$ and identification of the coefficient of the leading (double) NGLs was discussed in \cite{Hornig:2011tg}. The identification of the subleading (single) NGLs in this soft function was made in \cite{vonManteuffel:2013vja}. The separation of the soft and soft-collinear functions at the scales $2\Lambda$ and $2\Lambda R$ here is new. The full set of scales we have in the jet thrust cross section is illustrated in \fig{allscales}.

We will not provide here a formal derivation of \eqs{Skrefact}{Sksumrefact}. However, because an explicit two-loop calculation of the total jet thrust soft function in \eq{Sksumrefact} is available \cite{vonManteuffel:2013vja}, we can check if it does in fact obey the form \eq{Sksumrefact} in a consistent manner up to $\cO(\as^2)$, and proceed to compute the prediction arising from \eq{Skrefact} and the full factorized form of the cross section \eq{jetthrustcs} for the resummation of (global) logs to all orders in $\as$ at NLL and NNLL accuracy.

It will be most straightforward to work with the Laplace transform of \eq{Sksumrefact} in both $k$ and $E$.  We form the double transform:
\begin{align}
&\widetilde S(\nu,\lambda,R,\mu) = \int_0^\infty dk\, e^{-\nu k}\int_0^\infty dE \, e^{-\lambda E}  \\
&\times S_{\text{in}}(k/R,\mu)^2\!\otimes\! S_{s}(E,\mu) \!\otimes\! S_{sc}(ER,\mu)^2 \! \otimes\! S_{ng}(k,E,R) \,, \nn
\end{align}
which disentangles the convolutions, leaving us with
\be
\label{eq:SLaplacerefact}
\widetilde S(\nu,\lambda,R,\mu) = \widetilde S_{\text{in}}^2(\nu R,\mu) \widetilde S_s(\lambda,\mu)\widetilde S_{sc}^2(\lambda/R,\mu) \widetilde S_{ng}\,,
\ee
where
\be
\label{eq:SinSsSsc}
\begin{split}
\widetilde S_{\text{in}}(\nu R,\mu) &= \int_0^\infty dk\, e^{-\nu k} S_{\text{in}}(k/R,\mu) \\
\widetilde S_s(\lambda,\mu) &= \int_0^\infty dE\,  e^{-\lambda E} S_s(E,\mu) \\
\widetilde S_{sc}(\lambda/R,\mu) &= \int_0^\infty dE\,  e^{-\lambda E} S_{sc}(E R,\mu)
\end{split}
\ee
These functions should satisfy ordinary RGEs of a multiplicative form,
\be
\label{eq:SLaplaceRGE}
\begin{split}
\mu \frac{d}{d\mu} \widetilde S_{\text{in}}(\nu R,\mu) &= \gamma_{\text{in}}(\mu) \widetilde S_{\text{in}}(\nu R,\mu) \\
\mu\frac{d}{d\mu} \widetilde S_s(\lambda,\mu)  &= \gamma_{ss}(\mu) \widetilde S_s(\lambda,\mu)  \\
\mu\frac{d}{d\mu} \widetilde S_{sc}(\lambda R,\mu)  &= \gamma_{sc}(\mu) \widetilde S_{sc}(\lambda R,\mu)  \,, \\
\end{split}
\ee
where the anomalous dimensions take the forms
\be
\label{eq:softanomdims}
\begin{split}
\gamma_{\text{in}} (\mu) &= -2\Gamma_{\text{cusp}}[\as(\mu)]\ln\frac{\mu\nu R e^{\gamma_E}}{Q} + \gamma_{\text{in}}[\as(\mu)] \\
\gamma_{ss}(\mu) &= \phantom{-}4\Gamma_{\text{cusp}}[\as(\mu)] \ln\frac{\mu\lambda e^{\gamma_E}}{2} + \gamma_{ss}[\as(\mu)] \\
\gamma_{sc}(\mu) &= -2\Gamma_{\text{cusp}}[\as(\mu)]\ln\frac{\mu\lambda e^{\gamma_E}}{2R} + \gamma_{sc}[\as(\mu)]\,,
\end{split}
\ee
where the $\gamma[\as]$ terms on the right-hand sides are the non-cusp parts of the anomalous dimensions, and the proportionalities of the cusp parts to $\Gamma_{\text{cusp}}$ are deduced from the one-loop results \eqs{SsSsc1coeffs}{Sin1}.

Among the objects in the jet thrust cross section \eq{jetthrustcs}, the anomalous dimensions of the hard function \cite{Moch:2005id,Idilbi:2006dg, Becher:2006mr} and the jet function (which is the same as for the inclusive jet function)  \cite{Becher:2006mr} as well as the cusp anomalous dimension   \cite{Korchemsky:1987wg, Moch:2004pa} are all known to $\cO(\as^3)$. (We give these in \appx{anomdim}.) The jet thrust soft function for the thrust cone algorithm, and thus its anomalous dimension, were computed to $\cO(\as^2)$ in \cite{Kelley:2011aa,vonManteuffel:2013vja}. Thus far, whether it factorizes as \eq{Skrefact} and what the forms or the anomalous dimensions of each individual piece are remain unknown. Under the assumption that they obey \eqs{SLaplaceRGE}{softanomdims}, we can extract these individual anomalous dimensions from the form of the soft function in  \cite{vonManteuffel:2013vja}.
\\

\subsection{Two-Loop Soft Function}
\label{ssec:S2}

A function that obeys the multiplicative RGE \eq{SLaplaceRGE} with anomalous dimensions of the form \eq{softanomdims} whose expansions in $\as$ take the form \eq{gammaexpansion} must take the form given by \eqs{Fexp}{Fcoeffs} (see, \eg \cite{Almeida:2014uva,Hornig:2011tg}), that is,
\begin{align}
\label{eq:softexpansion}
&\wt S_i(x_i,\mu) = 1 + \frac{\as}{4\pi}\Bigl( \Gamma_i^0 \ln^2(\mu x_i e^{\gamma_E})+ \tilde c_i^1\Bigr) \\
&\quad + \Bigl(\frac{\as}{4\pi}\Bigr)^2\biggl[ \frac{1}{2}(\Gamma_i^0)^2 \ln^4(\mu x_i e^{\gamma_E}) + \frac{2}{3}\Gamma_i^0 \beta_0\ln^3 (\mu x_i e^{\gamma_E}) \nn \\
& \qquad\qquad  + \bigl( \Gamma_i^1 + \tilde c_i^1\Gamma_i^0\bigr)\ln^2 (\mu x_i e^{\gamma_E})  \nn\\
&\qquad\qquad  + \bigl( \gamma_i^1 + 2\tilde c_i^1\beta_0\bigr)\ln(\mu x_i e^{\gamma_E}) + \tilde c_i^2\biggr] \,, \nn
\end{align}
where we have already used that the $\cO(\as)$ non-cusp anomalous dimensions $\gamma_i^0 = 0$ for each function in \eq{SinSsSsc}. The arguments $x_i$ for each function are:
\begin{align}
x_{\text{in}} = \nu R\,,\quad x_{ss} = \frac{\lambda}{2}\,,\quad x_{sc} = \frac{\lambda}{2R}\,.
\end{align}
The cusp parts of the anomalous dimensions in \eq{softexpansion} are determined by
\begin{align}
\label{eq:softanomdim1}
\Gamma_{\text{in}} = \Gamma_{sc} = -\Gamma_{\text{cusp}}\,,\quad \Gamma_{ss} = 2\Gamma_{\text{cusp}}\,,
\end{align}
and the one-loop Laplace-space constants (see \eq{LPs}) are given by
\begin{align}
\tilde c_i^1 = c_i^1 + \Gamma_i^0 \frac{\pi^2}{6}\,,
\end{align}
where $c_{i}^1$ for $i=S_{\text{in}}$ was given in \eq{cin1}, and for $i=S_s$ and $S_{sc}$ in \eq{SsSsc1coeffs}.

Since the cusp anomalous dimension appearing in \eqs{softanomdim1}{Fexp} is known to two (in fact, three) loops, and the non-cusp anomalous dimensions and constants to one loop, up to NLL($'$) resummation (cf. \cite{Abbate:2010xh,Almeida:2014uva}) of (global) logs is already possible. To go to NNLL accuracy, the non-cusp anomalous dimensions are needed to two loops. For NNLL$'$ accuracy and higher, the two-loop constants are also needed (besides calculation and resummation of NGLs \cite{Larkoski:2015zka}). In this section, we will extract the two-loop non-cusp anomalous dimensions $\gamma_{\text{in},ss,sc}^1$ from known results for the full (but not fully refactorized) jet thrust soft function in the literature \cite{vonManteuffel:2013vja}.

Using the expansions \eqs{Fexp}{Fcoeffs} for $S_{\text{in},ss,sc}$ and multiplying them together according to \eq{Skrefact}, the fixed-order expansion of $\widetilde S(\nu,\lambda,R,\mu)$ in \eq{SLaplacerefact} up to $\cO(\as^2)$ can be written
\begin{widetext}
\begin{align}
\label{eq:SLaplace2}
\widetilde S(\nu,\lambda,R,\mu) &= 1 + \frac{\as}{4\pi} \Bigl[ 2\Gamma_0 \Bigl( -\ln^2(\mu R\nu e^{\gamma_E})+ \ln^2 \frac{\mu\lambda e^{\gamma_E}}{2} - \ln^2\frac{\mu\lambda e^{\gamma_E}}{2R} - \frac{\pi^2}{6}\Bigr) + 2c_{\text{in}}^1 + c_{ss}^1 + 2c_{sc}^1\Bigr] \\
&\quad + \Bigl( \frac{\as}{4\pi}\Bigr)^2 \biggl\{ 2(\Gamma_0)^2 \Bigl( -\ln^2(\mu R\nu e^{\gamma_E})+ \ln^2 \frac{\mu\lambda e^{\gamma_E}}{2} - \ln^2\frac{\mu\lambda e^{\gamma_E}}{2R} \Bigr)^2 \nn \\
&\qquad\qquad\quad   +  \frac{4}{3}\Gamma_0\beta_0 \Bigl( - \ln^3(\mu R\nu e^{\gamma_E}) + \ln^3\frac{\mu\lambda e^{\gamma_E}}{2} - \ln^3\frac{\mu\lambda e^{\gamma_E}}{2R}\Bigr) + \widetilde S_{\text{ng}}^{(2)}(\nu,\lambda,R\mu) \nn \\
&\qquad\qquad\quad   + 2\Bigl[ \Gamma_1 + \Gamma_0 \Bigl( 2c_{\text{in}} + c_{ss}^1 + 2c_{sc}^1 - \Gamma_0\frac{\pi^2}{3}\Bigr)\Bigr] \Bigl( -\ln^2(\mu R\nu e^{\gamma_E}) + \ln^2 \frac{\mu\lambda e^{\gamma_E}}{2} - \ln^2\frac{\mu\lambda e^{\gamma_E}}{2R} \Bigr) \nn \\
&\qquad\qquad\quad + 2(\gamma_{\text{in}}^1 + 2\beta_0\tilde c_{\text{in}}^1)\ln(\mu R\nu e^{\gamma_E}) + (\gamma_{ss}^1 + 2\beta_0\tilde c_{ss}^1) \ln\frac{\mu\lambda e^{\gamma_E}}{2} + 2(\gamma_{sc}^1 + 2\beta_0\tilde c_{sc}^1) \ln \frac{\mu\lambda e^{\gamma_E}}{2R} + \tilde c_{S_{\text{tot}}}^2\biggr\} \nn\,.
\end{align}
Our goal is now to extract the (thus far) unknown two-loop anomalous dimension coefficients $\gamma_{\text{in},ss,sc}^1$.
We will not extract the two-loop constant $\tilde c_{S_{\text{tot}}}^2$ in this paper.

To compare \eq{SLaplace2} to the expression for $S(k,\Lambda,R,\mu)$ given in \cite{vonManteuffel:2013vja}, we perform the inverse Laplace transform of \eq{SLaplace2} back to momentum space, using the formulae in \eq{iLPs},
\be
S_c(k,\Lambda,R,\mu) = \cL_\nu^{-1}\cL_\lambda^{-1} \biggl\{\frac{1}{\nu\lambda} \wt S(\nu,\lambda,R,\mu) \biggr\} \,,
\ee
for the cumulative soft function in $k$ and $\Lambda$, resulting in:
\begin{align}
\label{eq:Sk2}
S^\cu(k,\Lambda,R,\mu) &= 1 + \frac{\as}{4\pi} \Bigl[ 2\Gamma_0 \Bigl( - \ln^2 \frac{\mu R}{k} + \ln R \, \ln\! \frac{\mu^2}{4\Lambda^2 R}\Bigr) - \frac{\pi^2}{3}C_F\Bigr] + \Bigl(\frac{\as}{4\pi}\Bigr)^2 \biggl\{ 2(\Gamma_0)^2\Bigl( - \ln^2 \frac{\mu R}{k} + \ln R \, \ln\! \frac{\mu^2}{4\Lambda^2 R}\Bigr)^2  \nn \\
&\qquad \qquad + \frac{4}{3}\Gamma_0\beta_0\Bigl(-\ln^3\frac{\mu R}{k} + \ln^3\frac{\mu}{2\Lambda} - \ln^3\frac{\mu}{2\Lambda R}\Bigr) + S_{\text{ng}}^{(2)}(k,\Lambda,R,\mu) \\
&\qquad\qquad + 2\Bigl( \Gamma_1 - \Gamma_0\frac{\pi^2}{3}C_F\Bigr)\Bigl( - \ln^2 \frac{\mu R}{k} + \ln R \, \ln\! \frac{\mu^2}{4\Lambda^2 R}\Bigr) - \frac{4\pi^2}{3}(\Gamma_0)^2\Bigl( \ln^2\frac{\mu R}{k}  + \ln^2 R\Bigr) \nn \\
& \qquad\qquad + 2(\gamma_{\text{in}}^1 + 2\beta_0  c_{\text{in}}^1 - 8\zeta_3 \Gamma_0^2)\ln\frac{\mu R}{k} + (\gamma_{ss}^1 + 2\beta_0 c_{ss}^1) \ln\frac{\mu}{2\Lambda} + 2(\gamma_{sc}^1 + 2\beta_0 c_{sc}^1)\ln\frac{\mu}{2\Lambda R} + c_{S_{\text{tot}}}^2\,, \nn
\end{align}
where $c_{S_{\text{tot}}}^2$ is the constant term in this momentum-space total soft function.

We can separate the purely Abelian ($\as C_F$ and $\as^2 C_F^2$) terms and non-Abelian ($C_F C_A$, $C_F T_F n_F$) terms of \eq{Sk2},
\be
\label{eq:SCFnA}
S^\cu(k,\Lambda,R,\mu) = S_{C_F}(k,\Lambda,R,\mu)  + \Bigl(\frac{\as}{4\pi}\Bigr)^2 S_{nA}^{(2)}(k,\Lambda,R,\mu)\,,
\ee
where
\begin{align}
\label{eq:SCF}
S_{C_F}(k,\Lambda,R,\mu) &= 1  + \frac{\as}{4\pi} \Bigl[ 2\Gamma_0 \Bigl( - \ln^2 \frac{\mu R}{k} + \ln R \, \ln\! \frac{\mu^2}{4\Lambda^2 R}\Bigr) - \frac{\pi^2}{3}C_F\Bigr] + \Bigl(\frac{\as}{4\pi}\Bigr)^2 \biggl\{ 2(\Gamma_0)^2\Bigl( - \ln^2 \frac{\mu R}{k} + \ln R \, \ln\! \frac{\mu^2}{4\Lambda^2 R}\Bigr)^2  \nn \\
&\quad + 2 \Gamma_0\frac{\pi^2}{3}C_F\Bigl( \ln^2 \frac{\mu R}{k} - \ln R \, \ln\! \frac{\mu^2}{4\Lambda^2 R}\Bigr) - \frac{4\pi^2}{3}(\Gamma_0)^2\Bigl( \ln^2\frac{\mu R}{k}  + \ln^2 R\Bigr) -16\zeta_3 \Gamma_0^2 \ln\frac{\mu R}{k} + c_{C_F}^{(2)} \biggr\} \,,
\end{align}
and
\be
\label{eq:SnA}
\begin{split}
S_{nA}^{(2)}(k,\Lambda,R,\mu) &=  \frac{4}{3}\Gamma_0\beta_0\Bigl(-\ln^3\frac{\mu R}{k} + \ln^3\frac{\mu}{2\Lambda} - \ln^3\frac{\mu}{2\Lambda R}\Bigr) + 2 \Gamma_1 \Bigl( - \ln^2 \frac{\mu R}{k} + \ln R \, \ln\! \frac{\mu^2}{4\Lambda^2 R}\Bigr)  + S_{ng}^{c(2)}(k,\Lambda,R,\mu) \\
&\quad + 2(\gamma_{\text{in}}^1 + 2\beta_0  c_{\text{in}}^1 )\ln\frac{\mu R}{k} + (\gamma_{ss}^1 + 2\beta_0 c_{ss}^1) \ln\frac{\mu}{2\Lambda} + 2(\gamma_{sc}^1 + 2\beta_0 c_{sc}^1)\ln\frac{\mu}{2\Lambda R} + c_{nA}^{(2)}\,.
\end{split}
\ee
In \cite{vonManteuffel:2013vja}, an expression is given for the non-Abelian terms of the 2-loop soft function both for arbitrary $R$ and also in the limit that $R\to 0$. We are interested only in these terms that do not vanish as $R\to 0$. These terms can be directly compared to \eq{SnA} for extraction of the unknown anomalous dimensions $\gamma_{\text{in},ss,sc}^1$. We quote the formula from \cite{vonManteuffel:2013vja} \appx{soft}. In terms of known cusp anomalous dimension and beta functions coefficients, their result can be reorganized into the form:
\be
\label{eq:SnAknown}
\begin{split}
S_{nA\text{\cite{vonManteuffel:2013vja}}}^{(2)} &= \frac{4}{3}\Gamma_0\beta_0\Bigl(-\ln^3\frac{\mu R}{k} + \ln^3\frac{\mu}{2\Lambda} - \ln^3\frac{\mu}{2\Lambda R}\Bigr) + 2 \Gamma_1 \Bigl( - \ln^2 \frac{\mu R}{k} + \ln R \, \ln\! \frac{\mu^2}{4\Lambda^2 R}\Bigr)   \\
&\quad + \Bigl[\Bigl(56\zeta_3 - \frac{1616}{27}\Bigr) C_F C_A + \frac{448}{27} C_F T_F n_f  + \frac{4\pi^2}{3}\beta_0 C_F  \Bigr] \ln\frac{\mu R }{k} \\
&\quad + \Bigl[  \Bigl(\frac{1616}{27} - 56\zeta_3\Bigr) C_F C_A - \frac{448}{27} C_F T_F n_f  - \frac{8\pi^2}{3}\beta_0 C_F \Bigr] \ln\frac{\mu}{2\Lambda} \\
&\quad + \Bigl[ \Bigl(56\zeta_3 - \frac{1616}{27}\Bigr) C_F C_A + \frac{448}{27} C_F T_F n_f + \frac{4\pi^2}{3} \beta_0C_F \Bigr] \ln\frac{\mu}{2\Lambda R} + c_{nA}^{(2)} \\
&\quad - \frac{8}{3}\pi^2 C_F C_A\ln^2\frac{k}{2\Lambda R^2} + \Bigl[ \Bigl( - 16\zeta_3 - \frac{8}{3} \Bigr) C_F C_A + \frac{16}{3}C_F T_F n_f + \frac{8\pi^2}{3}\beta_0 C_F\Bigr] \ln\frac{k}{2\Lambda R^2} \,.
\end{split}
\ee
\end{widetext}
On the last line we have separated out those logs which are non-global in origin, coming from correlated emissions into two separated phase space regions. These correspond to the $S_{ng}^{(2)}$ term in  \eq{SnA}:
\begin{align}
\label{eq:Sng}
&S_{\text{ng}}^{\cu(2)}(k,\Lambda,R,\mu) = -\frac{8}{3}\pi^2 C_F C_A \ln^2 \frac{k}{2\Lambda R^2} \\
& + C_F \Bigl[ \Bigl( - 16\zeta_3 - \frac{8}{3} \Bigr)  C_A + \frac{16}{3} T_F n_f + \frac{8\pi^2}{3}\beta_0\Bigr] \ln\frac{k}{2\Lambda R^2} \,, \nn
\end{align}
the leading coefficient of which was computed in \cite{Hornig:2011tg}, and the single log coefficient in \cite{Kelley:2011aa,vonManteuffel:2013vja} . (Ref.~\cite{Hornig:2011tg}, however, only identified one power of $R$ in the argument of the NGLs). The computation of \eq{SnAknown} in \cite{Kelley:2011aa,vonManteuffel:2013vja} together with the factorization conjecture \eq{Skrefact} confirm that the argument of the NGL is, in fact, $k/(2\Lambda R^2)$ \cite{Banfi:2010pa}, which we notice is the ratio of the measured in-jet soft scale $k/R$ and the soft-collinear scale $2\Lambda R$ in \fig{allscales}.
From the ``in-out'' and ``in-in'' NGL coefficients computed in \cite{Hornig:2011tg} and the results of \cite{vonManteuffel:2013vja}, we can also form the corresponding non-global contribution to the double-differential jet thrust soft function in \eq{Skrefact},
\begin{align}
\label{eq:Sng2}
&S_{\text{ng}}^{c(2)}(k_n,\! k_\bn, \! \Lambda, \! R,\! \mu) = -\frac{4}{3}\pi^2 C_F C_A  \Bigl( \ln^2\! \frac{k_n}{2\Lambda R^2}  \plus \ln^2\! \frac{k_\bn}{2\Lambda R^2}\Bigr) \nn \\
&\qquad + C_F \Bigl[ \Bigl( - 16\zeta_3 - \frac{8}{3} \Bigr)  C_A + \frac{16}{3} T_F n_f +\frac{8\pi^2}{3}\beta_0\Bigr] \nn \\
&\qquad\qquad\times \Bigl(\ln\frac{k_n}{2\Lambda R^2} + \ln\frac{k_\bn}{2\Lambda R^2}\Bigr)\,.
\end{align}
In the limit $R\to 0$, no NGL of $k_n/k_\bn$ appears \cite{Hornig:2011tg}, as the phase space for soft gluons inside the two cones vanishes.
 A resummation of the NGLs requires a more advanced factorization theorem using technology such as that in \cite{Larkoski:2015zka,Becher:2015hka}.

Comparing \eqs{SnA}{SnAknown} and using \eqs{SsSsc1coeffs}{cin1} for the one-loop constants $c_{ss,sc,\text{in}}^1$, we are able to read off the non-cusp anomalous dimensions:
\be
\label{eq:softnoncusp}
\boxed{
\begin{split}
\gamma_{{ss}}^1 &= -2\gamma_{\text{in}}^1 = -2\gamma_{sc}^1 \\
&= C_F\Bigl[\Bigl(\frac{1616}{27} - 56\zeta_3\Bigr)  C_A - \frac{448}{27}  T_F n_f  - \frac{2\pi^2}{3}\beta_0\Bigr]\,.
\end{split}
}
\ee
The relation amongst the three anomalous dimensions in \eq{softnoncusp} should, in fact, generalize to all orders in $\as$. In the $R=1$ limit, the in-jet regions each become a whole hemisphere, so \eq{softanomdims} directly implies for the non-cusp anomalous dimensions:
\begin{align}
\label{eq:gammarelations}
\gamma_{\rm in}  &= \gamma_{\rm hemi} \nn\\
\gamma_{ss} &= -2 \gamma_{sc}
\,.\end{align}

We can derive several more useful relations amongst anomalous dimensions. To start, $\gamma_{ss}$ is the same anomalous dimension as for the timelike soft function that arises in threshold resummation in Drell-Yan, a property that was in fact already noted in \cite{Ellis:2010rw}. This follows from the definition of $S_{s}(E,\mu)$ in \eq{Ssdef}. The only difference with Drell-Yan is the direction of the Wilson lines (incoming \emph{vs.} outgoing), which does not affect the anomalous dimension (nor the value of the function itself to at least $\cO(\as^2)$ \cite{Kang:2015moa}) and so
\be
\label{eq:gamma_ss-DY}
\gamma_H + 2 \gamma_{qq} + \gamma_{ss} = 0
\,, \ee
where $ \gamma_{qq} $ is the non-cusp part of DGLAP evolution of quark parton distribution functions, which for example appears in the  Altarelli-Parisi splitting function as
\be
P_{qq}(z) = \frac{2 \Gamma_{\rm cusp}[\as]}{(1-z)_+} + \gamma_{qq}[\as] \delta(1-z) + \cdots
\,,\ee
where the ellipses denote terms that are non-singular  as $z \to 1$.
Then, we use the condition for the consistency of factorization in thrust (\eg \cite{Fleming:2007qr,Fleming:2007xt,Schwartz:2007ib,Bauer:2008dt}),
\begin{align}
\label{eq:tauconsistency}
\gamma_H + 2\gamma_J + 2 \gamma_{\rm hemi} = 0
\end{align}
and the relation satisfied by $\gamma_{H,J,qq}$ in DIS as $x\to 1$ \cite{Becher:2006mr},
\be
\label{eq:DISconsistency}
\gamma_H + \gamma_J + \gamma_{qq} = 0\,,
\ee
which together imply
\be
\label{eq:qqJS}
\gamma_J + 2\gamma_{\text{hemi}} = \gamma_{qq}\,.
\ee
Finally, \eqss{gamma_ss-DY}{tauconsistency}{qqJS} together imply
\be
2\gamma_{\text{hemi}} + \gamma_{ss} = 0\,.
\ee
Combining this with \eq{gammarelations}, we derive the all-orders relations among non-cusp anomalous dimensions,
\be
\label{eq:gammaequalities}
\boxed{
\gamma_{\rm hemi} = \gamma_{\text{in}}  = \gamma_{sc}  = -\frac{\gamma_{ss}}{2}}
\,.\ee
We give $\gamma_{\rm hemi} $  to three loops using the known results quoted in \eq{gammass2}.

The relations \eqs{gammarelations}{tauconsistency} imply the consistency of anomalous dimensions in our jet thrust factorization theorem \eqs{jetthrustcs}{Skrefact},
\be
\label{eq:HJSconsistency}
\gamma_H + 2\gamma_J +
2\gamma_{\text{in}} + \gamma_{ss} + 2\gamma_{sc} = 0 \,,
\ee
which in this form is satisfied by both the cusp and non-cusp parts.

\section{Two-Loop Fixed-Order Cross Sections}
\label{sec:twoloop}

With the results for the two-loop soft functions in \sec{S2} and the known results for the two-loop hard and jet functions in \appx{anomdim}, we can construct our prediction for the logarithmic terms in the full jet thrust cross section to two loops, and integrate it up to $\tau=\taumax$ to obtain our prediction for the logarithmic terms of the two-loop jet rate for the thrust cone algorithm.

\subsection{Jet Thrust Cross Section}

We have the ingredients to build the jet thrust cross section \eq{jetthrustcs} differential in each jet's thrust $\tau_{1,2}$ or the total thrust $\tau$ in \eq{jetthrustsum}. We will begin by working with the double distribution \eq{jetthrustcs}, though it is easier to express it in terms of its integral over $\tau_{1,2}$. Using the refactorization of the soft function \eq{Skrefact}, we have that
\begin{align}
\label{eq:jetthrustrefact}
&\sigma_\cu(\tau_1,\tau_2;\Lambda,R) \equiv \int_0^{\tau_1}\!\!\!d\tau_1' \int_0^{\tau_2}\!\!\!  d\tau_2'\frac{1}{\sigma_0}\frac{d\sigma(\Lambda,R)}{d\tau_1d\tau_2}   = H(Q^2,\mu) \nn \\
&\times \int\! dt_n dk_n \theta\Bigl(\tau_1 \minus \frac{t_n}{Q^2} \minus \frac{k_n}{Q}\Bigr)  J_n^\alg (t_n,R,\mu)  S_{\text{in}}\Bigl(\frac{k_n}{R},\mu\Bigr) \nn\\
&\times  \int\! dt_\bn dk_\bn \theta\Bigl(\tau_2 \minus \frac{t_\bn}{Q^2} \minus \frac{k_\bn}{Q}\Bigr)J_\bn^\alg (t_\bn,R,\mu)  S_{\text{in}}\Bigl(\frac{k_\bn}{R},\mu\Bigr) \nn \\
&\times   \int\!  dE_s dE_{sc} \theta(\Lambda - E_s - E_{sc} )  S_s(E_s,\mu) S_{sc}(E_{sc} R,\mu)^2  \nn \\
&\qquad  \otimes S_{ng}(k_n,k_\bn,E,R) \,. 
\end{align}
We will use this factorization to resum logs in the perturbative expansion of the cross section in \sec{NNLL}. For the purpose of computing the fixed-order cross section up to $\cO(\as^2)$, it is convenient to package pieces of the factorization formula \eq{jetthrustrefact} together, those contributing to $\tau_{1,2}$ and those depending on the soft veto $\Lambda$.

First we compute the convolution of $J_n^\alg$ and $S_{\text{in}}$. We will compute the product of their Laplace transforms and then inverse transform back to momentum space. The Laplace transforms of $J(t,R,\mu)$ and $S_{\text{in}}(k,R,\mu)$ take the form \eq{Fexp} to $\cO(\as^2)$. The jet function also includes an algorithm-dependent component $\Delta J^\alg$.

Up to $\cO(\as^2)$, the Laplace transform of the measured jet function is given by \eqs{Fexp}{Fcoeffs} with $\Gamma_J = 2\Gamma_{\text{cusp}}$ and $j_J = 2$:
\begin{widetext}
\begin{align}
\label{eq:Jmeasnu}
\tilde J_n^\alg(\nu,R,\mu) &\equiv \int_0^\infty dt \,e^{-\nu t/Q^2} J_n^\alg(t,R,\mu) \\
&= 1 + \frac{\as}{4\pi} \biggl[\frac{\Gamma_0}{2}\Bigl( \ln^2\frac{\mu^2 \nu e^{\gamma_E}}{Q^2} + \frac{\pi^2}{6}\Bigr) + \frac{\gamma_J^0}{2}\ln\frac{\mu^2 \nu e^{\gamma_E}}{Q^2} + c_J^1 + \Delta \tilde J_\alg^{(1)}(\nu,R)\biggr] \nn \\
&\quad + \Bigl(\frac{\as}{4\pi}\Bigr)^2\biggl\{ \frac{(\Gamma_0)^2}{8} \ln^4\frac{\mu^2 \nu e^{\gamma_E}}{Q^2} + \frac{\Gamma_0}{4}\Bigl(\gamma_J^0 + \frac{2}{3}\beta_0\Bigr)\ln^3\frac{\mu^2 \nu e^{\gamma_E}}{Q^2} \nn \\
&\qquad \qquad \quad  + \biggl[ \frac{\Gamma_1}{2} + \frac{(\gamma_J^0)^2}{8} + \frac{\gamma_J^0 \beta_0}{4} + \frac{\Gamma_0}{2} \Bigl( c_J^1 + \frac{\Gamma_0}{2}\frac{\pi^2}{6} + \Delta \tilde J_\alg^{(1)}(\nu,R)\Bigr)\biggr]\ln^2 \frac{\mu^2 \nu e^{\gamma_E}}{Q^2} \nn \\
&\qquad\qquad \quad + \biggl[ \frac{\gamma_J^1}{2} + \frac{\gamma_J^0+2\beta_0}{2}\Bigl(c_J^1 + \frac{\Gamma_0}{2}\frac{\pi^2}{6} + \Delta \tilde J_\alg^{(1)}(\nu,R)\Bigr) \biggr]\ln\frac{\mu^2 \nu e^{\gamma_E}}{Q^2} + \tilde c_J^2 + \Delta \tilde J_\alg^{(2)}(\nu,R)\biggr\}\,. \nn
\end{align}
All of the coefficients here are known (see \appx{anomdim}) except for the two-loop algorithm-dependent correction $\Delta J_\alg^{(2)}$. It is not expected to contribute to the singular part of the $\tau$ distribution (or later to logs of $R$ in the integrated 2-jet cross section) at $\cO(\as^2)$, just as it does not at one loop. The cross terms between the one-loop $\Delta J_\alg^{(1)}$ and logs from the one-loop jet and soft functions, however, do. So we keep them.
We will not compute the constant $\tilde c_J^2$ in this paper, nor compute explicitly the Laplace transform $\Delta\wt J_\alg^{(1)}$ but keep it in abstract form until we return to momentum space.

Meanwhile the Laplace transform of the measured soft (csoft) function $S_{\text{in}}$ is given to $\cO(\as^2)$ by
\be
\label{eq:Sinnu}
\begin{split}
\wt S_{\text{in}}(\nu R,\mu) &\equiv \int_0^\infty dk\, e^{-\nu k/Q} S_{\text{in}}(k/R,\mu) \\
&= 1 + \frac{\as}{4\pi} \Bigl[ -\Gamma_0 \Bigl(\ln^2 \frac{\mu R\nu e^{\gamma_E}}{Q} + \frac{\pi^2}{6}\Bigr) + c_{\text{in}}^1\Bigr]  + \Bigl(\frac{\as}{4\pi}\Bigr)^2 \biggl\{ \frac{(\Gamma_0)^2}{2} \ln^4\frac{\mu R\nu e^{\gamma_E}}{Q} - \frac{2}{3}\Gamma_0\beta_0\ln^3 \frac{\mu R\nu e^{\gamma_E}}{Q} \\
&\qquad \qquad \qquad - \Bigl[ \Gamma_1 + \Gamma_0 \Bigl(c_{\text{in}}^1- \Gamma_0\frac{\pi^2}{6}\Bigr) \Bigr]\ln^2\frac{\mu R \nu e^{\gamma_E}}{Q}  + \Bigl[ \gamma_{\text{in}}^1 + 2\beta_0\Bigl( c_{\text{in}}^1 - \Gamma_0\frac{\pi^2}{6}\Bigr)\Bigr] \ln \frac{\mu R\nu e^{\gamma_E}}{Q} + \tilde c_{{\rm in}}^2\biggr\}\,,
\end{split}
\ee
where all the coefficients are also known except the constant $\tilde c_{{\rm in}}^2$, which we do not compute in this paper. The two-loop non-cusp anomalous dimension $\gamma_{\text{in}}^1$ is now known through our extraction in \eq{softnoncusp} from the results of \cite{vonManteuffel:2013vja}.

Multiplying the Laplace transforms \eqs{Jmeasnu}{Sinnu}, we use \eqss{iLPs}{iLPL2L2}{iLPL1L2} to inverse transform the logs and products of logs of $\nu$.
The result of this exercise is
\begin{align}
\label{eq:JSconvolution}
&\iLP\Bigl \{\frac{1}{\nu} \tilde J^\alg(\nu,R,\mu) \wt S_{\text{in}}(\nu R,\mu) \Bigr\} = 1 + \frac{\as}{4\pi} \biggl[ \Gamma_0 \Bigl( \frac{1}{2}\ln^2\frac{\mu^2}{Q^2\tau} - \ln^2\frac{\mu R}{Q\tau}\Bigr) + \frac{\gamma_J^0}{2}\ln\frac{\mu^2}{Q^2\tau} + c_J^1 + c_{\text{in}}^1 + \Delta J_{\alg}^{c(1)}(\tau,R)\Bigr] \nn \\
&\quad + \Bigl(\frac{\as}{4\pi}\Bigr)^2 \Biggl\{\frac{(\Gamma_0)^2}{2} \Bigl( \frac{1}{2}\ln^2\frac{\mu^2}{Q^2\tau} - \ln^2\frac{\mu R}{Q\tau}\Bigr) ^2  +\frac{\Gamma_0\gamma_J^0}{2}\Bigl( \frac{1}{2}\ln^2\frac{\mu^2}{Q^2\tau} - \ln^2\frac{\mu R}{Q\tau}\Bigr) \ln\frac{\mu^2}{Q^2\tau} + \frac{2}{3}\Gamma_0\beta_0 \Bigl( \frac{1}{4}\ln^3\frac{\mu^2}{Q^2\tau} - \ln^3\frac{\mu R}{Q\tau}\Bigr) \nn \\
&\qquad\qquad\quad + \bigl[ \Gamma_1 + \Gamma_0(c_J^1 + c_{\text{in}}^1)\bigr] \Bigl( \frac{1}{2}\ln^2\frac{\mu^2}{Q^2\tau} - \ln^2\frac{\mu R}{Q\tau}\Bigr) + \Bigl[\frac{(\gamma_J^0)^2}{2} + \gamma_J^0 \beta_0\Bigr] \frac{1}{4}\ln^2\frac{\mu^2}{Q^2\tau}  \\
&\qquad\qquad\quad + \Gamma_0^2\Bigl( \frac{\pi^2}{12}\ln^2 \frac{\mu^2}{Q^2\tau} - \frac{\pi^2}{6}\ln^2\frac{\mu R}{Q\tau} - \frac{\pi^2}{6}\ln^2\frac{\mu}{QR}\Bigr)\nn \\
&\qquad\qquad\quad + \frac{1}{2}\Bigl[ \gamma_J^1 + c_J^1(\gamma_J^0 + 2\beta_0) +   c_{\text{in}}^1\gamma_J^0  - \frac{\pi^2}{6} \Gamma_0\gamma_J^0 + 2\zeta_3 \Gamma_0^2\Bigr]\ln\frac{\mu^2}{Q^2\tau} \nn \\
&\qquad\qquad\quad + \Bigl[ \gamma_{\text{in}}^1 + 2 c_{\text{in}}^1\beta_0  - 2\zeta_3 \Gamma_0^2 + \Gamma_0\gamma_J^0 \frac{\pi^2}{6}\Bigr] \ln\frac{\mu R}{Q\tau} + \Bigl( c_{\text{in}}^1 - \Gamma_0\frac{\pi^2}{12}\Bigr) \Delta J_\alg^{c(1)}(\tau,R) + \text{const.} + \Delta J_\alg^{c(2)}(\tau,R) \nn \\
&\qquad\qquad\quad +\int_0^\tau d\tau' \Delta J_\alg^{(1)}(\tau \minus \tau',R)\biggl\{\Gamma_0 \frac{Q^2}{\mu^2}\biggl[\frac{\ln(Q^2\tau'/\mu^2)}{Q^2\tau'/\mu^2}\biggr]_+  \minus  2\Gamma_0 \frac{Q}{\mu R}\biggl[ \frac{\ln(Q\tau'/(\mu R))}{Q\tau'/(\mu R)}\biggr]_+   \minus  \frac{\gamma_J^0+2\beta_0}{2}\frac{Q^2}{\mu^2}\biggl[ \frac{\mu^2}{Q^2\tau'}\biggr]_+  \biggr\} \Biggr\} \,. \nn
\end{align}
The algorithm-dependent one-loop contribution $\Delta J_\alg^{c(1)}$ is given by the integrals of \eqs{DeltaJcone}{DeltaJkT}.  For the cone algorithm,
\be
\Delta J_\cone^{c(1)} (\tau,R) = 6C_F \ln\Bigl(1 + \frac{\tau}{R^2}\Bigr).
\ee
The convolutions on the last line of \eq{JSconvolution} are given by \eq{DeltaJplus}, with the result
\be
\label{eq:DeltaJconvolution}
\begin{split}
&\int_0^\tau d\tau' \Delta J_\alg^{(1)}(\tau-\tau',R)\biggl\{\Gamma_0 \frac{Q^2}{\mu^2}\biggl[\frac{\ln(Q^2\tau'/\mu^2)}{Q^2\tau'/\mu^2}\biggr]_+ - 2\Gamma_0 \frac{Q}{\mu R}\biggl[ \frac{\ln(Q\tau'/(\mu R))}{Q\tau'/(\mu R)}\biggr]_+  - \frac{\gamma_J^0+2\beta_0}{2}\frac{Q^2}{\mu^2}\biggl[ \frac{\mu^2}{Q^2\tau'}\biggr]_+  \biggr\}  \\
&\qquad =6C_F\biggl\{ \frac{\Gamma_0}{2} \Bigl [ \ln^2 \frac{Q^2\tau}{\mu^2} \ln\Bigl(1+\frac{\tau}{R^2}\Bigr) - 2\ln\frac{Q^2\tau}{\mu^2}\Li_2\Bigl(\frac{\tau}{\tau+R^2}\Bigr) + 2\Li_3\Bigl(\frac{\tau}{\tau+R^2}\Bigr)\Bigr] \\
&\qquad\qquad\quad - \Gamma_0 \Bigl [ \ln^2 \frac{Q\tau}{\mu R} \ln\Bigl(1+\frac{\tau}{R^2}\Bigr) - 2\ln\frac{Q\tau }{\mu R}\Li_2\Bigl(\frac{\tau}{\tau+R^2}\Bigr) + 2\Li_3\Bigl(\frac{\tau}{\tau+R^2}\Bigr)\Bigr] \\
&\qquad\qquad\quad - \frac{\gamma_J^0 + 2\beta_0}{2} \Bigl[ \ln\frac{Q^2\tau}{\mu^2}\ln\Bigl(1+\frac{\tau}{R^2}\Bigr) - \Li_2\Bigl(\frac{\tau}{\tau+R^2}\Bigr)\biggr\}\,.
\end{split}
\ee

Meanwhile, the $\Lambda$-dependent parts $S_{s}\otimes S_{sc}^2$ in \eq{jetthrustrefact} can be combined into the function,
\be
\begin{split}
\label{eq:Sveto2}
S_{\text{veto}}^\cu(\Lambda,R,\mu) &\equiv \iLP\Bigl\{ \frac{1}{\lambda} \wt S_s(\lambda,\mu) \wt S_{sc}(\lambda /R,\mu)^2 \Bigl\} = 1 + \frac{\as}{4\pi}\Bigl[ 2\Gamma_0 \ln R \ln\frac{\mu^2}{4\Lambda^2 R} + c_{ss}^1 + 2c_{sc}^1\Bigr] \\
&\quad + \Bigl(\frac{\as}{4\pi}\Bigr)^2 \biggl\{ 2\Gamma_0^2\ln^2\! R \ln^2\! \frac{\mu^2}{4\Lambda^2 R}  + \frac{4}{3}\Gamma_0\beta_0 \Bigl( 3 \ln R \ln\frac{\mu}{2\Lambda}\ln\frac{\mu}{2\Lambda R} + \ln^3 R\Bigr) - \frac{4\pi^2}{3}(\Gamma_0)^2  \ln^2\! R\\
&\quad + 2\bigl[ \Gamma_1 + \Gamma_0 (c_{ss}^1 + 2c_{sc}^1)\bigr]\ln\! R \ln\! \frac{\mu^2}{4\Lambda^2R}   + 2\beta_0(c_{ss}^1 + 2c_{sc}^1)\ln\frac{\mu}{2\Lambda} + \bigl(\gamma_{ss}^1 - 4\beta_0 c_{sc}^1\bigr) \ln R + \text{consts.}\biggr\}\,,
\end{split}
\ee
\end{widetext}
where $c_{ss,sc}^1$ were given in \eq{SsSsc1coeffs}, and we have used \eq{softnoncusp} to relate $\gamma_{ss,sc}^1$ to each other. We note that $S_{\text{veto}}^c$ almost takes the form of product of two multiplicatively renormalized functions given by \eqs{Fexp}{Fcoeffs}, except for the $(4\pi^2/3)(\Gamma_0)^2 \ln^2 R$ term, which arises because $S_{\text{veto}}^c$ is actually built out of a convolution of pieces in the energy variable $E$, integrated up to $\Lambda$.

We note that $S_{\text{veto}}$ has no non-cusp anomalous dimension even at two loops, since from \eq{softnoncusp},
\be
\label{eq:vetononcusp}
\gamma_{\text{veto}}^1 \equiv \gamma_{ss}^1 + 2\gamma_{sc}^1 = 0\,.
\ee
Thus, without refactorizing $S_{\text{veto}}$ into soft and soft-collinear pieces as in \eq{Skrefact}, one would not know that the $\gamma_{ss}^1 \ln R$ term in \eq{Sveto2} is really built from the anomalous dimensions of two functions $S_{s,sc}$ which obey an RGE with the non-cusp anomalous dimension $\gamma_{ss}^1 = -2\gamma_{sc}^1$ and thus have no way to resum the associated logs of $R$.
Also in the form \eq{Sveto2}, one would not know that the coefficients of the double logs of $R$ are given by the cusp anomalous dimension; this can only be deduced from the refactorized form \eq{Skrefact} and the RG evolution of the single-scale dependent pieces $S_{s,sc}$.

Now, combining the pieces \eqs{JSconvolution}{Sveto2} together with the hard function given by \eq{Fexp} and $S_{ng}$ given by \eq{Sng2}, we can organize the cross section \eq{jetthrustrefact} into two $\mu$-independent factors. Expanding the cross section in $\as(Q)$, we obtain
\be
\label{eq:sigmatau1tau2}
\sigma_\cu(\tau_1,\tau_2,\Lambda,R) = \sigma_\cu(\tau_1;\Lambda,R)\sigma_\cu(\tau_2;\Lambda,R)\,,
\ee
where to $\cO(\as^2)$ for the cone algorithm,
\begin{widetext}
\begin{align}
\label{eq:jetthrustcs2}
\sigma_\cu^{\rm cone} (\tau;\Lambda,R) &= 1 + \frac{\as(Q)}{4\pi} \biggl[ \Gamma_0\Bigl( 2\ln R \ln\frac{Q}{2\Lambda} - \frac{1}{2}\ln^2\frac{\tau}{R^2}\Bigr) - \frac{\gamma_J^0}{2}\ln\tau + c_{\text{tot}}^1 + 6C_F\ln\Bigl(1 + \frac{\tau}{R^2}\Bigr)\biggr] \\
& + \Bigl(\frac{\as(Q)}{4\pi}\Bigr)^2 \biggl\{ \frac{(\Gamma_0)^2}{2}\Bigl( 2\ln R \ln\frac{Q}{2\Lambda} - \frac{1}{2}\ln^2\frac{\tau}{R^2}\Bigr)^2 - \frac{\Gamma_0\gamma_J^0}{2} \ln\tau \Bigl( 2\ln R \ln\frac{Q}{2\Lambda} - \frac{1}{2}\ln^2\frac{\tau}{R^2}\Bigr) \nn \\
&\qquad\qquad + 2\Gamma_0\beta_0 \Bigl( \ln R \ln\frac{Q}{2\Lambda}\ln\frac{Q}{2\Lambda R} + \frac{1}{4} \ln\tau \ln^2\frac{\tau}{R^2}\Bigr)- (\Gamma_0)^2 \frac{\pi^2}{12}\Bigl( \ln^2\frac{\tau}{R^2} + 8\ln^2 R\Bigr)   \nn  \\
&\qquad\qquad + \biggl[ \Gamma_1 + \Gamma_0 \biggl( c_{\text{tot}}^1 + 6C_F\ln\Bigl(1+\frac{\tau}{R^2}\Bigr)\biggr)\biggr]\Bigl( 2\ln R \ln\frac{Q}{2\Lambda} - \frac{1}{2}\ln^2\frac{\tau}{R^2}\Bigr) + \frac{(\gamma_J^0)^2 + 2\gamma_J^0\beta_0}{8} \ln^2\tau \nn  \\
&\qquad\qquad -\frac{4}{3}\pi^2 C_F C_A  \ln^2 \frac{Q\tau}{2\Lambda R^2} +  \Bigl[ \Bigl( - 8\zeta_3 - \frac{4}{3} \Bigr) C_F C_A + \frac{8}{3}C_F T_F n_f + \frac{4\pi^2}{3}\beta_0 C_F\Bigr] \ln\frac{Q\tau}{2\Lambda R^2} \nn \\
&\qquad\qquad + \biggl[- \frac{\gamma_J^1}{2} + \frac{\gamma_{ss}^1}{2}  - \frac{\gamma_J^0}{2}\biggl(c_{\text{tot}}^1 + 6C_F\ln\Bigl(1+\frac{\tau}{R^2}\Bigr)\biggr)\biggr]\ln\tau \nn  \\
&\qquad\qquad - \Bigl[ \Gamma_0\gamma_J^0\frac{\pi^2}{12} - \zeta_3\Gamma_0^2 - 6C_F\Gamma_0 \Li_2\Bigl(\frac{\tau}{\tau+R^2}\Bigr)\Bigr]\ln\frac{\tau}{R^2} \nn  \\
&\qquad\qquad - \beta_0 \Bigl[ \Bigl( c_J^1 + 6C_F\ln\Bigl(1+\frac{\tau}{R^2}\Bigr)\Bigr)\ln\tau + 2c_{\text{in}}^1 \ln\frac{\tau}{R} + c_{\text{veto}}^1\ln\frac{2\Lambda}{Q} + 2c_{sc}^1\ln R\Bigr] + \text{const.}\biggr\}  \,, \nn
\end{align}
\end{widetext}
where
\be
c_{\text{tot}}^1 = \frac{c_H^1}{2} + c_J^1 + c_{\text{in}}^1 + \frac{c_{\text{veto}}^1}{2}\,,\quad c_{\text{veto}}^1 = c_{ss}^1 + 2c_{sc}^1\,,
\ee
and where we have dropped purely non-singular terms in \eq{jetthrustcs2}.  We also used the consistency relation \eq{HJSconsistency} to replace $\gamma_H$ in the single log terms.

\subsection{Measured and Unmeasured Jet Functions}
\label{ssec:unmeasured2}

We observed in \sec{integrate} that integrating the one-loop jet thrust distribution up to $\tau=\taumax$ reproduces the total one-loop two-jet rate for the cone and \kT/Sterman-Weinberg algorithms. In fact we can identify the ``unmeasured jet function'' in \eq{naivefactorization} with the convolution of the measured jet function in \eq{jetthrustcs} and the $S_{\text{in}}$ part of the refactorized soft function in \eq{Skrefact}. Namely, we find that
\begin{align}
\label{eq:Junmeasfromtau}
J_{\text{un}}^\alg(QR,\mu) &= \int_0^{\taumax}\!\! d\tau \int\!  dt\, dk\,\delta\Bigl(\tau - \frac{t}{Q^2} - \frac{k}Q\Bigr)  \nn \\
&\qquad\times J_n^\alg(t,R,\mu)S_{\text{in}}(k/R,\mu)\,,
\end{align}
Explicitly at $\cO(\as)$,
\begin{align}
&\int_0^{\tau}\!\! d\tau'\int\! dt\, dk\,\delta\Bigl(\tau' \minus \frac{t}{Q^2} \minus \frac{k}Q\Bigr)   J_n^\alg(t,R,\mu)S_{\text{in}}(k/R,\mu) \nn \\
&= 1 + \frac{\as}{4\pi} \Bigl[ \Gamma_0 \Bigl( \frac{1}{2}\ln^2 \frac{\mu^2}{Q^2\tau} - \ln^2\frac{\mu R}{Q\tau}\Bigr) \\
&\qquad\qquad + \frac{1}{2}\gamma_J^0 \ln\frac{\mu^2}{Q^2\tau} + c_J^1 + c_{\text{in}}^1 + \Delta J_\alg^c(\tau,R)\Bigr]\,, \nn
\end{align}
where
\be
\Delta J_\alg^c (\tau,R) = \int_0^{Q^2\tau} dt\, \Delta J^\alg(t,R)\,.
\ee
At $\tau= \taumax$, we obtain for the cone ($\taumax = R^2$) and \kT ($\taumax=R^2/4$) algorithms,
\be
\label{eq:Juncone1}
\begin{split}
J_n^\cone\otimes S_{\text{in}} = 1 &+ \frac{\as}{4\pi}\Bigl[ \Gamma_0 \ln^2 \frac{\mu}{QR} + \gamma_J^0 \ln\frac{\mu}{QR} \\
&\qquad + \Bigl( 7 - \frac{5\pi^2}{6} + 6\ln 2\Bigr)C_F\Bigr]\,,
\end{split}
\ee
and
\be
\label{eq:JunkT1}
\begin{split}
J_n^\kT \otimes S_{\text{in}} = 1 &+ \frac{\as}{4\pi}\Bigl[ \Gamma_0  \ln^2 \frac{\mu}{QR}     + \gamma_J^0 \ln\frac{\mu}{QR} \\
&\qquad + \Bigl( 13 - \frac{3\pi^2}{2}\Bigr)C_F\Bigr]\,,
\end{split}
\ee
which equal the one-loop unmeasured jet functions given in \eq{J1}, with constants in \eq{cJalg}.

Let us continue the evaluation of the right-hand side of \eq{Junmeasfromtau} to $\cO(\as^2)$. Although an independent direct computation of $J_{\text{un}}$ to $\cO(\as^2)$ does not yet exist, we can construct the right-hand side of \eq{Junmeasfromtau} and check that it is consistent as a definition of $J_{\text{un}}$ at least to $\cO(\as^2)$. Namely, $J_{\text{un}}$ should obey the multiplicative RGE
\be
\label{eq:unmeasRGE}
\mu \frac{d}{d\mu} J_{\text{un}}(QR,\mu) = \gamma_{J\text{un}}(\mu)J_{\text{un}}(QR,\mu)\,,
\ee
with an anomalous dimension given by
\be
\label{eq:gammaJunmeas}
\gamma_{J\text{un}}(\mu) = 2\Gamma_{\text{cusp}}[\as(\mu)]\ln\frac{\mu}{QR} +  \gamma_{J\text{un}}[\as(\mu)]\,.
\ee
That is, $J_{\text{un}}$ must take the form \eq{Fexp}, which solves \eq{unmeasRGE}, with the cusp part of the anomalous dimension proportional to $\Gamma_{\text{cusp}}$ according to \eq{gammaJunmeas}.

We have evaluated the right-hand side of \eq{Junmeasfromtau} to $\cO(\as^2)$ for arbitrary $\tau$ in \eq{JSconvolution}.
Setting $\tau = \taumax=R^2$ there, including the contributions from $\Delta J_\alg$ in \eq{DeltaJconvolution}, gives $J_{\text{un}}^\cone$ constructed according to \eq{Junmeasfromtau} to $\cO(\as^2)$.  The relations that we need are given in \eq{DeltaJconerelations}. Using these, we obtain for the 2-loop unmeasured jet function,
\begin{widetext}
\begin{align}
\label{eq:Junmeas2}
J_{\text{un}}^{\cone} = 1 &+ \frac{\as}{4\pi}\Bigl[ \Gamma_0\ln^2 \frac{\mu}{QR} + \gamma_J^0 \ln\frac{\mu}{QR}  + \Bigl( 7 - \frac{5\pi^2}{6} + 6\ln 2\Bigr)C_F\Bigr] \\
& +  \Bigl(\frac{\as}{4\pi}\Bigr)^2 \biggl\{ \frac{(\Gamma_0)^2}{2} \ln^4 \frac{\mu}{QR} + \Gamma_0\Bigl(\gamma_J^0  + \frac{2}{3}\beta_0\Bigr) \ln^3\frac{\mu}{QR}   + \Bigl[ \Gamma_1 \plus \Gamma_0 \Bigl( 7\minus \frac{5\pi^2}{6} \plus 6\ln 2\Bigr) C_F + \frac{(\gamma_J^0)^2}{2} \plus \gamma_J^0 \beta_0\Bigr]\ln^2\frac{\mu}{QR} \nn \\
&\qquad\qquad + \Bigl[ \gamma_J^1 + \gamma_{\text{in}}^1 + (\gamma_J^0 + 2\beta_0)\Bigl( 7 - \frac{5\pi^2}{6} + 6\ln 2\Bigr)C_F\Bigr] \ln\frac{\mu}{QR}  + \text{consts.} \biggr\}. \nn
\end{align}
\end{widetext}
We note this is precisely the form of a function obeying a multiplicative RGE, as we expect from the na\"{\i}ve factorization in \eq{naivefactorization}, which is given by the expansion \eq{Fexp}, with
\be
\label{eq:GammaJunmeas}
\Gamma_{J\text{un}} = \Gamma_{\text{cusp}}\,,\quad \gamma_{J\text{un}}^1 = \gamma_J^1 + \gamma_{\text{in}}^1\,.
\ee
That $J_{\text{un}}$ constructed according to \eq{Junmeasfromtau} would simplify to this form obeying a simple multiplicative RGE was by no means obvious from the beginning, and is a strong consistency check of the notion of the unmeasured jet function. The non-cusp anomalous dimension in \eq{GammaJunmeas} can be constructed out of \eq{softnoncusp} and \eq{gammaJ},
\be
\label{eq:Junmeasnoncusp}
\boxed{
\gamma_{J\text{un}} \equiv  \gamma_J + \gamma_{\text{in}} = -\frac{\gamma_H}{2}\,,
}
\ee
where we used \eqs{gammaequalities}{HJSconsistency} to relate $\gamma_{\text{in}} = -\gamma_{ss}/2$ and $\gamma_H = -2\gamma_J + \gamma_{ss}$. This can be computed to $\cO(\as^3)$ using the results quoted in \appx{anomdim}.
This demonstrates that the unmeasured jet function introduced in \cite{Ellis:2010rw} does not simply have the same anomalous dimension as the inclusive jet function (as was already seen from the differing cusp parts of the anomalous dimensions). Physically this is because the unmeasured jet function includes energetic radiation at the edges of the jet that is treated as csoft when measured with a thrust or mass measurement, but becomes of the same scaling as the hard collinear radiation when integrated up to $\tau=R^2$, as we see in \eq{csscaling}.

Once contributions from $S_{\text{in}}$ are lumped together with $J_n^\alg$ when $\tau=R^2$ to make the $J_{\text{un}}$, the veto-dependent parts of the soft function $S_{s}\otimes S_{sc}^2$ in \eq{jetthrustrefact} do not contribute anything to the non-cusp anomalous dimension according to \eq{vetononcusp}, resulting in the simple proportionality between $\gamma_{J\text{un}}$ and $\gamma_H$ in \eq{Junmeasnoncusp}.

In the jet thrust cross section, we can still distinguish between the global logs coming from RG running between the scales in \fig{allscales}, and the NGL in the fixed-order total soft function \eq{Skrefact}. In the integrated total 2-jet rate, however, this NGL, in \eq{Sng}, becomes a log of $2\Lambda/Q$:
\be
\label{eq:NGL}
\ln^2\frac{Q\tau}{2\Lambda R^2} \overset{\tau=R^2}{\longrightarrow} \ln^2\frac{Q}{2\Lambda}\,,
\ee
which is indistinguishable from other global logs in the 2-jet rate. It will require a more complete dissection of the dynamics in the jets to fully resum both types of logs, such as those recently advanced in \cite{Larkoski:2015kga,Larkoski:2015zka,Becher:2015hka,Neill:2015nya}. Our treatment here provides key steps that must be part of such a complete analysis, allowing a resummation of all the global logs appearing in the jet thrust cross section, a clearer understanding of the relation of the integrated jet thrust and total 2-jet cross sections, and a determination of the evolution of the unmeasured jet function to two loops.

\subsection{2-Jet Rate}

The total 2-jet rate can be computed from the double integral of the double differential jet thrust distribution \eq{jetthrustcs}:
\be
\sigma_{\text{2-jet}}^\cone(\Lambda,R) = \int_0^{R^2}\!\! d\tau_1\int_0^{R^2}\!\!\ d\tau_2 \frac{1}{\sigma_0}\frac{d\sigma(\Lambda,R)}{d\tau_1d\tau_2}\,,
\ee
for the cone algorithm. The 2-jet rate is determined by this double integral instead of a single integral of the total jet thrust distribution \eq{jetthrustsum} since the size of both jets is limited to radius $R$, independent of the other jet. (The single integral would allow one jet to be fatter than $R$ and one narrower.) From \eq{jetthrustcs} and the construction of the unmeasured jet function in \eq{Junmeasfromtau}, we find the total 2-jet cross section is built from products of the unmeasured jet functions:
\begin{align}
\label{eq:conejetcs}
\sigma_{\text{2-jet}}^\cone(\Lambda,R)  &=  H(Q^2,\mu) J_{\text{un}}(QR,\mu)^2 \\
&\times S_s(\Lambda,\mu)\otimes S_{sc}(\Lambda R,\mu)^2\otimes S_{ng}^\cone(\Lambda,R)\,. \nn
\end{align}
We computed the combination $S_{\text{veto}}= S_s\otimes S_{sc}^2$ in \eq{Sveto2}.
The non-global part is given at $\cO(\as^2)$ by \eq{Sng} with $k\to QR^2$. The total jet rate can be obtained from \eqs{sigmatau1tau2}{jetthrustcs2} with $\tau=R^2$. We obtain
\be
\sigma_{\text{2-jet}}^\cone = 1 + \frac{\as(Q)}{4\pi}\sigma_\cone^{(1)} +  \Bigl(\frac{\as(Q)}{4\pi}\Bigr)^2\sigma_\cone^{(2)}\,,
\ee
where $\sigma_\cone^{(1)}$ is
\be
\sigma_\cone^{(1)} = 4\Gamma_0\ln R\ln\frac{Q}{2\Lambda} - 2\gamma_J^0 \ln R + c_\cone^1\,,
\ee
where $c_\cone^1 = (12 \ln 2- 2)C_F$.
As for the two-loop terms,
\begin{widetext}
\be
\label{eq:conerate2}
\begin{split}
\sigma_\cone^{(2)} &= 8 \Gamma_0^2 \ln^2 R \ln^2 \frac{2\Lambda}{Q} + 8\Gamma_0\gamma_J^0 \ln^2 R \ln\frac{2\Lambda}{Q} + 4\Gamma_0 \beta_0 \Bigl( \ln^2\frac{2\Lambda}{Q}\ln R + \ln \frac{2\Lambda}{Q}\ln^2 R\Bigr)  \\
&\quad - 4[ \Gamma_1 + \Gamma_0 c_\cone^1] \ln R\ln\frac{2\Lambda}{Q} + 2\Bigl[(\gamma_J^0)^2 + \gamma_J^0 \beta_0 - \frac{2\pi^2}{3}\Gamma_0^2\Bigr] \ln^2 R \\
&\quad  - \frac{8\pi^2}{3}C_F C_A \ln^2 \frac{2\Lambda}{Q} + C_F\Bigl[ \Bigl( 16\zeta_3 + \frac{8}{3}\Bigr) C_A - \frac{16}{3}T_F n_f - \frac{8\pi^2}{3}\beta_0\Bigr]\ln\frac{2\Lambda}{Q} \\
&\quad + \bigl[\gamma_H^1 + \gamma_{ss}^1 -2  \gamma_J^0 c_\cone^1  -4 (c_{J\text{un}}^1+c_{sc}^1)\beta_0\bigr ] \ln R  -  2(c_{ss}^1+2c_{sc}^1)\beta_0 \ln\frac{2\Lambda}{Q} + \text{consts.} \,,
\end{split}
\ee
where we used \eqs{softnoncusp}{Junmeasnoncusp} to relate the non-cusp anomalous dimensions in front of the single $\ln R$. Plugging in the explicit values of the coefficients in \eq{conerate2}, we obtain
\begin{align}
\sigma_\cone^{(2)} &= 4C_F^2 \biggl\{ \Bigl( 32 \ln^2\frac{2\Lambda}{Q}  + 48 \ln\frac{2\Lambda}{Q} + 18 - \frac{16\pi^2}{3}\Bigr)\ln^2 R  + \Bigl[ (8- 48\ln 2)\ln\frac{2\Lambda}{Q} + \frac{9}{2} + 2\pi^2 - 24\zeta_3 - 36\ln 2\Bigr]\ln R \biggr\} \nn \\
& + 4C_F C_A \biggl\{ \Bigl( \frac{44}{3} \ln\frac{2\Lambda}{Q} + 11\bigr) \ln^2 R + \Bigl[ \frac{44}{3}\ln^2 \frac{2\Lambda}{Q} + \Bigl( \frac{4\pi^2}{3} - \frac{268}{9}\Bigr)\ln\frac{2\Lambda}{Q} - \frac{57}{2} + 12\zeta_3 - 22\ln 2\Bigr]\ln R \\
&\qquad\qquad - \frac{2\pi^2}{3}\ln^2\frac{2\Lambda}{Q} + \Bigl( \frac{2}{3} + 4\zeta_3 - \frac{11\pi^2}{9}\Bigr)\ln\frac{2\Lambda}{Q}\biggr\} \nn \\
&+ 4C_F T_F n_f \biggl\{ \Bigl( -\frac{16}{3}\ln\frac{2\Lambda}{Q} - 4\Bigr)\ln^2 R + \Bigl( - \frac{16}{3}\ln^2\frac{2\Lambda}{Q} + \frac{80}{9} \ln\frac{2\Lambda}{Q} + 10 + 8\ln 2\Bigr) \ln R - \Bigl( \frac{4}{3} - \frac{4\pi^2}{9}\Bigr)\ln\frac{2\Lambda}{Q}\biggr\} \nn \\
& + \text{consts.} \nn
\end{align}
\end{widetext}
This agrees with the result given in \cite{Becher:2015hka} for all the $\ln^n R$ terms, as well as the $C_F T_F n_f \ln(2\Lambda/Q)$ terms, all of which can be accurately predicted by our factorization theorem \eq{conejetcs} with soft/soft-collinear emissions from one collinear direction. Emissions from two or more separate Wilson lines in the fundamental and adjoint representations will affect the logs of $2\Lambda/Q$, beginning with single logs in the $C_F^2$ and $C_F C_A$ color factors at $\cO(\as^2)$, as found in the results of \cite{Becher:2015hka}. The $n_f$ term still comes from emissions from one Wilson line and is thus accurately predicted.

\section{Comparison with Fixed-Order QCD Calculations}
\label{sec:EERAD}

\begin{figure*}[t]
\includegraphics[width=.675\columnwidth]{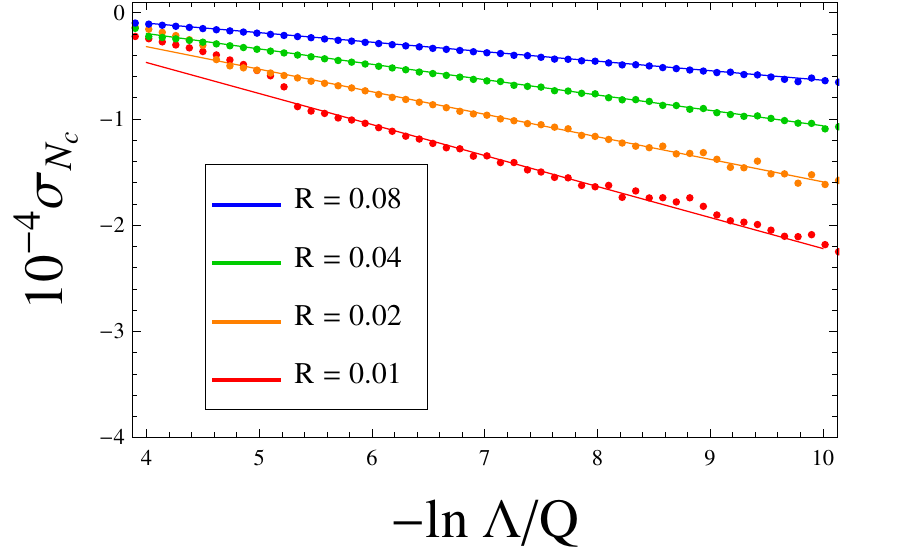} \includegraphics[width=.674\columnwidth]{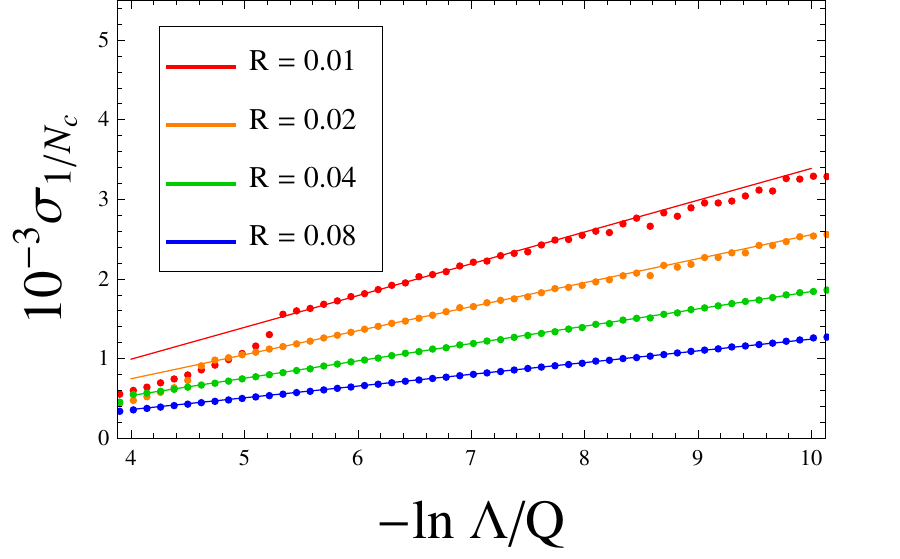} \includegraphics[width=.675\columnwidth]{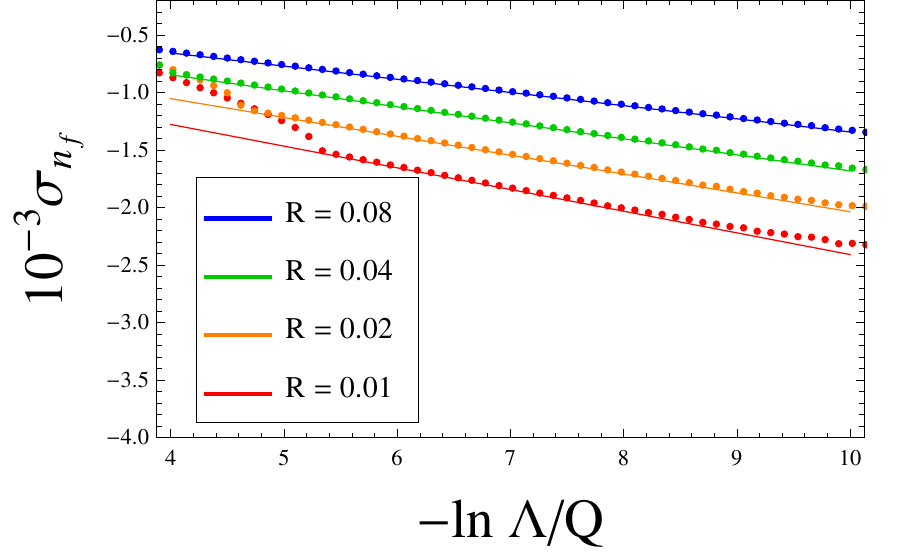}~\\[+5pt]
\includegraphics[width=.63\columnwidth]{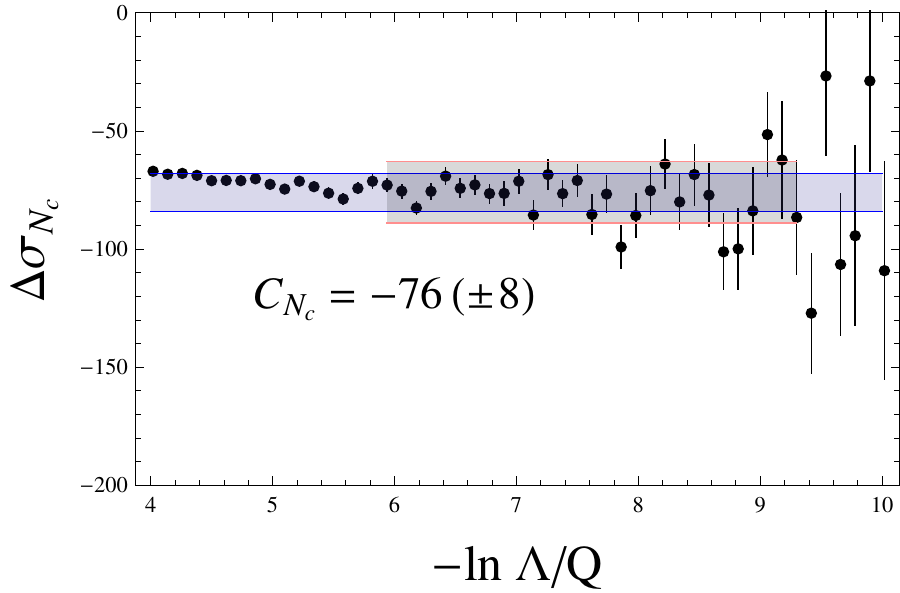}~~~~\includegraphics[width=.63\columnwidth]{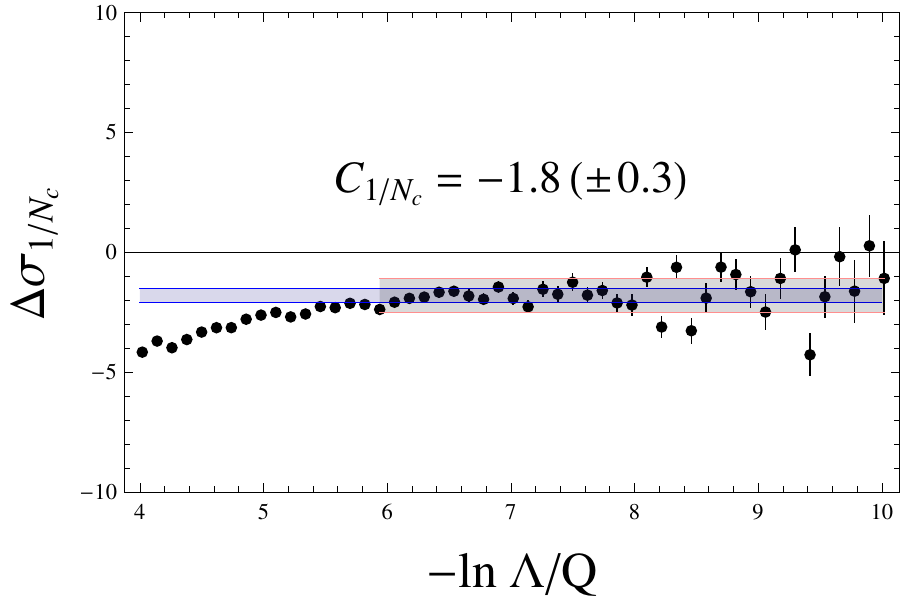}~~~~~~ \includegraphics[width=.63\columnwidth]{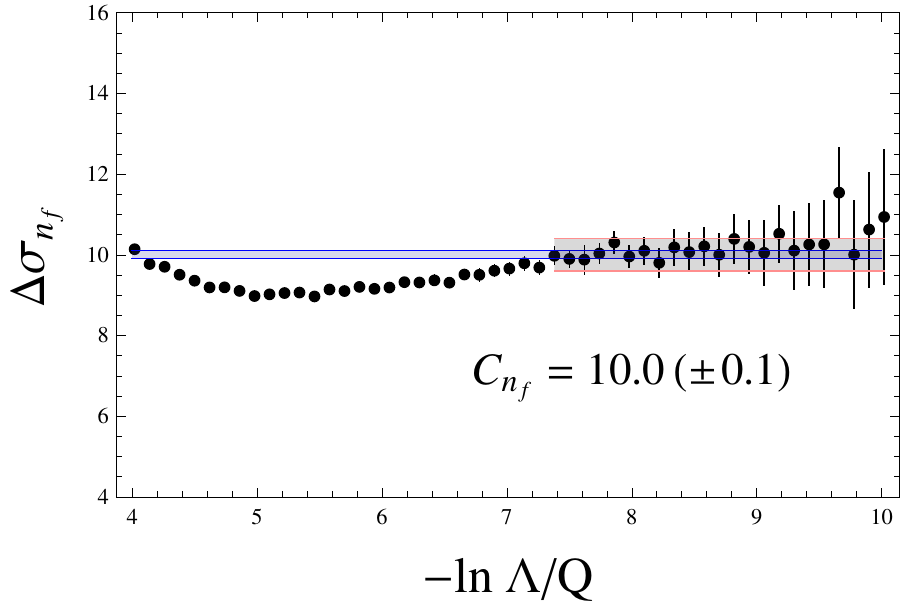}~~
\vspace{-2mm}
\caption{Comparison between \eq{conerate2} (solid lines) and the {\tt EERAD3} output (dots) for $\sigma_{N_c}$, $\sigma_{1/N_c}$ and $\sigma_{n_f}$, plotted (upper) as functions of $-\ln \Lambda/Q$ with fixed $R$. The linear dependence on $-\ln\Lambda/Q$ is shown as a check for the $\ln R$-enhanced terms predicted by our factorization theorem. Note the numerical sensitivity for small values of $R$ and $\Lambda/Q$. The differences (dots, with error bars) between \eq{conerate2} and the {\tt EERAD3} output are plotted (lower) as functions of $-\ln \Lambda/Q$ with $R =$ 0.32. Performing $\chi^2$ fits within the fit regions (red), we extract the coefficients $C_{N_c}=-76\pm 8$, $C_{1/N_c}=-1.8\pm0.3$ and $C_{n_f}=10.0\pm0.1$. }
\label{fig:eeradL}
\end{figure*}

\begin{figure*}
\begin{center}
\includegraphics[width=.68\columnwidth]{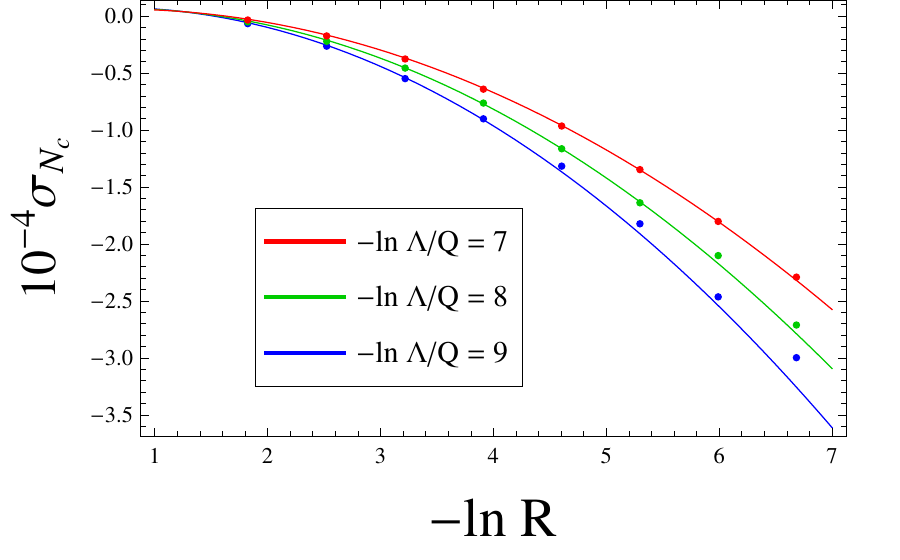} \includegraphics[width=.68\columnwidth]{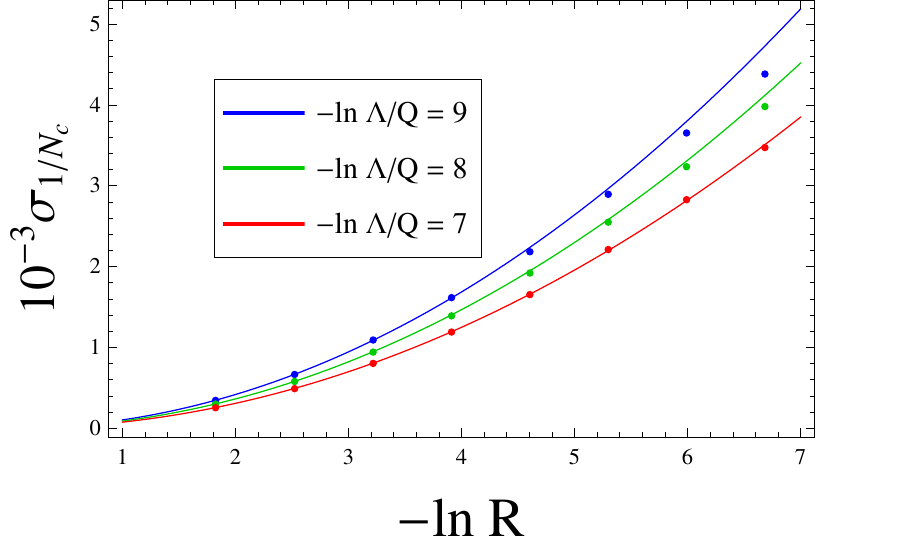} \includegraphics[width=.68\columnwidth]{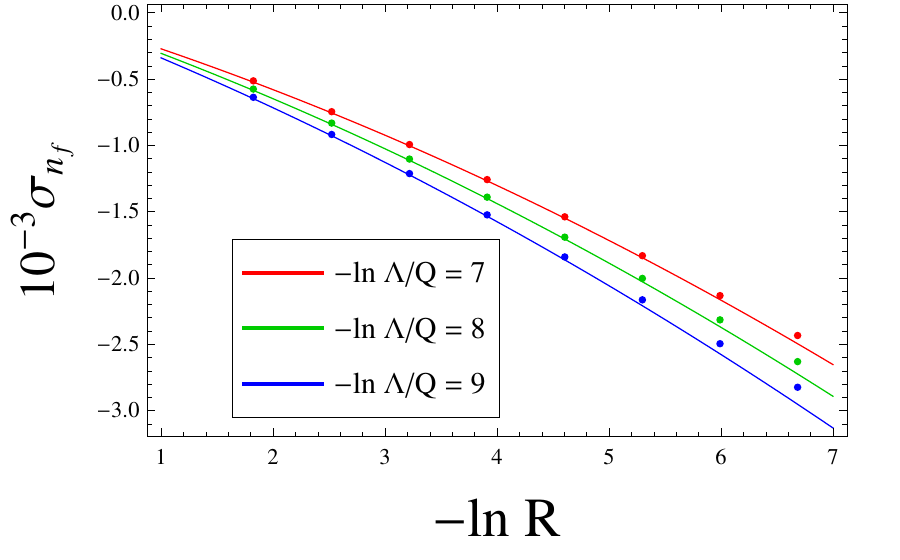}
\end{center}
\vspace{-5mm}
\caption{Comparison between Eq. (\ref{eq:conerate2}) (solid lines) and the output of {\tt EERAD3} (dots) for $\sigma_{N_c}$, $\sigma_{1/N_c}$ and $\sigma_{n_f}$, plotted as functions of $-\ln R$ with fixed $\Lambda/Q$. The quadratic dependence on $-\ln R$ is shown again as a check for the $\ln R$-enhanced terms predicted by our factorization theorem.}
\label{fig:eeradR}
\end{figure*}

We compare terms in our two-loop computation of the jet rate in the thrust cone algorithm to the full QCD fixed-order calculation implemented numerically in {\tt EERAD3} \cite{Ridder:2014wza}. We expect all the $\ln R$-enhanced terms as well as the $C_FC_A \ln^2(2\Lambda/Q)$ term, which is the well known leading NGL, and the $C_FT_Fn_f \ln(2\Lambda/Q)$ terms, which can be computed from soft emissions from a single Wilson line per jet, to be correct. To demonstrate, we subtract all the $\ln R$-enhanced terms and the $C_FC_A \ln^2(2\Lambda/Q)$ term from the numerical data and check whether the difference has at most single logs of $2\Lambda/Q$. We can only test those terms with an explicit $\ln(2\Lambda/Q)$ since we need to make a measurement of energy or other variable on the final state to be able to easily plot the \texttt{EERAD3} data. From the remainder, we can extract the coefficients of the $\ln(2\Lambda/Q)$ terms, whose accurate prediction requires additional operators in our factorization theorem.

Focusing on the terms that are $\Lambda$-dependent, we compare the differential distributions $d\sigma^{(2)}_{\rm cone}/dx$ plotted as functions of $x=-\ln \Lambda/Q$ for various choices of $R$ (\fig{eeradL}). After subtraction the differences in $d\sigma^{(2)}_{\rm cone}/dx$ should be constant in $x$, which allows us to extract the $\ln(2\Lambda/Q)$ coefficients. We also check the $R$-dependence by plotting with fixed $\Lambda/Q$ (\fig{eeradR}). To compare directly with the {\tt EERAD3} output, the distributions are decomposed into three color structures proportional to $N_c$, $1/N_c$ and $n_f$,
\be
    \frac{d\sigma^{(2)}_{\rm cone}}{dx}=\sigma_{N_c}+\sigma_{1/N_c}+\sigma_{n_f},
\ee
where $\sigma_{N_c}$ and $\sigma_{1/N_c}$ are linear combinations of the $C_F^2$ and $C_F C_A$ terms derived from \eq{conerate2}. The pure $R$-dependent terms can not be seen in $d\sigma^{(2)}_{\rm cone}/dx$, and their comparison is left for future work.

Power corrections are expected to become more significant when $QR^2 \gtrsim \Lambda$. Thus for smaller values of $R$, the agreement of the singular terms between QCD and SCET at two loops can only be seen at smaller values of $\Lambda/Q$. We evaluate the jet rate using {\tt EERAD3} for $R = 2^n \times 10^{-2}$ with $ -2 \leq n \leq 5$. The calculations require the technical cutoff in {\tt EERAD3} to be set to $10^{-15}$, and we have included six billion partonic events in the numerical integration. In the singular regions of $R$ and $\Lambda/Q\rightarrow0$, the upper plots in \fig{eeradL} show the linear dependence of $d\sigma^{(2)}_{\rm cone}/dx$ on $-\ln \Lambda/Q$, and the ones in \fig{eeradR} show the quadratic dependence on $-\ln R$. These show the agreement of the $\ln R$-enhanced terms predicted by our factorization theorem.

The differences between our analytic predictions and the {\tt EERAD3} outputs are shown in the lower plots in \fig{eeradL}. The differences are close to constant in $x=-\ln(\Lambda/Q)$, We perform $\chi^2$ fits within the fit regions (red) using constant functions to extract the coefficients $C_{N_c}$, $C_{1/N_c}$ and $C_{n_f}$ of the $\ln(2\Lambda/Q)$ terms in the jet rate. We use the {\tt EERAD3} output for $R=0.32$ in the fits, but the results do not change much when including more data with smaller $R$'s. The bands are fit uncertainties corresponding to the values of $\chi^2$ per degrees of freedom deviated from the minimum by at most 1. We obtain,
\begin{align}
\label{eq:fitEERAD}
C_{N_c}&=-76\pm 8 \nn \\
C_{1/N_c}&=-1.8\pm0.3 \nn \\
C_{n_f}&=10.0\pm0.1\,.
\end{align}
The coefficients extracted from the result given in \cite{Becher:2015hka} are
\begin{align}
\label{eq:fit}
C_{N_c}&=C_F N_c \Big(\frac{5}{3} \minus \frac{31\pi^2}{18} \minus 2\ln2 \minus 12\ln^22 \plus \zeta_3\Big)=-85.1 \nn \\
C_{1/N_c}&=\frac{C_F}{N_c}\Big(1-2\ln2+6\ln^22-5\zeta_3\Big)=-1.56 \nn \\
C_{n_f}&=C_F T_F n_f \Big(\frac{4\pi^2}{9}-\frac{4}{3}\Big)=10.2\,,
\end{align}
where $C_{n_f}$ is also predicted by our refactorization theorem. Note the good agreement of $C_{n_f}$ with the value extracted from {\tt EERAD3}. The values of $C_{N_c}$ and $C_{1/N_c}$ extracted from \cite{Becher:2015hka} are also consistent with the ones extracted from {\tt EERAD3} within the numerical uncertainties.

\section{Resummed Jet Thrust Cross Section}
\label{sec:NNLL}

\begin{figure*}[t]
\begin{center}
\includegraphics[width=1.95\columnwidth]{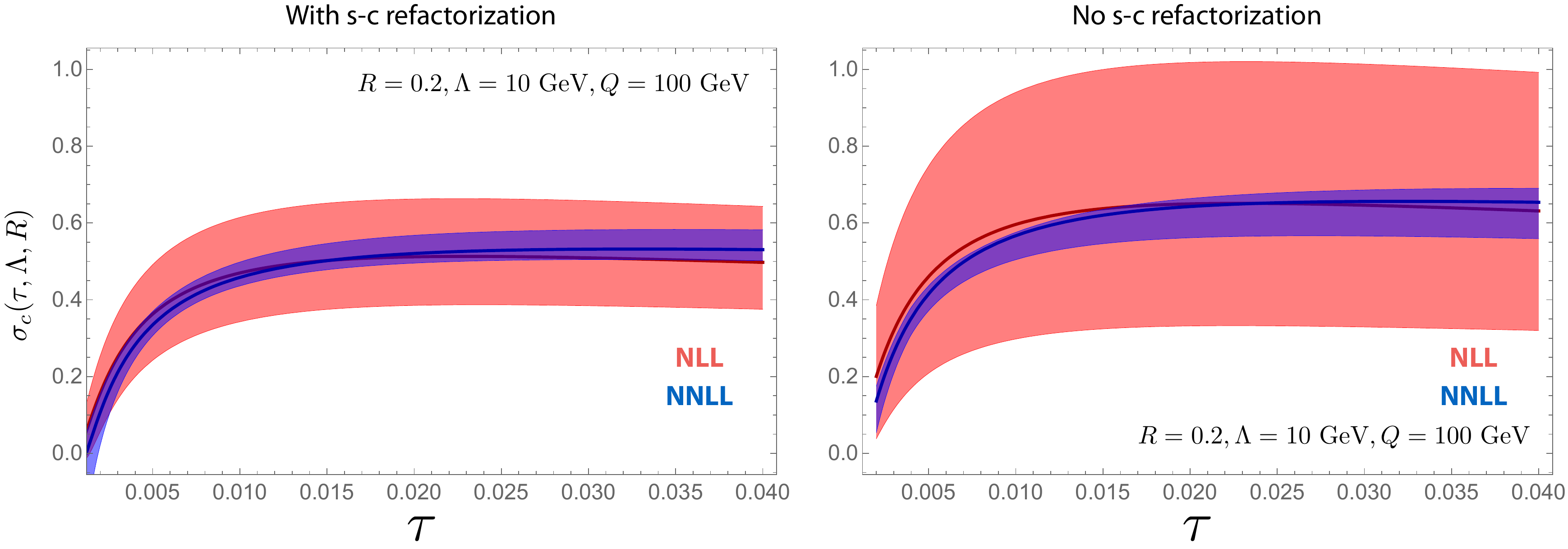}
\end{center}
\vspace{-7mm}
\caption{Resummed integrated jet thrust cross section. The integrated jet thrust cross section at $Q=100\GeV$, $R=0.3$ and $\Lambda=10\GeV$ using the thrust cone algorithm computed from \eq{resummedcs} is shown with (left) and without (right) the refactorization of soft- and soft-collinear modes in the soft function \eq{Skrefact}. The central values of scales on the left are $\mu_{ss}=2\Lambda$ and $\mu_{sc}=2\Lambda R$, and, on the right, $\mu_{ss}=\mu_{sc} = 2\Lambda\sqrt{R}$. The uncertainty bands come from the scale variations described in the main text. In both plots there is good convergence from NLL to NNLL accuracy (in global logs). Without refactorization, however, the overall scale variation is significantly larger, indicating better control of the perturbative series on the left. (The NGLs are not included in these plots.)}
\label{fig:NNLL}
\end{figure*}

The solutions of the RG equations obeyed by the hard, jet, and soft functions in the jet thrust cross section in \eq{jetthrustcs}, with the soft function refactorized as \eq{Skrefact}, give rise to the prediction for the resummed (integrated) jet thrust cross section (see, \eg \cite{Almeida:2014uva}),
\begin{align}
\label{eq:resummedcs}
&\sigma_\cu(\tau,\Lambda,R) = \frac{1}{\sigma_0}\int_0^\tau d\tau' \frac{d\sigma(\Lambda,R)}{d\tau'} = e^{\cK(\mu_H,\mu_J,\mu_{\text{in}},\mu_{ss},\mu_{sc},\mu)} \nn \\
&\times \Bigl(\frac{\mu_H}{Q}\Bigr)^{\omega_H(\mu_H,\mu)} \Bigl(\frac{\mu_J^2}{Q^2 \tau}\Bigr)^{2\omega_J(\mu_J,\mu)}  \Bigl(\frac{\mu_{\text{in}}R}{Q\tau}\Bigr)^{2\omega_{\text{in}}(\mu_{\text{in}},\mu)} \nn \\
&\times \Bigl(\frac{\mu_{ss}}{2\Lambda}\Bigr)^{\omega_{ss}(\mu_{ss},\mu)} \Bigl(\frac{\mu_{sc}}{2\Lambda R}\Bigr)^{2\omega_{sc}(\mu_{sc},\mu)} H(Q^2,\mu_H) \theta(\tau)\theta(\Lambda)  \nn \\
&\times \widetilde J\Bigl(\partial_\Omega + \ln\frac{\mu_J^2}{Q^2\tau},\mu_J\Bigr)^2\widetilde S_{\text{in}}\Bigl(\partial_\Omega + \ln\frac{\mu_{\text{in}}}{Q\tau},\mu_{\text{in}}\Bigr)^2 \frac{e^{\gamma_E \Omega}}{\Gamma(1\minus\Omega)}\nn \\
&\times \widetilde S_s\Bigl(\partial_\Upsilon + \ln\frac{\mu_{ss}}{2\Lambda},\mu_{ss}\Bigr) \widetilde S_{sc}\Bigl(\partial_\Upsilon+\ln\frac{\mu_{sc}}{2\Lambda R},\mu_{sc}\Bigr)^2 \frac{e^{\gamma_E \Upsilon}}{\Gamma(1 \minus\Upsilon)} \nn\\
&  \otimes S_{ng}(Q\tau/(2\Lambda R^2))\,,
\end{align}
where
\begin{subequations}
\label{eq:Keta}
\begin{align}
&\cK(\mu_H,\mu_J,\mu_{\text{in}},\mu_{ss},\mu_{sc},\mu) = K_H(\mu_H,\mu) + 2K_J(\mu_J,\mu) \nn \\
&\qquad + 2K_{\text{in}}(\mu_{\text{in}},\mu) + K_{ss}(\mu_{ss},\mu) + 2K_{sc}(\mu_{sc},\mu) \\
&\Omega \equiv\Omega(\mu_J,\mu_{\text{in}},\mu) = 2\omega_J(\mu_J,\mu) + 2\omega_{\text{in}}(\mu_{\text{in}},\mu) \\
&\Upsilon \equiv \Upsilon(\mu_{ss},\mu_{sc},\mu) = \omega_{ss}(\mu_{ss},\mu) + 2\omega_{sc}(\mu_{sc},\mu)\,,
\end{align}
\end{subequations}
and the individual evolution kernels are given by
\be
\begin{split}
K_F(\mu_F,\mu) &= -j_F \kappa_F K_\Gamma(\mu_F,\mu) + K_{\gamma_F}(\mu_F,\mu) \\
\omega_{F}(\mu_F,\mu) &= -\kappa_F \omega_\Gamma(\mu_F,\mu) \,,
\end{split}
\ee
where
\begin{align}
K_\Gamma(\mu_F,\mu) &= \int_{\mu_F}^\mu \frac{d\mu'}{\mu'} \Gamma_{\text{cusp}}[\as(\mu')]\ln\frac{\mu'}{\mu_F} \,, \\
\omega_\Gamma(\mu_F,\mu) &= \int_{\mu_F}^\mu \frac{d\mu'}{\mu'} \Gamma_{\text{cusp}}[\as(\mu')]\,, \nn \\
K_{\gamma_F}(\mu_F,\mu) &= \int_{\mu_F}^\mu \frac{d\mu'}{\mu'} \gamma_F[\as(\mu')]\,. \nn
\end{align}
Explicit expansions of these functions in $\as$ can be found in many places, \eg \cite{Almeida:2014uva}. The values of the coefficients $j_F,\kappa_F$ in \eq{Keta} for the various functions in \eq{resummedcs} are given by
\be
\begin{split}
j_H &= j_{\text{in}} = j_{ss}=j_{sc} = 1\,,\quad j_J = 2\,,\\
\kappa_H &= 4\,,\kappa _J = -2\,, \kappa_{\text{in}} = 2\,,\kappa_{ss} = -4\,,\kappa_{sc} = 2\,.
\end{split}
\ee
In \eq{resummedcs} we include the NGLs coming from $S_{\text{ng}}$ in \eq{Sng} simply at fixed order, here to $\cO(\as^2)$.

We have written \eq{resummedcs} using the formalism found in \cite{Becher:2006mr,Becher:2006nr} that turns the Laplace-transformed functions into differential operators with respect to $\Omega,\Upsilon$ to facilitate the transformation back to momentum space. With the results of \sec{S2}, especially
\eq{softnoncusp}, we have enough information to resum logs of $\tau$, $R$, and $\Lambda/Q$ to NNLL accuracy (modulo the NGLs).

In \fig{NNLL} we illustrate the integrated jet thrust cross section \eq{resummedcs} in the thrust cone algorithm at $Q=100\GeV$, $R=0.2$, and $\Lambda = 10\GeV$, with global log resummation at NLL and NNLL accuracy. At N$^k$LL accuracy, the cusp anomalous dimension and running of $\as$ are included to $\cO(\as^{k+1})$, the non-cusp anomalous dimensions to $\cO(\as^k)$, and the fixed-order hard, jet and soft matrix elements to $\cO(\as^{k-1})$. For N$^k$LL$'$ accuracy, the matrix elements would be included to $\cO(\as^k)$ \cite{Abbate:2010xh,Almeida:2014uva}, but we do not do this in this paper. We evaluate \eq{resummedcs} at the central values for the scales,
\be
\label{eq:centralscales}
\begin{split}
\mu=\mu_H &= Q\,,\ \mu_J = Q\sqrt{\tau}\,,\ \mu_{\text{in}} = \frac{Q\tau}{R}\,, \\
\mu_{ss} &= 2\Lambda\,, \ \mu_{sc} = 2\Lambda R\,.
\end{split}
\ee
To produce the result of the unrefactorized cross section, \eg with the soft function \eq{S1} at $\cO(\as)$, we set
\be
\label{eq:vetoscales}
\mu_{ss}= \mu_{sc} = 2\Lambda\sqrt{R}\,,
\ee
the choice one would make in an attempt to minimize logs in the $\cO(\as)$ soft function (which we saw in \eq{Sk2} does not work at higher order). For this exercise, we do not include the NGLs \eq{Sng} at unprimed NLL and NNLL accuracy, counting it as part of the fixed-order soft function, but of course in a full analysis we would have to include them.

For the estimation of perturbative uncertainty, we perform several sets of scale variations. First, we vary all scales in \eq{centralscales} together up and down by factors of 2. Next, we vary each of the $\tau$-dependent scales $\mu_{J,\text{in}}$ up and down using the formulae:
\be
\begin{split}
\mu_J(\tau;e_J) &= \Bigl[1 + e_J\Bigl(1 - \frac{\tau}{\taumax}\Bigr)^2\Bigr]Q\sqrt{\tau}\,, \\
\mu_{\text{in}}(\tau;e_{\text{in}}) &= \Bigl[1 + e_{\text{in}}\Bigl(1 - \frac{\tau}{\taumax}\Bigr)^2\Bigr]\frac{Q\tau}{R}\,,
\end{split}
\ee
and vary $e_{J,\text{in}}$ each between $(-0.5,0.5)$ (\emph{cf.}  \cite{Abbate:2010xh,Kang:2013nha}). Recall $\taumax =R^2$ for this cone algorithm.  Finally, we vary the veto-dependent scales $\mu_{ss,sc}$ each up and down by 2 (for the unrefactorized case, we vary the single soft veto scale in \eq{vetoscales}). We then add all these variations in quadrature.  These give the uncertainty bands plotted in \fig{NNLL}.

For a more robust uncertainty estimation, profile functions should be used for the jet and csoft scales \cite{Abbate:2010xh,Ligeti:2008ac} instead of the canonical scales in \eq{centralscales}, but this level of detail is not what we are after here. We simply illustrate the broad qualitative effect on the quality of the resummation of logs of $\tau$ and $R$ with and without refactorization using the soft-collinear mode. In both plots in \fig{NNLL} we see good convergence as uncertainty is reduced from NLL to NNLL. In the right-hand plots, without refactorization, the overall uncertainties for the same amount of scale variation are a bit larger. This indicates better control over the perturbative series in the left-hand plot with refactorization and resummed logs of $R$.

We do not include a numerical check against full QCD results for the jet thrust cross section here; the validity of the prediction of refactorization for the fixed-order global logs up to $\cO(\as^2)$ and the leading NGLs in cone and \kT-type algorithms was tested against EVENT2 \cite{Catani:1996jh,Catani:1996vz} in \cite{Hornig:2011tg}, where excellent agreement was found.

By integrating \eq{resummedcs} up to $\tau=\taumax$ we could obtain the jet rate with logs of global origin resummed to NNLL, although in this case accounting for the NGLs is more essential as they also turn into logs of $2\Lambda/Q$ as in \eq{NGL}, indistinguishable from the global logs. We leave the proper factorization and resummation of the NGLs, applying, \eg the formalism of  \cite{Larkoski:2015kga,Larkoski:2015zka,Neill:2015nya}, to future work.

\section{Summary of New Results}
\label{sec:results}

Here we collect, for convenience, the key new results of our paper, up to $\cO(\as^2)$, in terms of anomalous dimension and beta function coefficients given in \appx{anomdim}. All of these are in fact known to $\cO(\as^3)$.

The collinear-soft function appearing in \eq{Skrefact} for measured soft radiation in a two-jet event confined to be inside a cone of radius $R$, in integrated form from 0 to $k$, is given up to $\cO(\as^2)$ by
\begin{align}
\label{eq:Sink}
&S_\text{in}^\cu(k/R,\mu) = 1 + \frac{\as}{4\pi} \Bigl( -\Gamma_0\ln^2\frac{\mu R}{k} + c_\text{in}^1\Bigr) \\
&\quad + \Bigl(\frac{\as}{4\pi}\Bigr)^2 \biggl[ \frac{1}{2}\Gamma_0^2 \ln^4\frac{\mu R}{k} - \frac{2}{3}\Gamma_0\beta_0\ln^3\frac{\mu R}{k} \nn \\
&\qquad \qquad \quad + \Bigl( -\Gamma_1 - c_\text{in}^1\Gamma_0 - \frac{\pi^2}{3}(\Gamma_0)^2\Bigr)\ln^2\frac{\mu R}{k} \nn \\
&\qquad \qquad \quad + \bigl( \gamma_\text{in}^1 + 2 c_\text{in}^1\beta_0 - 4\zeta_3(\Gamma_0)^2\bigr) \ln\frac{\mu R}{k} + c_\text{in}^2 \biggr]\,, \nn
\end{align}
where $c_\text{in}^1 = (\pi^2/6)C_F$ and $\gamma_\text{in}^1$ is given in \eq{softnoncusp}. The Laplace transform of \eq{Sink} is given by the form \eq{softexpansion}.

The global soft function, defined by \eq{Ssdef}, for soft energy outside two back-to-back jets integrated up to the veto $E<\Lambda$, is given up to $\cO(\as^2)$ by
\begin{align}
\label{eq:SsLambda}
&S_s^\cu(\Lambda,\mu) = 1 + \frac{\as}{4\pi} \Bigl( 2\Gamma_0\ln^2\frac{\mu}{2\Lambda} + c_{ss}^1\Bigr) \\
&\quad + \Bigl(\frac{\as}{4\pi}\Bigr)^2 \biggl[ 2\Gamma_0^2 \ln^4\frac{\mu}{2\Lambda} + \frac{4}{3}\Gamma_0\beta_0\ln^3\frac{\mu}{2\Lambda} \nn \\
&\qquad \qquad \quad + \Bigl( 2\Gamma_1 +2 c_{ss}^1\Gamma_0 - \frac{4\pi^2}{3}(\Gamma_0)^2\Bigr)\ln^2\frac{\mu}{2\Lambda} \nn \\
&\qquad \qquad \quad + \bigl( \gamma_{ss}^1 + 2 c_{ss}^1\beta_0 - 16\zeta_3(\Gamma_0)^2\bigr) \ln\frac{\mu}{2\Lambda} + c_{ss}^2 \biggr]\,, \nn
\end{align}
where $c_{ss}^1 = -\pi^2 C_F$ and $\gamma_{ss}^1$ is given in \eq{softnoncusp}. We leave $c_{ss}^2$ undetermined. The Laplace transform of \eq{SsLambda} is given by the form \eq{softexpansion}.

The soft-collinear function, defined by \eq{Sscdef},  for soft energy outside two back-to-back cones of radius $R$ integrated up to the veto $E<\Lambda$ is given up to $\cO(\as^2)$ by
\begin{align}
\label{eq:SscLambda}
&S_{sc}^\cu(\Lambda R,\mu) = 1 + \frac{\as}{4\pi} \Bigl( -\Gamma_0\ln^2\frac{\mu}{2\Lambda R} + c_{sc}^1\Bigr) \\
&\quad + \Bigl(\frac{\as}{4\pi}\Bigr)^2 \biggl[ \frac{1}{2}\Gamma_0^2 \ln^4\frac{\mu}{2\Lambda R} - \frac{2}{3}\Gamma_0\beta_0\ln^3\frac{\mu}{2\Lambda R} \nn \\
&\qquad \qquad \quad + \Bigl( - \Gamma_1 -  c_{sc}^1\Gamma_0 - \frac{\pi^2}{3}(\Gamma_0)^2\Bigr)\ln^2\frac{\mu}{2\Lambda R} \nn \\
&\qquad \qquad \quad + \bigl( \gamma_{sc}^1 + 2 c_{sc}^1\beta_0 - 4\zeta_3(\Gamma_0)^2\bigr) \ln\frac{\mu}{2\Lambda R} + c_{sc}^2 \biggr]\,, \nn
\end{align}
where $c_{sc}^1 = (\pi^2/6) C_F$ and $\gamma_{sc}^1$ is given in \eq{softnoncusp}. We leave $c_{sc}^2$ undetermined. The Laplace transform of \eq{SscLambda} is given by the form \eq{softexpansion}.

The non-cusp parts of the anomalous dimensions of the csoft, global soft, and soft-collinear functions in \eqss{Sink}{SsLambda}{SscLambda} are all related to each other and the known anomalous dimensions of the hemisphere \cite{Fleming:2007xt,Becher:2008cf} and time-like Drell-Yan \cite{Becher:2007ty} soft functions. They are related to all orders by
\be
\gamma_\text{hemi} = \gamma_\text{in} = -\frac{\gamma_{ss}}{2} = \gamma_{sc} \,,
\ee
given explicitly to $\cO(\as^2)$ by
\begin{align}
\gamma_{sc}^0 &= 0 \\
\gamma_{sc}^1 &= C_F \Bigl[ \Bigl( \frac{1616}{27} - 56\zeta_3\Bigr)C_A - \frac{448}{27}T_F n_f - \frac{2\pi^2}{3}\beta_0\Bigr] \,, \nn
\end{align}
and to $\cO(\as^3)$ in \eq{gammass2}.

The convolution of $S_s \otimes S_{sc}^2$ to give the total unrefactorized veto soft function, integrated up to $E_{\text{out}}=\Lambda$, is given to $\cO(\as^2)$ by \eq{Sveto2}. It has no non-cusp anomalous dimension, but has leftover large logs of $R$.

The unmeasured jet function $J_\text{un}^\alg$ that appears in the total two-jet rate in \eq{naivefactorization} and is constructed according to \eq{Junmeasfromtau}, is given to $\cO(\as)$ for cone-type algorithms in \eq{Juncone1} and for S-W or \kT-type algorithms in \eq{JunkT1}, and to $\cO(\as^2)$ for the thrust cone algorithm by \eq{Junmeas2}:
\begin{align}
&J_\text{un}^\cone(QR,\mu) = 1 + \frac{\as}{4\pi}\Bigl( \Gamma_0 \ln^2\frac{\mu}{QR} + \gamma_J^0 \ln\frac{\mu}{QR} + c_\cone^1\Bigr) \nn \\
&+ \Bigl(\frac{\as}{4\pi}\Bigr)^2 \biggl[ \frac{1}{2}(\Gamma_0)^2\ln^4\frac{\mu}{QR} + \Gamma_0\Bigl(\gamma_J^0 + \frac{2}{3}\beta_0\Bigr)\ln^3\frac{\mu}{QR} \nn \\
&\qquad  + \Bigl( \Gamma_1 + c_\cone^1\Gamma_0  + \frac{1}{2}(\gamma_J^0)^2 + \gamma_J^0 \beta_0\Bigr)\ln^2\frac{\mu}{QR} \nn \\
&\qquad + \Bigl( \gamma_{J\text{un}}^1 + c_\cone^1(\gamma_J^0 + 2\beta_0)\Bigr)\ln\frac{\mu}{QR} + c_\cone^2\biggr] ,
\end{align}
where $c_\cone^1 = (7+ 6\ln2 -5\pi^2/6 )C_F$, and we leave $c_\cone^2$ undetermined. $J_\text{un}$ has a non-cusp anomalous dimension given by
\be
\gamma_{J\text{un}} = \gamma_J + \gamma_\text{in} = -\frac{\gamma_H}{2}\,,
\ee
which is given to $\cO(\as^3)$ by \eq{gammaH}.

The $\cO(\as^2)$ fixed-order jet thrust cross section for the thrust cone algorithm predicted by these results is given in \eq{jetthrustcs2}, and the formula that resums global logs of $\tau$ and $R$ in this cross section is given by \eq{resummedcs}. \eq{conerate2} gives the terms in the fixed-order $\cO(\as^2)$ total two-jet rate for the thrust cone algorithm predicted by our formulas. All the $\log^n R$-enhanced terms and the $\ln (2\Lambda/Q)$ term for the $n_f$ color structure are predicted accurately by this result. The pure $\ln(2\Lambda/Q)$ terms for the other color structures require inclusion of additional subjet operators in our factorization theorem \eq{conejetcs} to be completely predicted.

\section{Conclusions}
\label{sec:summary}

In this paper we have worked out several ramifications of introducing a new mode, the soft-collinear mode, into SCET, allowing the separation of scales $2\Lambda$ and $2\Lambda R$ created by the measurement of a soft veto energy outside cones of radius $R$ in a jet cross section, as well as the resummation of the logs of $R$ due to the ratio of these scales. In the jet thrust cross sections, we found the presence of hard-collinear, collinear-soft (csoft), global (veto) soft, and soft-collinear scales, and leading to the theory \SCETplusplus\ to factorize and resum logarithms of ratios of these scales.

Studying the jet thrust cross section using the thrust cone algorithm, we extracted for the first time the two- and three-loop non-cusp anomalous dimensions of the refactorized soft functions for in-jet measured (csoft) radiation, global soft radiation at the veto scale, and soft-collinear radiation at the veto scale in the cones.

We related the jet thrust cross section and the total two-jet rate, deriving a relation between the measured jet and in-jet soft functions and the ``unmeasured jet function'' in the total two-jet cross section, and as a bonus extracted its two-loop and three-loop non-cusp anomalous dimensions for the first time.

We compared the predictions of the refactorized jet cross section with the output of \texttt{EERAD3} to $\cO(\as^2)$. Finally, we provided a formula the resummed jet thrust cross section (modulo NGLs) and performed global log resummation to NNLL accuracy.

We should emphasize once again, as we began, that our treatment here was not aimed at a resummation of the NGLs that necessarily appear in these exclusive jet thrust or total $n$-jet cross sections. Technology to do this in the framework of SCET is rapidly developing, especially with the recent work of \cite{Larkoski:2015kga,Larkoski:2015zka,Neill:2015nya,Becher:2015hka}. These developments have made clear the necessity to identify subjets or multiple collinear Wilson lines emitting soft radiation into other regions in order to resum NGLs.

What we have explored here is the necessity of introduction of a soft-collinear mode to factorize and resum even the global logs of $R$ arising from the ratio of soft- and soft-collinear scales at $2\Lambda$ and $2\Lambda R$ and various ramifications and insights following from this step. It is the first step towards a full resummation that would include the NGLs. Not only have we factorized and resummed the global logs in the jet thrust and total 2-jet cross sections, we tied together some loose ends and made new connections among ideas in the existing literature. Namely, as mentioned, we clarified the relation between measured jet and in-jet soft functions for jet thrust with the unmeasured jet function of  \cite{Ellis:2009wj,Ellis:2010rw} appearing in total jet cross sections. We identified the in-jet measured soft radiation as actually the csoft modes of \cite{Bauer:2011uc} and its merging with the hard-collinear mode in total jet cross sections as $\tau\to \taumax\sim R^2$. We established on firmer footing earlier conjectures about the proportionality  to the cusp anomalous dimension of $\ln R$ terms in soft anomalous dimensions for jet cross sections, and extracted from the computations of \cite{Kelley:2011aa,vonManteuffel:2013vja} the two-loop anomalous dimensions of the individual in-jet measured soft function, global veto soft function, and soft-collinear functions. We believe the insights here even about the global logs in jet cross sections provide firmer ground on which to build frameworks to resum all logs, global and non-global, in jet cross sections.

\acknowledgments{CL would like to thank the organizers of the SCET 2010 workshop at Ringberg Castle in Germany where the idea of the soft-collinear mode for jet rates was conceived; Simon Freedman, Z. Ligeti, M. Luke, J. Walsh, and S. Zuberi for subsequent discussions that helped fuel its further development; and the Institute for Nuclear Theory at the University of Washington for hospitality during the Program ``Intersections of BSM Phenomenology and QCD for New Physics Searches'' while this paper was being completed. We also thank J. Chay, C. Kim, and I. Kim for discussions on \cite{Chay:2015ila} and the organizers of the SCET 2015 workshop where some of those discussions took place. This work was supported by the U.S. Department of Energy through the Office of Science, Office of Nuclear Physics under Contract DE-AC52-06NA25396 and by an Early Career Research Award, and through the LANL/LDRD Program.}

\appendix

\section{Plus Distributions}
\label{app:plus}

The plus distribution prescription can be used to make a singular function into an integrable distribution. For a function $f$, its plus distribution is defined by (\eg \cite{Ligeti:2008ac}),
\be
\label{eq:plusdef}
\begin{split}
[f(x)]_+ &= \lim_{\epsilon\to 0}\frac{d}{dx} \bigl[ \theta(x-\epsilon)F(x)\bigr] \\
 &= \lim_{\epsilon\to 0} \bigl[ \theta(x-\epsilon)f(x) - \delta(x-\epsilon) F(x)\bigr]\,,
\end{split}
\ee
where
\be
F(x) = \int_1^x dx' f(x')\,.
\ee
Thus, integrals of regular functions against plus distributions obey, for $x\geq 0$,
\be
\begin{split}
&\int_{-\infty}^x dx' [\theta(x')f(x')]_+ g(x) \\
&\quad = \int_0^x dx' f(x')[g(x') - g(0)]\, + g(0) F(x)\,,
\end{split}
\ee
so, \eg
\be
\begin{split}
&\int_{-\infty}^x dx'\Bigl[ \frac{\theta(x')\ln^n x'}{x'}\Bigr]_+ g(x') \\
&\quad = \int_0^x dx' \frac{\ln^n x'}{x'}[g(x') - g(0)] + \frac{\ln^{n+1}x}{n+1}g(0)\,.
\end{split}
\ee

\section{Laplace Transforms}
\label{app:Laplace}

Here we collect results for the Laplace transforms and inverse Laplace transforms ($\iLP$) between the logs
\be
L\equiv \ln \frac{1}{\tau} \,,\qquad \tilde L \equiv \ln(\nu e^{\gamma_E})
\ee
The Laplace transform is defined by
\be
\widetilde F(\nu) \equiv \LP \{ F\} (\nu) = \int_0^\infty d\tau e^{-\nu\tau} F(\tau)\,.
\ee
The transforms of the logs we need in this paper are given by
\begin{subequations}
\label{eq:LPs}
\begin{align}
\LP\bigl\{ 1\bigr\} &= \frac{1}{\nu} \\
\LP\bigl\{ L\bigr\} &= \frac{1}{\nu} \, \tilde L \\
\LP\bigl\{ L^2\bigr\}&= \frac{1}{\nu} \biggl\{ \tilde L^2 + \frac{\pi^2}{6} \biggr\} \\
\LP\bigl\{ L^3\bigr\}  &= \frac{1}{\nu} \biggl\{ \tilde L^3 + \frac{\pi^2}{2} \tilde L + 2\zeta_3 \biggr\} \\
\LP\bigl\{ L^4\bigr\} &= \frac{1}{\nu} \biggl\{ \tilde L^4 + \pi^2 \tilde L^2 + 8\zeta_3 \tilde L  + \frac{3\pi^4}{20} \biggr\}
\,.
\end{align}
\end{subequations}
The inverse Laplace transforms are given by
\be
\iLP \{ \widetilde F\}(\tau) = \int_{\gamma-i\infty}^{\gamma+i\infty}\frac{d\nu}{2\pi i} e^{\nu\tau} \widetilde F(\nu)\,,
\ee
where $\gamma$ lies to the right of all the poles of $\widetilde F$ in the complex plane.
The inverse transforms we need are
\begin{subequations}
\label{eq:iLPs}
\begin{align}
\iLP\Bigl\{\frac{1}{\nu} \Bigr\} &= 1 \\
\iLP\Bigl\{\frac{1}{\nu}\tilde L\Bigr\} &= L \\
\iLP\Bigl\{\frac{1}{\nu}\tilde L^2\Bigr\} &= L^2 - \frac{\pi^2}{6} \\
\iLP\Bigl\{\frac{1}{\nu}\tilde L^3\Bigr\}  &= L^3 - \frac{\pi^2}{2} L - 2\zeta_3 \\
\iLP\Bigl\{\frac{1}{\nu}\tilde L^4\Bigr\}  &= L^4 - \pi^2 L^2 - 8\zeta_3 L  + \frac{\pi^4}{60}
\,.
\end{align}
\end{subequations}
Some other particular transforms we need in the main text that can be derived from \eq{iLPs} are
\begin{align}
\label{eq:iLPL2L2}
&\iLP\biggl\{\frac{1}{\nu} \ln^2 \frac{\mu^2 \nu e^{\gamma_E}}{Q^2}\ln^2 \frac{\mu R \nu e^{\gamma_E}}{Q}\biggr\} \\
&\quad= \ln^2\frac{\mu^2}{Q^2\tau}\ln^2\frac{\mu R}{Q\tau}  - \frac{\pi^2}{2}\Bigl(\ln^2\frac{\mu^2}{Q^2\tau} + \ln^2\frac{\mu R}{Q\tau}\Bigr) \nn \\
&\qquad - 4\zeta_3 \Bigl(\ln\frac{\mu^2}{Q^2\tau} + \ln\frac{\mu R}{Q\tau}\Bigr)  + \frac{\pi^2}{3}\ln^2\frac{\mu}{QR} + \frac{\pi^4}{60} \,, \nn
\end{align}
and
\begin{align}
\label{eq:iLPL1L2}
&\iLP\biggl\{\frac{1}{\nu} \ln \frac{\mu^2 \nu e^{\gamma_E}}{Q^2}\ln^2 \frac{\mu R \nu e^{\gamma_E}}{Q}\biggr\} \\
&\quad= \ln\frac{\mu^2}{Q^2\tau}\ln^2\frac{\mu R}{Q\tau}  - \frac{\pi^2}{6}\Bigl(\ln\frac{\mu^2}{Q^2\tau} + 2\ln\frac{\mu R}{Q\tau}\Bigr)  - 2\zeta_3 \,. \nn
\end{align}

Convolutions involving $\Delta J_\cone^1$ that we need are
\begin{subequations}
\label{eq:DeltaJconerelations}
\begin{align}
&\int_0^{R^2} d\tau \Delta J_\cone^1(\tau,R) = 6\ln 2 \,C_F\,, \\
&\int_0^{R^2}d\tau \iLP\biggl\{ \Delta \tilde J_\cone^1(\nu,R) \ln\frac{\mu^2 \nu e^{\gamma_E}}{Q^2}\biggr\}  \\
&\qquad = \Bigl(12\ln 2 \ln\frac{\mu}{QR} - 3\ln^2 2 + \frac{\pi^2}{2}\Bigr) C_F\,, \nn \\
&\int_0^{R^2}d\tau \iLP\biggl\{ \Delta \tilde J_\cone^1(\nu,R) \ln^2\frac{\mu^2 \nu e^{\gamma_E}}{Q^2}\biggr\}  \\
&\qquad = \Bigl[24\ln 2 \ln^2\frac{\mu}{QR} + (2\pi^2 - 12\ln^2 2)\ln\frac{\mu}{QR} \nn \\
&\qquad\quad  - 2\pi^2 \ln^2 2 + 2\ln^3 2  + \frac{21\zeta_3}{2}\Bigr]C_F\,, \nn \\
&\int_0^{R^2}d\tau \iLP\biggl\{ \Delta \tilde J_\cone^1(\nu,R) \ln^2\frac{\mu R \nu e^{\gamma_E}}{Q}\biggr\}  \\
&\qquad = \Bigl[6\ln 2 \ln^2\frac{\mu}{QR} + (\pi^2 - 6\ln^2 2)\ln\frac{\mu}{QR} \nn \\
&\qquad\quad  - 2\pi^2 \ln^2 2 + 2\ln^3 2  + \frac{21\zeta_3}{2}\Bigr]C_F\,. \nn
\end{align}
\end{subequations}
To derive the last three relations, we used the results for the convolutions in momentum space,
\begin{subequations}
\begin{align}
&\int_0^{R^2}d\tau\int_0^\tau d\tau'\Delta J_\cone^1(\tau-\tau') \Bigl[\frac{1}{\tau'}\Bigr]_+ \\
&\qquad = C_F\Bigl[ -\frac{\pi^2}{2} + 3 \ln^2 2 + 12\ln 2\ln R\Bigr] \nn \\
&\int_0^{R^2}d\tau\int_0^\tau d\tau'\Delta J_\cone^1(\tau-\tau') \Bigl[\frac{\ln\tau'}{\tau'}\Bigr]_+ \\
&\qquad = C_F\Bigl[\ln^3 2 + 6\ln 2\ln R(\ln 2 + 2\ln R) \nn \\
&\qquad\qquad - \frac{\pi^2}{2}(\ln 2 + 2\ln R) + \frac{21\zeta_3}{4}\Bigr]\,. \nn
\end{align}
\end{subequations}
For arbitrary upper limits on the $\tau$ integrals (for any $\tau \leq R^2$), we have the convolutions
\begin{subequations}
\label{eq:DeltaJplus}
\begin{align}
&\int_0^\tau d\tau' \int_0^{\tau'} d\tau'' \Delta J_\cone^1(\tau'-\tau'') \lambda\biggl[\frac{\ln(\lambda\tau'')}{\lambda \tau''}\biggr]_+ \\
&\quad = 3C_F\Bigl[ \ln^2(\lambda\tau)\ln\Bigl( 1+ \frac{\tau}{R^2}\Bigr) \nn \\
&\qquad - 2 \ln(\lambda\tau) \Li_2\Bigl(\frac{\tau}{\tau+R^2}\Bigr) + 2\Li_3\Bigl(\frac{\tau}{\tau+R^2}\Bigr)\Bigr] \nn \\
&\int_0^\tau d\tau' \int_0^{\tau'} d\tau'' \Delta J_\cone^1(\tau'-\tau'') \lambda\biggl[\frac{1}{\lambda \tau''}\biggr]_+ \\
&\quad = 6C_F\Bigl[ \ln(\lambda\tau) \ln\Bigl( 1+ \frac{\tau}{R^2}\Bigr) - \Li_2\Bigl(\frac{\tau}{\tau+R^2}\Bigr) \Bigr] \nn
\end{align}
\end{subequations}

\section{Jet Thrust Soft Function}
\label{app:soft}

The non-Abelian terms of the two-loop jet thrust soft function in \eq{SCFnA} that survive in the limit $R\to 0$ , given in \cite{vonManteuffel:2013vja}, are
\begin{widetext}
\begin{align}
S_{nA\text{\cite{vonManteuffel:2013vja}}}^{(2)}(k,\Lambda, R,\mu) &= C_A C_F\biggl[ -\frac{176}{9} \ln^3\frac{\mu}{k} + \Bigl(-\frac{176\ln R}{3} + \frac{8\pi^2}{3} - \frac{536}{9}\Bigr)\ln^2\frac{\mu}{k} \\
&\qquad\quad + \Bigl( - \frac{176}{3}\ln^2 R + \frac{16}{3}\pi^2\ln R - \frac{1072}{9}\ln R + 56\zeta_3 + \frac{44\pi^2}{9} - \frac{1616}{27}\Bigr)\ln\frac{\mu}{k} \nn \\
&\qquad\quad  +\Bigl( - \frac{176}{3}\ln^2 R - \frac{16}{3}\pi^2\ln R + \frac{1072}{9}\ln R - \frac{44\pi^2}{9} \Bigr)\ln\frac{\mu}{2\Lambda} + \frac{176}{3}\ln R \ln^2\frac{\mu}{2\Lambda} - \frac{8}{3}\pi^2 \ln^2\frac{k}{2\Lambda R^2} \nn \\
&\qquad\quad + \Bigl(-16\zeta_3 - \frac{8}{3} + \frac{88\pi^2}{9}\Bigr)\ln\frac{k}{2\Lambda R^2} - \frac{682\zeta_3}{9} + \frac{109\pi^4}{45}-\frac{1139\pi^2}{54} - \frac{1636}{81} \biggr] \nn \\
&+ C_F T_F n_f \biggl[ \frac{64}{9} \ln^3\frac{\mu}{k} + \Bigl(\frac{64\ln R}{3} + \frac{160}{9}\Bigr)\ln^2\frac{\mu}{k} + \Bigl(  \frac{64}{3}\ln^2 R + \frac{320}{9}\ln R - \frac{16\pi^2}{9} + \frac{448}{27}\Bigr)\ln\frac{\mu}{k} \nn \\
&\qquad\qquad  +\Bigl( \frac{64}{3}\ln^2 R - \frac{320}{9}\ln R + \frac{16\pi^2}{9} \Bigr)\ln\frac{\mu}{2\Lambda}  - \frac{64}{3}\ln R \ln^2\frac{\mu}{2\Lambda} + \Bigl( \frac{16}{3} - \frac{32\pi^2}{9}\Bigr)\ln\frac{k}{2\Lambda R^2}  \nn \\
&\qquad\qquad + \frac{248\zeta_3}{9} + \frac{218\pi^2}{27} - \frac{928}{81} \biggr] - 4\Gamma_1\ln^2 R \,. \nn
\end{align}
\end{widetext}
(We have translated the notation \cite{vonManteuffel:2013vja} from $Q\tau_\omega\to k$, $r\to R^2$, $\omega\to \Lambda$.)
In \ssec{S2} we reorganize this formula into a refactorized form that matches \eq{Skrefact}.

\section{Anomalous Dimensions}
\label{app:anomdim}

The hard function and Laplace transform of the jet and soft functions in \eq{jetthrustcs} have the expansions
\be
\label{eq:Fexp}
 F(\mu) = \sum_{n=0}^\infty \Bigl(\frac{\as}{4\pi}\Bigr)^n F_n\,,
\ee
where
\begin{align}
\label{eq:Fcoeffs}
F_0 &= 1 \\
F_1 &= \frac{\Gamma_F^0}{j_F^2} L_F^2 + \frac{\gamma_F^0}{j_F} L_F + c_F^1 \nn \\
F_2 &= \frac{1}{2j_F^4}(\Gamma_F^0)^2L_F^4  + \frac{\Gamma_F^0}{j_F^3}\Bigl(\gamma_F^0 + \frac{2}{3}\beta_0\Bigr) L_F^3   \nn \\
&\quad + \frac{1}{j_F^2}\Bigl( \Gamma_F^1 + \frac{1}{2}(\gamma_F^0)^2 + \gamma_F^0\beta_0 +  c_F^1 \Gamma_F^0\Bigr)L_F^2   \nn \\
&\quad + \frac{1}{j_F}(\gamma_F^1 +  c_F^1\gamma_F^0 + 2 c_F^1\beta_0)L_F +  c_F^2\,. \nn
\end{align}
For $F=H,J,S$, the logs $L_F$ are of the form
\be
\label{eq:LF}
L_H = \ln\frac{\mu}{Q}\,,\quad L_J = \ln\frac{\mu^{2} \nu e^{\gamma_E}}{Q_J^{2}}\,, \quad L_S = \ln\frac{\mu \nu  e^{\gamma_E}}{Q_S}
\ee
so  $j_J = 2$ and $j_H =j_S= 1$. The denominators $Q_{J,S}$ in the logs depend on the precise definition of the measurement in the jet or soft function.

The expansions in \eq{Fcoeffs} contain the coefficients in the expansions of cusp and non-cusp anomalous dimensions in $\as$, given by
\be
\label{eq:gammaexpansion}
\Gamma_{\text{cusp}}[\as] = \sum_{n=0}^\infty \Bigl(\frac{\as}{4\pi}\Bigr)^n \Gamma_n\,,\ \gamma[\as] = \sum_{n=0}^\infty \Bigl(\frac{\as}{4\pi}\Bigr)^n \gamma_n\,.
\ee
The first few coefficients for the cusp anomalous dimension are given by \cite{Korchemsky:1987wg,Moch:2004pa}
\begin{align}
\label{eq:cuspcoeffs}
\Gamma_0 &= 4C_F \,, \\
\Gamma_1 &= 4C_F\Bigl[ \Bigl(\frac{67}{9} - \frac{\pi^2}{3}\Bigr) C_A - \frac{20}{9}T_F n_f\Bigr]\,, \nn \\
\Gamma_2 &= 4C_F \Bigl[ \Bigl( \frac{245}{6} - \frac{134\pi^2}{27} + \frac{11\pi^4}{45} + \frac{22\zeta_3}{3}\Bigr)C_A^2 \nn \\
&\qquad + \Bigl( - \frac{418}{27} + \frac{40\pi^2}{27} - \frac{56\zeta_3}{3}\Bigr) C_A T_F n_f \nn \\
&\qquad + \Bigl( -\frac{55}{3}+ 16 \zeta_3 \Bigr)C_F T_F n_f-\frac{16}{27} T_F^2 n_f^2\Bigr]\,, \nn
\end{align}
The non-cusp anomalous dimension coefficients we need in this paper are given up to three-loop order for the hard and jet functions by \cite{Moch:2005id,Becher:2006mr}
\begin{widetext}
\begin{align}
\label{eq:gammaH}
\gamma_H^0 &= -12C_F \\
\gamma_H^1 &= -2C_F \Bigl[\Bigl(\frac{82}{9}-52\zeta_3\Bigr) C_A + (3-4\pi^2 + 48\zeta_3)C_F
+ \Bigl(\frac{65}{9} + \pi^2\Bigr)\beta_0\Bigr]\,, \nn\\
\gamma_H^2 &= -4 C_F \Big[ \Big( \frac{66167}{324}- \frac{686\pi^2}{81}- \frac{302 \pi^4}{135}- \frac{782 \zeta_3}{9}+ \frac{44\pi^2 \zeta_3}{9}+136 \zeta_5  \Big) C_A^2 \nn\\
&\qquad\quad +  \Big(\frac{151}{4}-\frac{205\pi^2}{9}-\frac{247\pi^4}{135}+\frac{844\zeta_3}{3}+\frac{8\pi^2 \zeta_3}{3}+120 \zeta_5\Big) C_F C_A \nn\\
& \qquad\quad+ \Big(  \frac{29}{2}+3\pi^2 +\frac{8\pi^4}{5}+68 \zeta_3 -\frac{16\pi^2\zeta_3}{3}-240 \zeta_5\Big) C_F^2 +  \Big(-\frac{10781}{108} +\frac{446\pi^2}{81} +\frac{449\pi^4}{270} -\frac{1166\zeta_3}{9}  \Big)C_A \beta_0 \nn\\
 & \qquad\quad  +  \Big(\frac{2953}{108} -\frac{13\pi^2}{18} -\frac{7\pi^4}{27} +\frac{128 \zeta_3}{9}  \Big)\beta_1 +  \Big( -\frac{2417}{324} +\frac{5\pi^2}{6} +\frac{2 \zeta_3}{3} \Big) \beta_0^2 \Big] \,, \nn
\end{align}
and
\begin{align}
\label{eq:gammaJ}
\gamma_J^0 &= 6C_F \\
\gamma_J^1 &= C_F \Bigl[\Bigl(\frac{146}{9}-80\zeta_3\Bigr) C_A + (3-4\pi^2 + 48\zeta_3)C_F
+ \Bigl(\frac{121}{9} + \frac{2\pi^2}{3}\Bigr)\beta_0\Bigr]\,. \nn\\
\gamma_J^2 &= 2 C_F \Big[ \Big( \frac{52019}{162} -\frac{841\pi^2}{81} -\frac{82\pi^4}{27} -\frac{2056\zeta_3}{9} +\frac{88\pi^2 \zeta_3}{9} +232 \zeta_5 \Big) C_A^2 \nn\\
&\qquad\quad+  \Big( \frac{151}{4}-\frac{205\pi^2}{9}-\frac{247\pi^4}{135}+\frac{844\zeta_3}{3}+\frac{8\pi^2 \zeta_3}{3}+120 \zeta_5\Big) C_F C_A \nn\\
& \qquad\quad+ \Big( \frac{29}{2}+3\pi^2 +\frac{8\pi^4}{5}+68 \zeta_3 -\frac{16\pi^2\zeta_3}{3}-240 \zeta_5 \Big) C_F^2 +  \Big( - \frac{7739}{54}+ \frac{325 \pi^2}{81}+ \frac{617\pi^4}{270}- \frac{1276 \zeta_3}{9}\Big)C_A \beta_0 \nn\\
 & \qquad \quad +  \Big(- \frac{3457}{324}+ \frac{5\pi^2}{9}+ \frac{16 \zeta_3}{3} \Big)\beta_0^2 +  \Big(  \frac{1166}{27}- \frac{8\pi^2}{9}- \frac{41\pi^4}{135}+ \frac{52 \zeta_3}{9}\Big) \beta_1 \Big] \,. \nn
\end{align}
The anomalous dimensions of the csoft, global soft, and soft-collinear functions are all related by \eq{gammaequalities},  $\gamma_{\rm hemi} = \gamma_{\text{in}}=-\gamma_{ss}/2 = \gamma_{sc}$, and can be determined through \eq{tauconsistency} or \eq{HJSconsistency} to be equal to $\gamma_{sc} =  - \gamma_J-\frac{\gamma_H}{2}$, or, explicitly,
\begin{align}
\label{eq:gammass2}
\gamma_{sc}^0  &= 0 \\
\gamma_{sc}^1 &= -C_F \Bigl[ \Big(\frac{64}{9} - 28\zeta_3\Bigr) C_A + \Bigl(\frac{56}{9} - \frac{\pi^2}{3}\Bigr)\beta_0 \Bigr ] \nn \\
\gamma_{sc}^2
&= - C_F\bigg[ C_A^2 \bigg( \frac{37871}{162} - \frac{310 \pi^2}{81}-\frac{8 \pi^4}{5} -\frac{2548 \zeta_3}{9}+ \frac{88\pi^2 \zeta_3}{9} +192 \zeta_5\bigg)
+ C_A \beta_0 \bigg( -\frac{4697}{54}-\frac{242\pi^2}{81}+\frac{56\pi^4}{45}-\frac{220 \zeta_3}{9}\bigg) \nn
\\ & \qquad\quad
+ \beta_1 \bigg( \frac{1711}{54}-\frac{\pi^2}{3}-\frac{4\pi^4}{45}-\frac{152 \zeta_3}{9}\bigg)
+ \beta_0^2  \bigg(-\frac{520}{81}-\frac{5\pi^2}{9}+\frac{28 \zeta_3}{3} \bigg)\bigg] \,. \nn
\end{align}
\end{widetext}

Meanwhile $\as$ itself satisfies
\be
\mu\frac{d\as}{d\mu} = \beta[\as]\,,
\ee
where the beta function has the expansion
\be
\label{eq:betaexpansion}
\beta[\as] = -2\as\sum_{n=0}^\infty \Bigl(\frac{\as}{4\pi}\Bigr)^{n+1} \beta_n\,.
\ee
whose first few coefficients are given by \cite{Tarasov:1980au,Larin:1993tp}
\begin{align}
\label{eq:betacoeffs}
\beta_0 &= \frac{11}{3}C_A - \frac{4}{3}T_F n_f \,, \\
\beta_1 &= \frac{34}{3} C_A^2 - \Bigl(\frac{20}{3}C_A + 4C_F\Bigr)T_F n_f\,, \nn \\
\beta_2 &= \frac{2857}{54}C_A^3 + \Bigl( C_F^2 - \frac{205}{18}C_F C_A - \frac{1415}{54} C_A^2\Bigr) 2T_F n_f \nn \\
&\qquad + \Bigl(\frac{11}{9}C_F + \frac{79}{54}C_A\Bigr) 4 T_F^2 n_f^2\,. \nn
\end{align}
The running coupling $\as(\mu)$ itself is given to three-loop order by
\begin{align}
\as(\mu) &= \as(Q) \biggl\{ X + \as(Q)\frac{\beta_1}{4\pi\beta_0}\ln X \\
&+ \frac{\as^2(Q)}{(4\pi)^2} \biggl[ \frac{\beta_2}{\beta_0}\Bigl( 1- \frac{1}{X}\Bigr) + \frac{\beta_1^2}{\beta_0^2}\Bigl(\frac{\ln X}{X} + \frac{1}{X} - 1\Bigr)\biggr]\biggr\}^{-1} \,, \nn
\end{align}
where $X \equiv 1 + \frac{\as(Q)}{2\pi}\beta_0\ln\frac{\mu}{Q}$.

\bibliography{Rlogs}

\end{document}